\documentclass[twocolumn,showpacs,preprintnumbers,amsmath,amssymb,prb,superscriptaddress,footinbib]{revtex4-2}

\usepackage{mathrsfs}
\usepackage{graphicx}
\usepackage{dcolumn}
\usepackage{bm}
\usepackage{color}
\usepackage{multirow}

\usepackage[colorlinks,bookmarks=false,citecolor=blue,linkcolor=red,urlcolor=blue]{hyperref}
\usepackage{multirow}
\usepackage{float}
\usepackage{bbm}
\usepackage{tablefootnote}

\usepackage{bbold}
\DeclareMathAlphabet{\mymathbb}{U}{BOONDOX-ds}{m}{n}

\usepackage{epsf}
\usepackage{here,times}
\usepackage{amsmath}
\usepackage{makecell}

\newcommand{\RNum}[1]{\uppercase\expandafter{\romannumeral #1\relax}}

\newcommand{\Rom}[1]{ \uppercase\expandafter{\romannumeral#1}}

\usepackage{epsf}
\usepackage{here,times}

\usepackage{amsthm}
\usepackage{thmtools}
\usepackage{blindtext}
\usepackage{physics, bm}
\usepackage{amsfonts, mathtools, amsmath,amssymb,dsfont}
\usepackage{esvect}
\usepackage{enumerate,dcolumn,multirow,bbold, enumitem}
\usepackage{longtable}
\usepackage{array}
\usepackage{tikz-cd}
\usepackage{bbding}
\usepackage[normalem]{ulem}
\usepackage{calligra}

\usepackage{algorithm, algorithmic}
\usepackage{amsmath}

\newcommand{\ttau}{\bm{\tau}}

\definecolor{darkblue}{rgb}{0.15,0.25,0.6}
\definecolor{ZYcolor}{rgb}{0.1,0.5,0.4}
\definecolor{darkred}{rgb}{0.65,0.2,0.15}

\newcommand{\bbb}[1]{\mathbf{#1}}
\newcommand{\ccc}[1]{\mathcal{#1}}
\newcommand{\tutabf}[1]{\tilde{\mathbf{#1}}}
\begin{document}

\title{Superconducting Order Parameters in Spin Space Groups: Methodology and Application}
\author{Xilin Feng}
\affiliation{Department of Physics, Hong Kong University of Science and Technology, Clear Water Bay, Hong Kong, China}
\author{Zhongyi Zhang}
\email{zyzhang@iphy.ac.cn}
\affiliation{Department of Physics, Hong Kong University of Science and Technology, Clear Water Bay, Hong Kong, China}

\begin{abstract}
    We employ a systematic approach to construct superconducting order parameters based on the spin space group.
    Compared to magnetic space groups where spatial and spin rotation of elements are completely locked, the superconducting channels and the forms of basis functions in spin space groups can be different.
    Furthermore, we utilize the obtained basis functions to construct the Ginzburg-Landau free energy and discuss the possible superconducting ground states as well as the interplay between superconducting states and magnetic orders.
    By virtue of those results, we discuss the application of our theory to realistic systems.
    Our work provides a rigorous description of superconducting gap functions in the case of spin space group, and enables further study of the interplay between superconductivity and magnetism.
\end{abstract}
\maketitle

\section{Introduction}
Pairing symmetry, which refers to the symmetry properties of the superconducting gap function, is related to many intriguing properties of the superconductivity, such as magnetic response, nodal properties, topological properties and so on~\cite{sigrist1991phenomenological,van1995phase,sigrist1998time,tsuei2000pairing,frigeri2004spin,mazin2009pairing,qi2011topological,wen2011materials,samokhin2015symmetry,chiu2016classification,sato2017topological,smidman2017superconductivity,leggett2021symmetry,chubukov2012pairing,fischer2023superconductivity}.
At the mean-field level, the gap function is the eigenstate of the linearized gap equation~\cite{sigrist1991phenomenological,frigeri2004superconductivity,sigrist2009introduction}, and the corresponding eigenvalue is related to the transition temperature of the superconductivity represented by this gap function. Due to the property of the eigenvalue equation, the gap functions corresponding to each transition temperature form a set of bases for an irreducible representation (irrep) of the equation's symmetry group.
In particular, when taking crystalline symmetry and spin-orbital coupling (SOC) into account, one should classify the pairing symmetry and construct the gap function according to the irreps of the little co-group at the original point in the Brillouin zone ~\cite{altmann1977induced,woodman1970symmetry,blount1985,ueda1985p,cvetkovic2013space,yang2024symmetry,yang2024representation}.

Furthermore, when the superconductivity coexists with the ferromagnetism~\cite{saxena2000superconductivity,aoki2001coexistence,huxley2001uge2,nishioka2002unusual,samokhin2002possible,huy2007superconductivity,akazawa2004pressure}, it's necessary to include anti-unitary operations to describe the symmetry of these systems~\cite{bradley1968magnetic,lifshitz1998symmetry,litvin2013magnetic}. On the one hand, the gap functions are classified according to the irreducible co-representations (coirreps) of the magnetic point group ~\cite{vpmineev2002,garcia2010pairing,mineev2014superconductivity,mineev2017superconductivity}.
On the other hand, the non-zero net magnetization in a ferromagnetic material splits the Kramers degeneracy of the Fermi surface, naturally favoring nonunitary pairing states~\cite{balian1963superconductivity,leggett1975theoretical,mineev1999introduction}.
Recently, the breaking of Kramers degeneracy has been observed in some exotic materials that exhibit zero net magnetization and lack SOC~\cite{PhysRevB.102.014422,vsmejkal2020crystal,reichlova2020macroscopic,yuan2021prediction,feng2022anomalous,bai2022observation,karube2022observation,vsmejkal2022emerging,mazin2022altermagnetism,bose2022tilted,liu2022chiral,mazin2023altermagnetism,betancourt2023spontaneous,steward2023dynamic,zhou2024crystal,hariki2024x,lee2024broken,vsmejkal2022beyond}.
These materials, termed altermagnets, can exhibit distinct spin textures on their Fermi surfaces compared to ferromagnetic materials.
The symmetry group of altermagnets includes operations in real space, spin space, and combinations of operations in both. This type of symmetry group is called the spin space group~\cite{brinkman1966theory,litvin1977spin}, 
where the pairing symmetry and gap function should be determined by the corresponding spin point group at the original point in the Brillouin zone. 

In the past few years, the unconventional superconductivity in materials described by spin space group has garnered significant interest~\cite{mazin2022notes,papaj2023andreev,fernandes2024topological,chourasia2024thermodynamic,mazin2024altermagnetism,reimers2024direct,zyuzin2024magnetoelectric} due to its potential to exhibit a variety of intriguing properties, such as the topological properties~\cite{zhu2023topological,Yang2024,ghorashi2024altermagnetic,brekke2023two,heung2024probing,hong2024unconventional,wei2024gapless,maeland2024many,zhu2024field}, superconducting diode effect~\cite{banerjee2024altermagnetic,chakraborty2024perfect,sim2024pair}, and so on.
However, while superconductivity in altermagnets and other materials described by spin space groups has been widely discussed theoretically, a systematic symmetry classification of superconducting order parameters in these materials is still lacking.
In this work, we employ a systematic method to construct basis functions for each coirrep of the three types of spin point groups, including: collinear, coplanar, and noncoplanar~\cite{litvin1974spin,litvin1977spin,xiao2023spin,ren2023enumeration,jiang2023enumeration,schiff2023spin,PhysRevX.12.021016,chen2023spin}. Compared to the superconducting channels and basis functions discussed in point groups~\cite{sigrist1991phenomenological} and magnetic point groups~\cite{mineev2017superconductivity}, those in spin point groups can exhibit distinct characteristics.
With the free energy constructed by the corresponding basis functions, 
the possible superconducting ground states will also be distinct. Furthermore, the coupling terms between magnetic orders and superconductivity in the free energy can also be constructed, reflecting the interplay between superconductivity and magnetic order.
In addition, the methodology outlined in this paper is applicable to realistic systems, such as electric-field-gated monolayer FeSe with checkerboard magnetic order~\cite{mazin2023induced} and the kagome lattice with all-in-all-out magnetic order~\cite{sachdev1992,hayami2020spontaneous,hayami2020bottom,Yang2024}. 
Our work provides a rigorous description of superconducting gap functions among a wide variety of materials, and facilitates further investigation of superconducting properties in magnetic materials described by spin space groups.

This paper is organized as follows: In Sec.~\ref{method}, we introduce a general approach for constructing basis functions in spin point groups. In Sec.~\ref{collinear}, we use two collinear spin point groups as examples and construct basis functions according to their coirreps. In Sec.~\ref{coplanar}, we construct basis functions for a coplanar spin point group and compare them with those from the magnetic point group that is isomorphic to the chosen coplanar spin point group. In Sec.~\ref{noncoplanar}, we analyze two noncoplanar spin point groups by using the same approach as in Sec.~\ref{coplanar}. In Sec.~\ref{GL}, we use the basis functions from one of the collinear spin point groups to construct the Ginzburg-Landau free energy and discuss the possible superconducting ground states as well as the interplay between superconducting states and magnetism. In Sec.~\ref{application}, we discuss the applications of our theory to two realistic systems. In Sec.~\ref{con}, we present a brief conclusion and outline directions for further exploration based on our work.

\section{Constructing the superconducting order parameters in spin point group} \label{method}

We start by introducing the transformation properties of the gap function and the symmetry properties of the normal state Hamiltonian in the spin space group $\ccc{X}$. The group elements in spin space group are generally written as $[Y\hat{g}_{s}||\hat{g}_{r}]$. To the left of the double bar, $Y$ can be the identity operation $E$ or the time-reversal operation $\ccc{T}$. The time-reversal operation $\ccc{T}$ reverses both spin and momentum simultaneously. The $\hat{g}_{s}$ acts on the spin degree of freedom and it is a spin-rotation operation. To the right of the double bar, $\hat{g}_{r}$ acts on the spatial degree of freedom and can be written as $\{R| {\ttau}\}$, where $R$ is a point group operation and $ {\ttau}$ represents a translation by a vector $\ttau$. 

The Hamiltonian for superconductivity in materials with symmetry described by the spin space group $\ccc{X}$ can be expressed as: $H_{sc}=H_{n}+\hat{\Delta}$.
$H_{n}$ represents the normal state Hamiltonian including magnetic orders and can be written as:
\begin{equation}
    H_{n}=\sum_{\bbb{k},\sigma,\sigma'}c_{\bbb{k}\sigma}^{\dagger}H_{0}(\bbb{k})_{\sigma,\sigma'}c_{\bbb{k}\sigma'},
\end{equation}
where $\hat{c}_{\bbb{k},\sigma}^{\dagger}$ and $\hat{c}_{\bbb{k},\sigma}$ are creation and annihilation operators for an electron with spin $\sigma$ and momentum $\bbb{k}$. For any group element $[Y\hat{g}_{s}||\hat{g}_{r}]$ belonging to $\ccc{X}$, the transformation properties of $H_{0}(\bbb{k})$ under this element can be derived from the transformation properties of the electron creation (annihilation) operator $\hat{c}_{\bbb{k},\sigma}^{\dagger} (\hat{c}_{\bbb{k},\sigma})$ under the same group element. Since the symmetry operation $\hat{g}_{r}$ only acts on the spatial degree of freedom, it only affects the momentum $\bbb{k}$ of the electron creation operator $\hat{c}_{\bbb{k},\sigma}^{\dagger}$ and annihilation operator $\hat{c}_{\bbb{k},\sigma}$:
\begin{equation}
\begin{aligned}
        \hat{g}_{r}\hat{c}_{\bbb{k},\sigma}^{\dagger}\coloneq\{{R}|\ttau\}\hat{c}_{\bbb{k},\sigma}^{\dagger}=e^{i\bbb{k}\cdot\ttau}\hat{c}^{\dagger}_{D({R})\bbb{k},\sigma},\\
        \hat{g}_{r}\hat{c}_{\bbb{k},\sigma}\coloneq\{{R}|\ttau\}\hat{c}_{\bbb{k},\sigma}=e^{-i\bbb{k}\cdot\ttau}\hat{c}_{D({R})\bbb{k},\sigma},
\end{aligned}
\end{equation}
where $D({R})$ is the faithful representation of the point group element ${R}$. 

On the other hand, the spin operation $\hat{g}_s$ acts on the spin $\sigma$ of the electron creation operator $\hat{c}_{\bbb{k},\sigma}^{\dagger}$ and annihilation operator $\hat{c}_{\bbb{k},\sigma}$:
\begin{equation}
\begin{aligned}
    \hat{g}_{s}\hat{c}_{\bbb{k},\sigma}^{\dagger}\coloneq D_{1/2}(\hat{g}_{s})_{\sigma\sigma'} \hat{c}_{\bbb{k},\sigma'}^{\dagger},\\
    \hat{g}_{s}\hat{c}_{\bbb{k},\sigma}\coloneq [D_{1/2}(\hat{g}_{s})_{\sigma\sigma'}]^{*} \hat{c}_{\bbb{k},\sigma'},
\end{aligned}
\end{equation}
where $D_{1/2}(\hat{g}_{s})$ represents the spin-$\frac{1}{2}$ rotation about the axis $\bbb{n}$ by an angle of $\phi$ degrees:  $D_{1/2}(\hat{g}_{s})=e^{-i\frac{\phi}{2}\bbb{n}\cdot\bm{\sigma}}$. Here, $\sigma_{i}~(i=1,2,3)$ are the Pauli matrices. The time-reversal operator $\ccc{T}$ is expressed as: $i\sigma_{2} \ccc{K}$, where $\ccc{K}$ is the conjugate operator.

Therefore, we conclude that $H_{0} (\bbb{k})$ satisfies the symmetry condition under the group elements $[Y\hat{g}_{s}||\hat{g}_{r}]$ as:
\begin{equation}\label{symmH0}
\begin{aligned}
        D_{1/2}(\hat{g}_{s}) H_{0}(D^{-1}(\hat{g}_{r})\bbb{k})& D_{1/2}^{\dagger}(\hat{g}_{s}) \\=& \begin{cases}
     H_{0}(\bbb{k}) & \text{ if $Y=E$} \\  \sigma_{2}H_{0}^{*}(-\bbb{k})\sigma_{2} & \text{ if $Y=\ccc{T}$.}
    \end{cases}
\end{aligned}
\end{equation}
Additionally, the magnetic order part of the normal state Hamiltonian can be expressed as:
\begin{equation}
H_{\text{mag}}=\sum_{\bbb{k},s s'}c^{\dagger}_{\bbb{k},s}(\bbb{m}(\bbb{k})\cdot \bm{\sigma})_{s s'} c_{\bbb{k},s'}.
\end{equation}
For a given spin point group, all the symmetry-allowed magnetic order parameter $\bbb{m}(\bbb{k})$ can be derived by using Eq.~(\ref{symmH0}).

$\hat{\Delta}$ is the pairing term in the Hamiltonian for superconductivity and can be written as:
\begin{equation}\label{pterm}
  \hat{\Delta}=\sum_{\bbb{k},\sigma\sigma'}\hat{c}_{\bbb{k},\sigma}^{\dagger}\Delta(\bbb{k})_{\sigma\sigma'}\hat{c}_{-\bbb{k},\sigma'}^{\dagger}+h.c..
\end{equation}
The transformation properties of the gap function $\Delta(\bbb{k})$ under group elements $[Y\hat{g}_{s}||\hat{g}_{r}]$ can also be derived from the transformation properties of the the electron creation operator $\hat{c}_{\bbb{k},\sigma}^{\dagger}$ under the same group elements. From Eq.~(\ref{pterm}), we find that the translation operation $\ttau$ does not change the gap function because the gap function $\Delta(\bbb{k})$ is the coefficient of two electron creation operators with opposite momenta. Therefore, throughout this paper, we classify the pairing symmetry and construct the gap function of the superconductivity based on the little-group at original point in the Brillouin zone, which is isomorphic to a spin point group. Unless otherwise specified, the spatial symmetry operation $\hat{g}_{r}$ should be understood as a point group operation hereafter.

In the Nambu basis, the gap function $\Delta(\bbb{k})$ can be parameterized by the vector $\bbb{d}(\bbb{k})$ and the scalar $\psi(\bbb{k})$:
\begin{equation}\label{eq_gappar}
    \Delta(\bbb{k})=\left(\bbb{d}(\bbb{k})\cdot \bm{\sigma}+\psi(\bbb{k})\right)i\sigma_{2}.
\end{equation}
The vector $\bbb{d}(\bbb{k})$ and the scalar $\psi(\bbb{k})$ are referred to as superconducting order parameters. The transformations of the gap function $\Delta(\bbb{k})$ under time-reversal, spin-rotation, and point group operations can be expressed through the corresponding transformations of the superconducting order parameters (see Appendix.~\ref{app-a} for details). Accordingly, the transformations of the gap function under spin point group elements can be expressed as the transformations of the superconducting order parameters:
\begin{equation}
    \begin{aligned}\label{d-phi}
        [Y\hat{g}_{s}||\hat{g}_{r}]\bbb{d}(\bbb{k})&=\begin{cases}R_{\bbb{n}}(\phi)\bbb{d}(D^{-1}(\hat{g}_{r})\bbb{k}) &\text{ if $Y=E$} \\ [R_{\bbb{n}}(\phi)\bbb{d}(D^{-1}(\hat{g}_{r})\bbb{k})]^{*} &\text{ if $Y=\ccc{T}$} \end{cases}\\
        [Y\hat{g}_{s}||\hat{g}_{r}]\psi(\bbb{k})&=\begin{cases}
        \psi(D^{-1}(\hat{g}_{r})\bbb{k})    &\qquad\quad\ \text{ if $Y=E$}\\
        \psi^{*}(D^{-1}(\hat{g}_{r})\bbb{k})   &\qquad\quad\ \text{ if $Y=\ccc{T}$}
        \end{cases}.
    \end{aligned}
\end{equation}
In Eq.~(\ref{d-phi}), $R_{\bbb{n}}(\phi)$ denotes the three-dimensional rotation about the axis $\bbb{n}$ by an angle of $\phi$ degrees, where $\bbb{n}$ and $\phi$ are the same as those in spin-$\frac{1}{2}$ rotation $D_{1/2}(\hat{g}_{s})=e^{-i\frac{\phi}{2}\bbb{n}\cdot\bm{\sigma}}$.

In addition, the gap functions corresponding to each transition temperature are eigenstates of the linearized gap equation, thus forming a set of bases for a coirrep of the corresponding spin point group.
Therefore, each coirrep corresponds to an individual superconducting channel, allowing us to label the superconducting channels by the coirreps $\Gamma^{i}$ of the spin point group $X$.

To construct the superconducting order parameters for each superconducting channel, we define the basis function $\xi(\bbb{k},\bbb{r})$ as:
\begin{equation}\label{basisf}
    \xi(\bbb{k},\bbb{r})= \bbb{d}(\bbb{k}) \cdot \bbb{r}+\psi(\bbb{k}),
\end{equation}
where $\bbb{r}=(\tutabf{x},\tutabf{y},\tutabf{z})$ is a vector formed by axial-vectors $\tutabf{x},\tutabf{y}$ and $\tutabf{z}$. 
According to the transformation properties of the superconducting order parameters in Eq.~(\ref{d-phi}), the transformation of the the basis function $\xi(\bbb{k},\bbb{r})$ under the linear operator $\hat{P}_{[Y\hat{g}_{s}||\hat{g}_{r}]}$ is written as:
\begin{equation}\label{basistrans}
\begin{aligned}
    \hat{P}_{[Y\hat{g}_{s}||\hat{g}_{r}]}\xi(\bbb{k},&\bbb{r})\\
    =&\begin{cases}\xi(D^{-1}(\hat{g}_{r})\bbb{k},R_{\bbb{n}}^{-1}(\phi)\bbb{r}) &\text{ if $Y=E$}\\
\xi^{*}(D^{-1}(\hat{g}_{r})\bbb{k},R_{\bbb{n}}^{-1}(\phi)\bbb{r}) &\text{ if $Y=\ccc{T}$}
\end{cases}.
\end{aligned}
\end{equation}
Next, we will define the projection operator for each coirrep of the spin point groups by using the operators $\hat{P}_{[\hat{g}_{s}||\hat{g}_{r}]}$ and the representation matrices $D([\hat{g}_{s}||\hat{g}_{r}])$ of the unitary group elements. With the transformation properties shown in Eq.~(\ref{basistrans}), the basis functions for each coirrep can be obtained. Then, the superconducting order parameters for each coirrep are expressed as linear combinations of the basis functions for the corresponding coirrep.

\begin{figure*}
        \centering
        \includegraphics[width=0.95\textwidth]{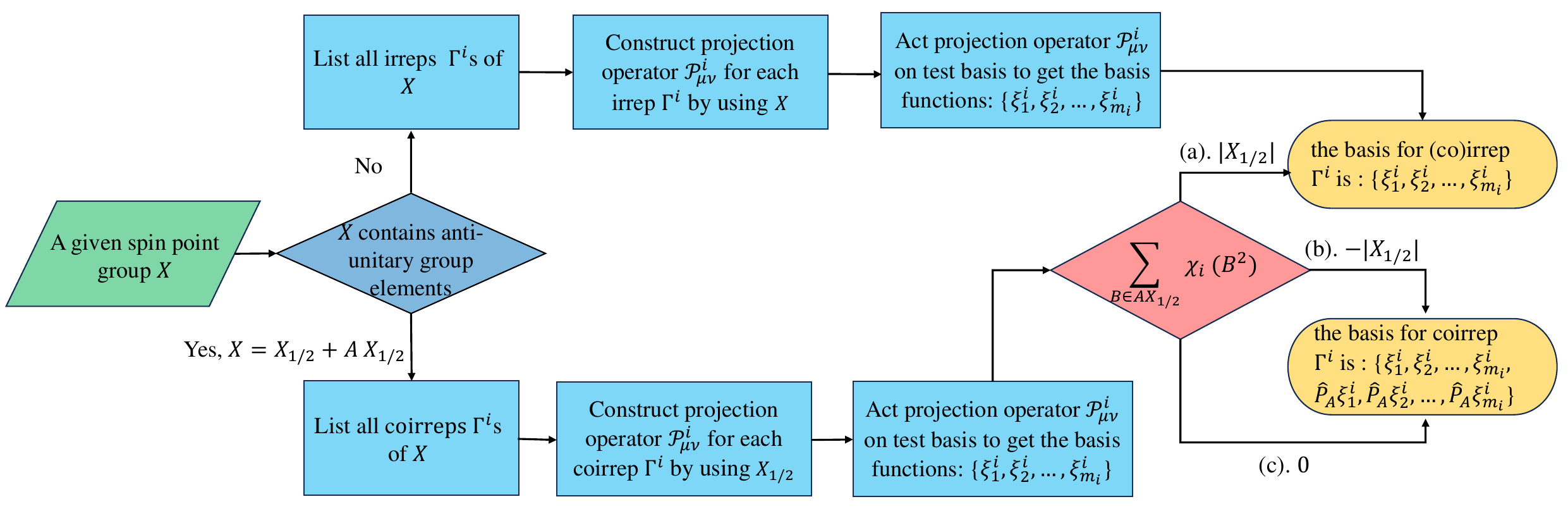}
        \caption{
        The flowchart for constructing the basis functions for each coirrep $\Gamma^{i}$ of the spin point group $X$. When $X$ contains anti-unitary group elements, $X_{1/2}$ represents the unitary part of $X$, and $A$ is an element belonging to the anti-unitary coset. $\hat{\ccc{P}}^{i}_{\mu\nu}$ is the projection operator corresponding to the coirrep $\Gamma^{i}$, as shown in Eq.~(\ref{proj}) or Eq.~(\ref{proj2}).}
        \label{flowchart}
    \end{figure*}

Spin point groups can be divided into two categories based on whether they contain anti-unitary group elements.
When the spin point group $X$ does not contain any anti-unitary group elements, $X$ must be noncoplanar spin point group which describes the symmetry of systems where the magnetic moments $\bbb{M}(\bbb{r})$ are not confined to a single plane. And $X$ is isomorphic to a point group. To find the basis functions for each irrep $\Gamma^{i}$ of $X$, we define a projection operator~\cite{hamermesh2012group}:
\begin{equation}\label{proj}
    \hat{\ccc{P}}^{i}_{\mu\nu}=\frac{m_{i}}{|X|}\sum_{S\in X}[D_{\mu\nu}^{i}(S)]^{*}\hat{P}_{S},
\end{equation}
where $|X|$ is the order of the group $X$, $D^{i}(S)$ is the representation matrix of the group element $S$ for the $i$-th irrep $\Gamma^{i}$, and $\mu$, $\nu$ are the row and column indices, respectively. $\hat{P}_{S}$ is the linear operator for group element $S$, and the transformation of the basis functions under this linear operator is defined in Eq.~(\ref{basistrans}). $m_{i}$ is the dimension of the $i$-th irrep. Then, we generate the test basis functions, which are polynomials formed by the components of the vectors $\bbb{k}$ and $\bbb{r}$, such as $k_{x}^2$, $k_{y}^{3}\tutabf{x}$, and $k_{x}\tutabf{y}+k_{y}\tutabf{x}$.
By acting the projection operator $\hat{\ccc{P}}^{i}_{\mu\nu}$ on these test basis functions, we obtain the basis functions $\{\xi^{i}_{1},\xi^{i}_{2},...,\xi^{i}_{m_{i}}\}$ for the irrep $\Gamma^{i}$ of $X$. These basis functions satisfy the condition:
\begin{equation}
\hat{P}_{S}\xi^{i}_{\nu}=\sum_{\mu}\xi^{i}_{\mu}D^{i}_{\mu\nu}(S),
\end{equation}
where the $\xi^{i}_{\nu}$ is the $\nu$-th basis function of the $i$-th irrep $\Gamma^{i}$.

When the spin point group $X$ contains anti-unitary group elements, it can be written as:
\begin{equation}
    X=X_{1/2}+AX_{1/2},
\end{equation}
where the $X_{1/2}$ is the subgroup of $X$, consisting of all unitary group elements that belong to $X$, and $A$ is an arbitrary anti-unitary element belonging to the group $X$. In this case, we define the projection operator for each coirrep $\Gamma^{i}$ of $X$ as:
\begin{equation}\label{proj2}
    \hat{\ccc{P}}^{i}_{\mu\nu}=\frac{m_{i}}{|X_{1/2}|}\sum_{S\in X_{1/2}}[D_{\mu\nu}^{i}(S)]^{*}\hat{P}_{S},
\end{equation}
where $m_{i}$ is the dimension of the coirrep $\Gamma^{i}$. Note that if $X$ belongs to the collinear spin point groups which describes the symmetry of systems where the magnetic moments $\bbb{M}(\bbb{r})$ are either aligned or anti-aligned along the same direction, the order of the group is infinite. The reason is that rotating the system about the axis aligned with the magnetic moments by any angle $\theta$ does not change the system. The summation in Eq.~(\ref{proj2}) should be replaced by integration over the angle $\theta$ and summation of the discrete elements of $X_{1/2}$, then normalized by $2\pi$. 

Moreover, when $X$ contains anti-unitary group elements, to find the basis functions for $\Gamma^{i}$, in addition to acting the projection operators on the test basis functions, we need to determine the type of coirrep $\Gamma^{i}$ by using the Dimmock test~\cite{dimmock1963representation}:
\begin{equation}\label{Dimmock}
    \sum_{B\in AX_{1/2}}\chi(B^{2})=\begin{cases}
        |X_{1/2}| &\text{(a)}\\
        -|X_{1/2}| &\text{(b)}\\
        0 &\text{(c)}\\
    \end{cases},
\end{equation}
where $\chi(R)$ represents the character of the unitary group element $R$. For the case that the coirrep $\Gamma^{i}$ belongs to type (a), the basis functions are $\{\xi^{i}_{1},\xi^{i}_{2},...,\xi^{i}_{m_{i}}\}$ which are only determined by the unitary part of the group. While for the case that coirrep $\Gamma^{i}$ belongs to type (b) or (c), the basis functions are $\{\xi^{i}_{1},\xi^{i}_{2},...,\xi^{i}_{m_{i}},\hat{P}_{A}\xi^{i}_{1},\hat{P}_{A}\xi^{i}_{2},...,\hat{P}_{A}\xi^{i}_{m_{i}}\}$, where $A$ is an anti-unitary element in the group $X$, and $\hat{P}_{A}$ is the corresponding operator. Thus, for the coirreps belonging to type (b) or (c), the number of the basis functions is doubled compared to that obtained from the unitary part of the group, and the basis functions are determined by both unitary and anti-unitary parts of the group. The flowchart for constructing the basis functions is shown in Fig.~\ref{flowchart}.

\section{Superconductivity in collinear spin point group}\label{collinear}
Firstly, we discuss the superconducting order parameters in collinear spin space group. As mentioned in Sec.~\ref{method}, we can classify the pairing symmetry and construct superconducting order parameters according to the corresponding collinear spin point group. The collinear spin point group is the symmetry group that characterizes systems in which the spins (or magnetic moments $\bbb{M}(\bbb{r})$) are arranged such that their directions are all parallel to one another, either in the same or opposite directions. In general, the collinear spin point group $X$ can be written as the direct product of the spin-only group $G_{\text{SS}}^{l}$ and the non-trivial spin point group $G_{\text{NS}}$~\cite{xiao2023spin,ren2023enumeration,jiang2023enumeration,schiff2023spin}:
\begin{equation}
    X=G_{\text{SS}}^{l} \times G_{\text{NS}} ,
\end{equation}
where the non-trivial spin point group $G_{\text{NS}}$ does not contain any pure spin operations $[Y\hat{g}_{s}||E]$,
while the spin-only group $G_{\text{SS}}^{l}$ contains only pure spin operations. Since all the symmetry operations in a collinear spin point group preserve the relative orientations of the magnetic moments $\bbb{M}(\bbb{r})$ along a single axis, the group structure of $G_{\text{SS}}^{l}$ can be expressed as:
\begin{equation}\label{Gss}
    G_{\text{SS}}^{l}=\{[E||E],[\ccc{T}R_{\bbb{n}_{\bot}}(\pi)||E]\}\ltimes \text{SO}(2).
\end{equation}
In Eq.~(\ref{Gss}), $\ccc{T}$ is the time-reversal operation, $\bbb{n}_{\bot}$ is an arbitrary axis which is perpendicular to the direction of $\bbb{M}(\bbb{r})$, and $R_{\bbb{n}_{\bot}}(\pi)$ is the rotation about the axis $\bbb{n}_{\bot}$ by an angle of $\pi$. $\text{SO}(2)$ represents the two-dimensional rotation group in the plane which is perpendicular to the direction of $\bbb{M}(\bbb{r})$. Without loss of generality, we fix the direction of $\bbb{M}(\bbb{r})$ along the $\pm \bbb{z}$ axis for collinear spin point groups throughout this paper. Thus, the elements in $\text{SO}(2)$ are represented by $R_{\bbb{z}}(\theta)$, where $\theta \in [0,2\pi)$. Additionally, we choose the aixs $\bbb{n}_{\bot}=~\bbb{x}$ in this paper.

The collinear spin point groups can be divided into two types based on their non-trivial spin point group part $G_{\text{NS}}$. If $G_{\text{NS}}$ does not contain any anti-unitary group elements, $X$ is classified as a type-I collinear spin point group; otherwise, it is classified as a type-II collinear spin point group~\cite{xiao2023spin,schiff2023spin}. Next, we will discuss the superconducting channels and basis functions by taking one example from each type of collinear spin point group.

\begin{table}[t]
    \centering
    \begin{tabular}{>{\centering\arraybackslash}m{1.5cm}|>{\centering\arraybackslash}m{6.5cm}}
    \hline\hline
    \multicolumn{2}{c}{$G_{\text{SS}}^{l}\times ~^{1}4/ ^{1}m^{1}m^{1}m$}\\
    \hline
     coirrep    & basis functions \\ \hline
     \multicolumn{2}{c} {$\sigma=0$} \\\hline
     $\Gamma^{1}_{\sigma=0}$  &   $k_{x}^2+k_{y}^{2}$; $k_{z}^2$.  \\
     $\Gamma^{2}_{\sigma=0}$  &   $k_{x}k_{y}(k_{x}^2-k_{y}^2)k_{z}\tutabf{z}$.  \\
     $\Gamma^{3}_{\sigma=0}$  &   $k_{x}k_{y}(k_{x}^2-k_{y}^2)$.  \\
     $\Gamma^{4}_{\sigma=0}$  &   $k_{z}\tutabf{z}$. \\
     $\Gamma^{5}_{\sigma=0}$  &   $k_{x}^2-k_{y}^2$.  \\
     $\Gamma^{6}_{\sigma=0}$  &   $k_{x}k_{y}k_{z}\tutabf{z}$.  \\
     $\Gamma^{7}_{\sigma=0}$  &   $k_{x}k_{y}$.  \\
     $\Gamma^{8}_{\sigma=0}$  &   $(k_{x}^2-k_{y}^2)k_{z}\tutabf{z}$.  \\
     $\Gamma^{9}_{\sigma=0}$ &  \{$k_{x}k_{z}+ik_{y}k_{z}$, $-k_{x}k_{z}+ik_{y}k_{z}$\}. \\
     $\Gamma^{10}_{\sigma=0}$  &  \makecell{ \{$(k_{x}+ik_{y})\tutabf{z}$, $(k_{x}-ik_{y})\tutabf{z}$\}.}   \\ \hline
     \multicolumn{2}{c} {$\sigma=1$} \\\hline
     $\Gamma^{1}_{\sigma=1}$  &   -  \\
     $\Gamma^{2}_{\sigma=1}$  &   $k_{x}k_{y}(k_{x}^2-k_{y}^2)k_{z}(\tutabf{x}-i\tutabf{y})$.  \\
     $\Gamma^{3}_{\sigma=1}$  &   -  \\
     $\Gamma^{4}_{\sigma=1}$  &   $k_{z}(\tutabf{x}-i\tutabf{y})$. \\
     $\Gamma^{5}_{\sigma=1}$  &   -  \\
     $\Gamma^{6}_{\sigma=1}$  &   $k_{x}k_{y}k_{z}(\tutabf{x}-i\tutabf{y})$.  \\
     $\Gamma^{7}_{\sigma=1}$  &   -  \\
     $\Gamma^{8}_{\sigma=1}$  &   $(k_{x}^2-k_{y}^2)k_{z}(\tutabf{x}-i\tutabf{y})$.  \\
     $\Gamma^{9}_{\sigma=1}$ &   -  \\
     $\Gamma^{10}_{\sigma=1}$  &   \makecell{\{$(k_{x}+ik_{y})(\tutabf{x}-i\tutabf{y})$, $(k_{x}-ik_{y})(\tutabf{x}-i\tutabf{y})$\}.} \\
      \hline
          \multicolumn{2}{c} {$\sigma=-1$} \\\hline
     $\Gamma^{1}_{\sigma=-1}$  &   -  \\
     $\Gamma^{2}_{\sigma=-1}$  &   $k_{x}k_{y}(k_{x}^2-k_{y}^2)k_{z}(\tutabf{x}+i\tutabf{y})$.  \\
     $\Gamma^{3}_{\sigma=-1}$  &   -  \\
     $\Gamma^{4}_{\sigma=-1}$  &   $k_{z}(\tutabf{x}+i\tutabf{y})$. \\
     $\Gamma^{5}_{\sigma=-1}$  &   -  \\
     $\Gamma^{6}_{\sigma=-1}$  &   $k_{x}k_{y}k_{z}(\tutabf{x}+i\tutabf{y})$.  \\
     $\Gamma^{7}_{\sigma=-1}$  &   -  \\
     $\Gamma^{8}_{\sigma=-1}$  &   $(k_{x}^2-k_{y}^2)k_{z}(\tutabf{x}+i\tutabf{y})$.  \\
     $\Gamma^{9}_{\sigma=-1}$ &   -  \\ 
     $\Gamma^{10}_{\sigma=-1}$  &   \makecell{\{$(k_{x}+ik_{y})(\tutabf{x}+i\tutabf{y})$, $(k_{x}-ik_{y})(\tutabf{x}+i\tutabf{y})$\}.} \\
     \hline
    \end{tabular}
    \caption{The basis functions for the collinear spin point group $G_{\text{SS}}^{l}\times ~^{1}4/ ^{1}m^{1}m^{1}m$. In this table, $\sigma~(\in \mathbb{Z})$ is an integer that labels the $z$-component of spin angular momentum of the Cooper pairs in the superconducting state. Thus, for charge-2e case, the value of $\sigma$ can only be $0,\pm 1$. For each coirrep, we present the basis functions with the lowest orbital angular momentum in this channel and omit the higher-order terms.}
    \label{bases-coll}
\end{table}

\subsection{The type-I collinear spin point group $G_{\text{SS}}^{l}\times ~^{1}4/ ^{1}m^{1}m^{1}m$}
\label{type-I-coll}
We choose $G_{\text{SS}}^{l}\times ~^{1}4/ ^{1}m^{1}m^{1}m$ as an example of type-I collinear spin point groups.
The generators and coirreps of the spin point group $G_{\text{SS}}^{l}\times ~^{1}4/ ^{1}m^{1}m^{1}m$ are shown in Table.~\ref{co-irrep-coll}. With the approach shown in the flowchart Fig.~\ref{flowchart}, we can derive the basis functions for each coirrep. The results are presented in Table.~\ref{bases-coll}. Note that in the tables presented in this paper, we list the basis functions with the lowest orbital angular momentum in each channel and omit the higher-order terms.

From Table.~\ref{bases-coll}, we can find that, similar to the case in centrosymmetric space groups, the spin-singlet and triplet pairing states are decoupled due to the presence of the inversion symmetry $[E||I]$. However, there is no magnetic point group which is isomorphic to the type-I collinear spin point group. Therefore, the classification of the superconducting channels and basis functions are different from that of any magnetic point group. One important distinction between type-I collinear spin point groups and magnetic point groups is that, in type-I collinear spin point groups, the basis functions representing superconducting states with Cooper pairs having different $z$-components of spin angular momentum belong to different superconducting channels. These different $z$-components of spin angular momentum are labeled by $\sigma$, as shown in Table.~\ref{bases-coll} (see Appendix.~\ref{magnetic_moment} for detail). Here, we only classify the pairing symmetry for charge-2e superconducting states, where the Cooper pairs' spin angular momentum can only be $0$ or $\hbar$. Correspondingly, the value of $\sigma$ can only be $0$ or $\pm 1$.

In the $\sigma=0$ case, the representation matrices of the group element $[R_{z}(\theta)||E]$ have a trivial phase factor, indicating the $z$-component of the Cooper pair's spin angular momentum is zero~\cite{volovik1984gyromagnetism,fang2017topological,venderbos2018pairing,machida2021nonunitary,brison2021p,yarzhemsky2021multiplicity} (see Appendix.~\ref{magnetic_moment} for detail). Moreover, the representation matrices of the group elements $[R_{z}(\theta)||E]$ are identity matrices, which means the operator $\hat{P}_{[R_{z}(\theta)||E]}$ does not change the basis functions of these channels with $\sigma=0$. Therefore, the superconductivity in channels with $\sigma=0$ cannot be nonunitary triplet pairing states. In addition, for each coirrep of spin point group $G_{\text{SS}}^{l}\times ~^{1}4/ ^{1}m^{1}m^{1}m$ in $\sigma=0$ case, there is a corresponding coirrep of the type-II magnetic point group $4/mmm1'$, which is homomorphic to it. However, as will be discussed later, basis functions for some coirreps of $G_{\text{SS}}^{l}\times ~^{1}4/ ^{1}m^{1}m^{1}m$ differ from those for the corresponding coirreps of the group $4/mmm1'$. For contrast, the coirreps and generators of $4/mmm1'$ are listed in Table.~\ref{co-irrep-4mmm1} and the basis functions for each coirrep of group $4/mmm1'$ are presented in Table.~\ref{bases-4mmm1'}. In this subsection, we label the coirreps of group $4/mmm1'$ as $\Gamma^{i}~(i=1,2...,10)$.
\begin{table}
    \centering
    \begin{tabular}{>{\centering\arraybackslash}m{1.5cm}|>{\centering\arraybackslash}m{6.5cm}}
        \hline\hline
    \multicolumn{2}{c} {$4/mmm1'$}\\
    \hline
     coirrep    & basis functions\\ \hline
     $\Gamma^{1}$   & \makecell{$k_{x}^2+k_{y}^2$;$k_{z}^2$.}\\ 
     $\Gamma^{2}$   & \makecell{$k_{x}\tutabf{x}+k_{y}\tutabf{y}$;$k_{z}\tutabf{z}$.} \\
     $\Gamma^{3}$   & $k_{x}k_{y}(k_{x}^2-k_{y}^{2})$ \\
     $\Gamma^{4}$   & $k_{x}\tutabf{y}-k_{y}\tutabf{x}$. \\
     $\Gamma^{5}$   & $k_{x}^2-k_{y}^2$. \\
     $\Gamma^{6}$   & $k_{x}\tutabf{x}-k_{y}\tutabf{y}$.\\
     $\Gamma^{7}$   & $k_{x}k_{y}$.  \\
     $\Gamma^{8}$   & $k_{x}\tutabf{y}+k_{y}\tutabf{x}$. \\
     $\Gamma^{9}$   & \{$k_{x}k_{z}+i k_{y}k_{z}$, $-k_{x}k_{z}+i k_{y}k_{z}$\}. \\ \hline
     $\Gamma^{10}$  & \makecell{\{$(k_{x}+i k_{y})\tutabf{z}$,$(k_{x}-i k_{y})\tutabf{z}$\};~\\~\{$k_{z}(\tutabf{x}+i\tutabf{y})$,$k_{z}(\tutabf{x}-i\tutabf{y})$\}.}  \\ \hline
    \end{tabular}
    \caption{The basis functions for each coirrep of type-II magnetic point group $4/mmm1'$.}
    \label{bases-4mmm1'}
\end{table}

We first discuss the spin-singlet pairing states in $\sigma=0$ case. As shown in Table.~\ref{bases-coll}, the basis functions for coirreps $\Gamma^{i}_{\sigma=0}$ with $i=1,3,5,7,9$ are all scalar functions which represent spin-singlet pairing states. For coirrep $\Gamma^{1}_{\sigma=0}$, the basis functions represent $s$-wave superconducting states, which remain invariant under all spatial and spin operations. The coirrep $\Gamma^{1}_{\sigma=0}$ in $G_{\text{SS}}^{l}\times ~^{1}4/ ^{1}m^{1}m^{1}m$ is homomorphic to the coirrep $\Gamma^{1}$ in $4/mmm1'$, and the basis functions for these two coirreps are the same. The reason is as follows: the homomorphism between these two coirreps can be defined through the representation matrices for generators of $4/mmm1'$ and the representation matrices for corresponding group elements of $G_{\text{SS}}^{l}\times ~^{1}4/ ^{1}m^{1}m^{1}m$:
\begin{equation}\label{homo}
    \begin{aligned}
        &D^{1}([C_{4z}||C_{4z}])~\rightarrow~D^{1}_{\sigma=0}([R_{\bbb{z}}(\theta)||C_{4z}]),\\
        &D^{1}([C_{2z}||IC_{2z}])~\rightarrow~D^{1}_{\sigma=0}([R_{\bbb{z}}(\theta)||IC_{2z}]),\\
        &D^{1}([C_{2y}||IC_{2y}])~\rightarrow~D^{1}_{\sigma=0}([R_{\bbb{z}}(\theta)||IC_{2y}]),\\
        &D^{1}([\ccc{T}||E])~\rightarrow~D^{1}_{\sigma=0}(([\ccc{T}R_{\bbb{n}_{\bot}}(\pi)R_{\bbb{z}}(\theta)||E]),\\
    \end{aligned}
\end{equation}
where $D^{1}(g)$ is the representation matrix of generator $g$ for coirrep $\Gamma^{1}$, and $D^{1}_{\sigma=0}(g')$ is the representation matrix of group element $g'$ for coirrep $\Gamma^{1}_{\sigma=0}$. According to the Dimmock test in Eq.~(\ref{Dimmock}), both $\Gamma^{1}_{\sigma=0}$ and $\Gamma^{1}$ belong to type (a). Thus, the basis functions for both coirreps can be obtained by using the projection operator defined in Eq.~(\ref{proj2}) and are only determined by the unitary group elements. In Eq.~(\ref{homo}), the representation matrices of generators belonging to the unitary part of $4/mmm1'$ are identical to those of the corresponding group elements of $G_{\text{SS}}^{l}\times ~^{1}4/ ^{1}m^{1}m^{1}m$, and the representation matrices of the group elements $[R_{z}(\theta)||E]$ with any $\theta$ for $\Gamma^{1}$ are identity matrices. Moreover, the group elements belonging to the unitary part of $G_{\text{SS}}^{l}\times ~^{1}4/ ^{1}m^{1}m^{1}m$ share the same spatial operations with the corresponding group elements of $4/mmm1'$, differing only in their spin-rotation operations. However, these spin-rotation operations do not affect spin-singlet superconducting states due to the zero spin angular momentum of spin-singlet Cooper pairs. By virtue of those results, one can conclude that the basis functions for coirrep $\Gamma^{1}_{\sigma=0}$ are identical to those for the coirrep $\Gamma^{1}$. 

Comparing to the superconducting states represented by the basis functions for the $\Gamma^{1}_{\sigma=0}$ channel, the spin-singlet pairing states in other superconducting channels exhibit higher orbital angular momentum. For instance, the basis function $\xi^{3,\sigma=0}=k_{x}k_{y}(k_{x}^2-k_{y}^2)$ for coirrep $\Gamma^{3}_{\sigma=0}$ represents a $g$-wave superconducting state. The spatial transformation operator $\hat{P}_{[E||IC_{2y}]}$ changes its sign but $\hat{P}_{[E||C_{4z}]}$ does not:
\begin{equation}\label{g-wave-basis}
\begin{aligned}
    \hat{P}_{[E||IC_{2y}]} \xi^{3,\sigma=0} &=& -\xi^{3,\sigma=0},\\
    \hat{P}_{[E||C_{4z}]} \xi^{3,\sigma=0} &=& \xi^{3,\sigma=0}.\\
\end{aligned}
\end{equation}
Moreover, the basis function $\xi^{5,\sigma=0}=k_{x}^2-k_{y}^2$ for coirrep $\Gamma^{5}_{\sigma=0}$ and the basis function $\xi^{7,\sigma=0}=k_{x}k_{y}$ for coirrep $\Gamma^{7}_{\sigma=0}$ both correspond to $d$-wave superconducting states, the spatial transformation operator $\hat{P}_{[E||C_{4z}]}$ changes their signs:
\begin{equation}
    \hat{P}_{[E||C_{4z}]} \xi^{5 / 7,\sigma=0} = -\xi^{5 / 7,\sigma=0}.
\end{equation}
In addition, the basis functions \{$\xi^{9,\sigma=0}_{1}$, $ \xi^{9,\sigma=0}_{2}$\} for the two-dimensional coirrep $\Gamma^{9}_{\sigma=0}$ represent $d\pm id$ superconducting states, respectively, as shown in Table.~\ref{bases-coll}. These two basis functions can be related to each other by the spatial transformation operator $\hat{P}_{[E||IC_{2y}]}$:
\begin{equation}
     \hat{P}_{[E||IC_{2y}]} \xi^{9,\sigma=0}_{1}=-\xi^{9,\sigma=0}_{2}.
\end{equation}
For the same reason as for coirrep $\Gamma^{1}_{\sigma=0}$, the basis functions for coirreps $\Gamma^{3}_{\sigma=0}$, $\Gamma^{5}_{\sigma=0}$, $\Gamma^{7}_{\sigma=0}$, and $\Gamma^{9}_{\sigma=0}$ are identical to those for their corresponding coirreps $\Gamma^{3}$, $\Gamma^{5}$, $\Gamma^{7}$, and $\Gamma^{9}$, respectively.

We then discuss the spin-triple pairing states in the $\sigma=0$ case. In Table.~\ref{bases-coll}, the basis functions for the coirreps $\Gamma^{i}_{\sigma=0}$ with $i=2,4,6,8,10$ are all vector functions which represent unitary spin-triplet superconducting states. The lowest-order basis function for coirrep $\Gamma^{2}_{\sigma=0}$ is $k_{x}k_{y}(k_{x}^2-k_{y}^2)k_{z}\tutabf{z}$, which represents a unitary $h$-wave superconducting state. The Cooper pairs in this superconducting state exhibit no spin polarization. In contrast to the spin-singlet pairing cases, although the coirrep $\Gamma^{2}_{\sigma=0}$ is homomorphic to the coirrep $\Gamma^{2}$, the superconducting states for these two coirreps are different. The basis functions for $\Gamma^{2}$ are $k_{x}\tutabf{x}+k_{y}\tutabf{y}$ and $k_{z}\tutabf{z}$, which represent unitary $p$-wave superconducting states.

The basis function for coirrep $\Gamma^{4}_{\sigma=0}$ is $k_{z}\tutabf{z}$. It represents the unitary $p$-wave superconducting state with only out-of-plane pairing components. In contrast, the basis function for its corresponding coirrep $\Gamma^{4}$ of $4/mmm1'$ is $k_{x}\tutabf{y}-k_{y}\tutabf{x}$, which contains only in-plane pairing components. For coirreps $\Gamma^{6}_{\sigma=0}$ and $\Gamma^{8}_{\sigma=0}$, the lowest-order basis functions $k_{x}k_{y}k_{z}\tutabf{z}$ and $(k_{x}^2-k_{y}^2)k_{z}\tutabf{z}$ represent unitary $f$-wave superconducting states. However, the basis functions for their corresponding coirreps $\Gamma^{6}$ and $\Gamma^{8}$ are $k_{x}\tutabf{x}- k_{y}\tutabf{y}$ and $k_{x}\tutabf{y}+ k_{y}\tutabf{x}$, respectively, and both of them represent unitary $p$-wave superconducting states. 

Interestingly, for the two-dimensional coirrep $\Gamma^{10}_{\sigma=0}$, the set of basis functions \{$(k_{x}+ik_{y})\tutabf{z}$, $(k_{x}-ik_{y})\tutabf{z}$\} is identical to one set of basis functions for $\Gamma^{10}$. It means that the superconducting states represented by the basis functions for $\Gamma^{10}_{\sigma=0}$ are equivalent to those represented by one set of basis functions for $\Gamma^{10}$.
However, for the coirrep $\Gamma^{10}$, there is another set of basis functions \{$k_{z}(\tutabf{x}+i \tutabf{y})$, $k_{z}(\tutabf{x}- i\tutabf{y})$\} that is not equivalent to any of the basis functions for coirrep $\Gamma^{10}_{\sigma=0}$. This implies that there are fewer possible superconducting states for $\Gamma^{10}_{\sigma=0}$ compared to $\Gamma^{10}$. The reason is as follows: the representation matrices of the group elements $[R_{\bbb{z}}(\theta)||E]$ for $\Gamma^{10}_{\sigma=0}$ are identity matrices for any $\theta$. However, under the operators $\hat{P}_{[R_{\bbb{z}}(\theta)||E]}$ corresponding to these group elements, the axial-vectors $\tutabf{x}$ and $\tutabf{y}$ are transformed into their linear combinations with coefficients depending on the parameter $\theta$:
\begin{equation}
    \begin{aligned}
        \hat{P}_{[R_{\bbb{z}}(\theta)||E]}\tutabf{x}&=\cos(\theta)\tutabf{x}+\sin(\theta)\tutabf{y},\\
        \hat{P}_{[R_{\bbb{z}}(\theta)||E]}\tutabf{y}&=-\sin(\theta)\tutabf{x}+\cos(\theta)\tutabf{y}.
    \end{aligned}
\end{equation}
Therefore, the basis functions containing $\tutabf{x}$ or $\tutabf{y}$ are forbidden for coirrep $\Gamma^{10}_{\sigma=0}$. On the other hand, the spin operations in type-II magnetic point group $4/mmm1'$ are completely locked with the spatial operations and are all discrete rotation operations. Under spin-rotation operators, the transformations of axial-vectors $\tutabf{x}$, $\tutabf{y}$, and $\tutabf{z}$ are identical to the transformations of the momentum components $k_{x}$, $k_{y}$, and $k_{z}$ under the corresponding spatial rotation operators, respectively. Therefore, since the set of basis functions \{$(k_{x}+i k_{y})\tutabf{z}$, $(k_{x}-i k_{y})\tutabf{z}$\} is allowed for coirrep $\Gamma^{10}$, the set of basis functions \{$k_{z}(\tutabf{x}+i \tutabf{y})$, $k_{z}(\tutabf{x}-i \tutabf{y})$\} is also allowed.

Unlike the $\sigma = 0$ case, the representation matrices of the group elements $[R_{z}(\theta)]||E]$ contain a nontrivial phase factor $e^{i\theta}$ in the $\sigma = 1$ case, indicating that in superconducting states represented by the basis functions for these coirreps, the $z$-component of the Cooper pairs' spin angular momentum is $-\hbar$ (see Appendix.~\ref{magnetic_moment} for detail).
Due to the constraint on the Cooper pairs' spin angular momentum, basis functions representing spin-singlet pairing states are not allowed in the $\sigma=1$ case. As shown in Table.~\ref{bases-coll}, there is no basis function for coirreps $\Gamma^{i}_{\sigma=1}~(i=1,3,5,7,9)$. The reason is as follows: the representation matrices of $[E||I]$ for coirreps $\Gamma^{i}_{\sigma=1}~(i=1,3,5,7,9)$ are identity matrices. Therefore, if any basis functions exist for these coirreps, they must be even functions. Furthermore, according to the Eq.~(\ref{Aext}), if the basis functions are even functions, they must be scalar functions which represent spin-singlet pairing states (see Appendix.~\ref{app-a} for detail). Consequently, the requirements for basis functions of coirreps $\Gamma^{i}_{\sigma=1}~(i=1,3,5,7,9)$ contradict the constraint on the Cooper pairs' spin angular momentum in the $\sigma=1$ case. However, basis functions can be found for coirreps $\Gamma^{i}_{\sigma=1}~(i=2,4,6,8,10)$. This is because the representation matrices of $[E||I]$ for these coirreps are minus identity matrices. Therefore, the basis functions for coirreps $\Gamma^{i}_{\sigma=1}~(i=2,4,6,8,10)$ are odd vector functions which represent spin-triplet pairing states. In spin-triplet pairing states, the $z$-component of the Cooper pairs' spin angular momentum can be $-\hbar$, which is consistent with the constraint on the Cooper pairs' spin angular momentum in the $\sigma=1$ case.
In addition, the coefficients preceding the axial-vectors in the basis functions for $\Gamma^{i}_{\sigma=1}~(i=2,4,6,8,10)$ are identical to those for $\Gamma^{i}_{\sigma=0}~(i=2,4,6,8,10)$. The reason is that the coefficients are unaffected by spin-rotation operations.

For the same reason as in the $\sigma=1$ case, in the $\sigma=-1$ case, the basis functions vanish for coirreps $\Gamma^{i}_{\sigma=-1}~(i=1,3,5,7,9)$ and only exist for coirreps $\Gamma^{i}_{\sigma=-1}~(i=2,4,6,8,10)$. Additionally, for coirreps $\Gamma^{2}_{\sigma=-1}$, $\Gamma^{4}_{\sigma=-1}$, $\Gamma^{6}_{\sigma=-1}$, $\Gamma^{8}_{\sigma=-1}$, and $\Gamma^{10}_{\sigma=-1}$, the coefficients preceding the axial-vectors of the basis functions are identical to those for coirreps $\Gamma^{2}_{\sigma=1}$, $\Gamma^{4}_{\sigma=1}$, $\Gamma^{6}_{\sigma=1}$, $\Gamma^{8}_{\sigma=1}$, and $\Gamma^{10}_{\sigma=1}$, respectively. However, the axial-vector parts represent opposite spin polarizations in the $\sigma=-1$ case and $\sigma=1$ case. This difference arises from the opposite $z$-component of the Cooper pairs' spin angular momentum in these two cases.

\subsection{The type-II collinear spin point group $G_{\text{SS}}^{l}\times ~^{1}4/ ^{1}m^{\overline{1}}m^{\overline{1}}m$}
\label{typ-II-coll}
In this subsection, we use the type-II collinear spin point group $G_{\text{SS}}^{l}\times ~^{1}4/ ^{1}m^{\overline{1}}m^{\overline{1}}m$ as an example to illustrate its superconducting channels and basis functions. The generators and coirreps of this spin point group are listed in Table.~\ref{co-irrep-coll2}. Using the approach illustrated in the flowchart Fig.~\ref{flowchart}, we can obtain the basis functions for each superconducting channel. The results are shown in Table.~\ref{bases-coll2}. Due to the presence of the inversion symmetry $[E||I]$, the spin-singlet and triplet pairing states are decoupled. In addition, similar to the type-I collinear spin point group, there is no magnetic point group which is isomorphic to the type-II collinear spin point group. Therefore, the classification of the superconducting channels and basis functions in a type-II collinear spin point group differs from that in any magnetic point group.
\begin{table}[b]
    \centering
        \begin{tabular}{>{\centering\arraybackslash}m{1.5cm}|>{\centering\arraybackslash}m{6.5cm}}
    \hline\hline
    \multicolumn{2}{c}{$G_{\text{S}}^{cl}\times ~^{1}4/ ^{1}m^{\overline{1}}m^{\overline{1}}m$}\\
    \hline
     coirrep    & basis functions \\ \hline
     $\Gamma^{1}$  &   $k_{x}^2+k_{y}^{2}$; $k_{z}^2$.  \\
     $\Gamma^{2}$  &   $k_{z}\tutabf{z}$. \\
     $\Gamma^{3}$  &    $k_{x}k_{y}(k_{x}^2-k_{y}^2)$.  \\
     $\Gamma^{4}$  &    $k_{x}k_{y}(k_{x}^2-k_{y}^2)k_{z}\tutabf{z}$.  \\
     $\Gamma^{5}$  &   $k_{x}^2-k_{y}^2$.  \\
     $\Gamma^{6}$  &   $(k_{x}^2-k_{y}^2)k_{z}\tutabf{z}$.  \\
     $\Gamma^{7}$  &   $k_{x}k_{y}$.  \\
     $\Gamma^{8}$  &   $k_{x}k_{y}k_{z}\tutabf{z}$. \\
     $\Gamma^{9}$  &    \{$k_{x}k_{z}+ik_{y}k_{z}$, $-k_{x}k_{z}+ik_{y}k_{z}$\}.   \\
     $\Gamma^{10}$  &   \makecell{ \{$(k_{x}+ik_{y})\tutabf{z}$, $-(k_{x}-ik_{y})\tutabf{z}$\}.}   \\
      \hline
     \multicolumn{2}{c}{$\nu=1$}\\ \hline
     $\Gamma^{11}_{\nu=1}$  &   -   \\ \hline
     $\Gamma^{12}_{\nu=1}$  &   \makecell{$\{k_{z}(\tutabf{x}-i\tutabf{y})$, $k_{z}(\tutabf{x}+i\tutabf{y})$\}.}   \\ \hline
     $\Gamma^{13}_{\nu=1}$ &    -   \\ \hline
     $\Gamma^{14}_{\nu=1}$ &    \{$(k_{x}^2-k_{y}^2)k_{z}(\tutabf{x}-i\tutabf{y})$, $(k_{x}^2-k_{y}^2)k_{z}(\tutabf{x}+i\tutabf{y})$\}.   \\ \hline
     $\Gamma^{15}_{\nu=1}$ &    -  \\ \hline
     $\Gamma^{16}_{\nu=1}$ &    \makecell{\{$(k_{x}+ik_{y})(\tutabf{x}-i\tutabf{y})$, $-(k_{x}-ik_{y})(\tutabf{x}+i\tutabf{y})$, \\ $(k_{x}-ik_{y})(\tutabf{x}-i\tutabf{y})$, $-(k_{x}+ik_{y})(\tutabf{x}+i\tutabf{y})$ \}.}  \\ \hline
     
    \end{tabular}
    \caption{The basis functions for the type-II collinear spin point group $G_{\text{SS}}^{l}\times ~^{1}4/ ^{1}m^{\overline{1}}m^{\overline{1}}m$. Here $\nu \in \mathbb{N}/\{0\}$ is related to the $z$-components of spin angular momentum of the Cooper pairs in the superconducting states represented by the basis functions of these channels. Because we focus on the charge-2e superconducting states, $\nu=1$ is the only appropriate choice. In addition, by using Dimmock test, we find that both coirreps $\Gamma^{15}_{\nu=1}$ and $\Gamma^{16}_{\nu=1}$ belong to the type (c). Therefore, the number of their basis functions is doubled compared to the number of basis functions obtained from the unitary part of the group.}
    \label{bases-coll2}
\end{table}

First of all, we discuss the basis functions for the first ten superconducting channels of $G_{\text{SS}}^{l}\times ~^{1}4/ ^{1}m^{\overline{1}}m^{\overline{1}}m$ listed in Table.~\ref{bases-coll2}, by comparing them with the basis functions for coirreps of $G_{\text{SS}}^{l}\times ~^{1}4/ ^{1}m^{1}m^{1}m$ in the $\sigma=0$ case (see Table.~\ref{bases-coll}). The first ten superconducting channels in Table.~\ref{bases-coll2} do not include the $\nu$ index. In this subsection, unless otherwise specified, the $\Gamma^{i}~(i=1,2,...,10)$ represents the coirreps of $G_{\text{SS}}^{l}\times ~^{1}4/ ^{1}m^{\overline{1}}m^{\overline{1}}m$, and the $\Gamma^{i}_{\sigma=0}~(i=1,2,...,10)$ represents the coirreps of $G_{\text{SS}}^{l}\times ~^{1}4/ ^{1}m^{1}m^{1}m$. 

Since the representation matrices of $[R_{z}(\theta)||E]$ with any $\theta$ are identity matrices for coirreps $\Gamma^{i}~(i=1,2,...,10)$ and $\Gamma^{i}_{\sigma=0}~(i=1,2,...,10)$, the isomorphism between $\Gamma^{i}$ and $\Gamma^{i}_{\sigma=0}$ with $i=1,2,...10$ can be defined through the representation matrices of the generators:
\begin{equation}\label{isomo1}
    \begin{aligned}
        &D^{i}([R_{z}(\delta\theta)||C_{4z}])\rightarrow~D^{i}_{\sigma=0}([R_{z}(\delta\theta)||C_{4z}]),\\
        &D^{i}([R_{z}(\delta\theta)||IC_{2z}])\rightarrow~D^{i}_{\sigma=0}([R_{z}(\delta\theta)||IC_{2z}]),\\
        &D^{i}([R_{\bbb{n}_{\bot}}(\pi)R_{z}(\delta\theta)||IC_{2y}])\rightarrow~D^{i}_{\sigma=0}([R_{z}(\delta\theta)||IC_{2y}]),\\
        &D^{i}([\ccc{T}R_{\bbb{n}_{\bot}}(\pi)R_{z}(\delta\theta)||E])\rightarrow~D^{i}_{\sigma=0}([\ccc{T}R_{\bbb{n}_{\bot}}(\pi)R_{z}(\delta\theta)||E]),
    \end{aligned}
\end{equation}
where $\delta\theta$ is an infinitesimal angle. By virtue of this isomorphism, the sameness and differences between basis functions for $\Gamma^{i}~(i=1,2,...,10)$ and those for corresponding coirreps $\Gamma^{i}_{\sigma=0}~(i=1,2,...,10)$ can be understood.

We start by discussing the spin-singlet pairing states. For coirreps $\Gamma^{i}$ with $i=1,3,5,7,9$, the basis functions are all scalar functions representing spin-singlet pairing states, as shown in Table.~\ref{bases-coll2}. 
In addition, the basis functions for coirreps $\Gamma^{1}$, $\Gamma^{3}$, $\Gamma^{5}$, $\Gamma^{7}$, and $\Gamma^{9}$ are identical to the basis functions for the corresponding coirreps $\Gamma^{1}_{\sigma=0}$, $\Gamma^{3}_{\sigma=0}$, $\Gamma^{5}_{\sigma=0}$, $\Gamma^{7}_{\sigma=0}$, and $\Gamma^{9}_{\sigma=0}$, respectively. The reason is as follows: according to the Dimmock test Eq.~(\ref{Dimmock}), all the coirreps $\Gamma^{i}$ and $\Gamma^{i}_{\sigma=0}$ with $i=1,3,5,7,9$ belong to type (a).
Thus, the forms of the basis functions are determined by the unitary parts of the groups. In Eq.~(\ref{isomo1}), the spatial operations and representation matrices of generators belonging to the unitary part of $G_{\text{SS}}^{l}\times ~^{1}4/ ^{1}m^{\overline{1}}m^{\overline{1}}m$ are the same as those of the corresponding generators of $G_{\text{SS}}^{l}\times ~^{1}4/ ^{1}m^{1}m^{1}m$, but their spin-rotation operations are different. Moreover, spin-singlet pairing states are not affected by spin-rotation operations. Therefore, the basis functions for coirreps $\Gamma^{i}~(i=1,3,5,7,9)$ are identical to those for the corresponding coirreps $\Gamma^{i}_{\sigma=0}~(i=1,3,5,7,9)$.

We then discuss the spin-triplet pairing states. As shown in Table.~\ref{bases-coll2}, the basis functions for coirreps $\Gamma^{i}$ with $i=2,4,6,8,10$ are all vector functions representing spin-triplet pairing states. Similar to the case of spin-singlet pairing states, all coirreps $\Gamma^{i}$ with $i=2,4,6,8,10$ belong to type (a) according to the Dimmock test Eq.~(\ref{Dimmock}). However, unlike the case for coirreps $\Gamma^{i}$ with $i=1,3,5,7,9$, the basis functions for coirreps $\Gamma^{i}$ with $i=2,4,6,8,10$ are not identical to those for their corresponding coirreps $\Gamma^{i}_{\sigma=0}$ with $i=2,4,6,8,10$.

For coirrep $\Gamma^{2}$, the basis function is identical to the basis function for coirrep $\Gamma^{4}_{\sigma=0}$, rather than the corresponding coirrep $\Gamma^{2}_{\sigma=0}$. The reason is as follows: for coirreps $\Gamma^{i}$ and $\Gamma^{i}_{\sigma=0}$ with $i=2,4,6,8,10$, the spin-rotation operators $\hat{P}_{[R_{\bbb{z}}(\theta)||E]}$ with any angle $\theta$ do not change the basis functions. Thus, the axial-vector part of the basis functions for these coirreps can only be $\tutabf{z}$. The spin-rotation operator $\hat{P}_{[R_{\bbb{n}_{\bot}}(\pi)R_{z}(\delta\theta)||IC_{2y}]}$ changes the sign of $\tutabf{z}$, but $\hat{P}_{[R_{z}(\delta\theta)||IC_{2y}]}$ does not. Additionally, the representation matrix of $[R_{\bbb{n}_{\bot}}(\pi)R_{z}(\delta\theta)||IC_{2y}]$ for coirrep $\Gamma^{2}$ is identical to the representation matrix of $[R_{z}(\delta \theta)||IC_{2y}]$ for coirrep $\Gamma^{2}_{\sigma=0}$. Therefore, the basis functions for these two coirreps, $\Gamma^{2}$ and $\Gamma^{2}_{\sigma=0}$, are different. However, the representation matrix of $[R_{\bbb{n}_{\bot}}(\pi)R_{z}(\delta\theta)||IC_{2y}]$ for coirrep $\Gamma^{2}$ has the opposite sign with the representation matrix of $[R_{z}(\delta\theta)||IC_{2y}]$ for coirrep $\Gamma^{4}_{\sigma=0}$. In addition, the representation matrices of other generators in Eq.~(\ref{isomo1}) belonging to the unitary part of group $G_{\text{SS}}^{l}\times ~^{1}4/ ^{1}m^{\overline{1}}m^{\overline{1}}m$ for coirrep $\Gamma^{2}$ are the same as those for the corresponding generators in $G_{\text{SS}}^{l}\times ~^{1}4/ ^{1}m^{1}m^{1}m$ for coirrep $\Gamma^{4}_{\sigma=0}$. Thus, the basis functions for these two coirreps are identical. For the same reason, the basis functions for coirreps $\Gamma^{4}$, $\Gamma^{6}$, and $\Gamma^{8}$ are identical to those for coirreps $\Gamma^{2}_{\sigma=0}$, $\Gamma^{8}_{\sigma=0}$, and $\Gamma^{6}_{\sigma=0}$ rather than coirreps $\Gamma^{4}_{\sigma=0}$, $\Gamma^{6}_{\sigma=0}$, and $\Gamma^{8}_{\sigma=0}$, respectively. 

Similarly, for two-dimensional coirrep $\Gamma^{10}$, the set of basis functions is different from that of the corresponding coirrep $\Gamma^{10}_{\sigma=0}$. The second basis function $-(k_{x}-ik_{y})\tutabf{z}$ for $\Gamma^{10}$ has an opposite sign with respect to the second basis function $(k_{x}-ik_{y})\tutabf{z}$ for $\Gamma^{10}_{\sigma=0}$. The reason is as follows: on the one hand, the representation matrices of generators in Eq.~(\ref{isomo1}) for two coirreps $\Gamma^{10}$ and $\Gamma^{10}_{\sigma=0}$ are identical. On the other hand, the additional spin-rotation operation $R_{\bbb{n}_{\bot}}(\pi)$ changes the sign of the axial-vector $\tutabf{z}$. It is worth to note that although the forms of the basis functions for two coirreps $\Gamma^{10}$ and $\Gamma^{10}_{\sigma=0}$ are different, the possible superconducting states for these two coirreps are the same. This is because superconducting order parameters are linear combinations of the basis functions, and basis functions for these two coirreps are linearly dependent on each other.

In Table.~\ref{bases-coll2}, besides the first ten superconducting channels without the $\nu$ index, there are six superconducting channels with $\nu=1$ in the spin point group $G_{\text{SS}}^{l}\times ~^{1}4/ ^{1}m^{\overline{1}}m^{\overline{1}}m$. They correspond to the coirreps $\Gamma^{i}_{\nu=1}~(i=11,12,13,14,15,16)$. Specifically, four of them correspond to two-dimensional coirreps and two of them correspond to four-dimensional coirreps. Similar to the $\sigma \neq 0$ cases in type-I collinear spin point groups, a nonzero $\nu$ index indicates that the Cooper pairs’ spin angular momenta of the superconducting states are nonzero.
For coirreps with $\nu=1$, due to the constraint on the Cooper pairs' spin angular momentum, spin-singlet paring states are not allowed. As shown in Table.~\ref{bases-coll2}, there is no basis function for coirreps $\Gamma^{11}_{\nu=1}$, $\Gamma^{13}_{\nu=1}$ and $\Gamma^{15}_{\nu=1}$. The reason is that the representation matrices of the inversion symmetry $[E||I]$ for $\Gamma^{11}_{\nu=1}$, $\Gamma^{13}_{\nu=1}$ and $\Gamma^{15}_{\nu=1}$ are all identity matrices, which means the basis functions for these coirreps are even scalar functions representing spin-singlet pairing states. This requirement contradicts the constraint on the Cooper pairs' spin angular momentum in $\nu=1$ case. Suitable basis functions can only be found in superconducting channels corresponding to the coirreps $\Gamma^{12}_{\nu= 1}$, $\Gamma^{14}_{\nu=1}$ and $\Gamma^{16}_{\nu= 1}$. For these coirreps, the representation matrices of inversion symmetry $[E||I]$ are minus identity matrices. Thus, the basis functions for these coirreps are all odd vector functions representing spin-triplet pairing states. Specifically, the basis functions \{$k_{z}(\tutabf{x}-i\tutabf{y})$, $k_{z}(\tutabf{x}+i\tutabf{y})$\} for coirrep $\Gamma^{12}_{\nu= 1}$ represent $p$-wave superconducting states with opposite spin polarizations. Additionally, the basis functions \{$(k_{x}^2-k_{y}^2)k_{z}(\tutabf{x}-i\tutabf{y})$, $(k_{x}^2-k_{y}^2)k_{z}(\tutabf{x}+i\tutabf{y})$\} for the coirrep $\Gamma^{14}_{\nu=1}$ represent nonunitary $f$-wave superconducting states with opposite spin polarizations. Interestingly, the coirrep $\Gamma^{16}_{\nu= 1}$ belongs to the type (c) according to the Dimmock test. Thus, the number of its basis functions is doubled compared to that obtained from the unitary part of the spin point group. All four basis functions for $\Gamma^{16}_{\nu=1}$ represent nonunitary $p$-wave superconducting states, with both orbital and spin magnetic moments being nonzero.

\section{Superconductivity in coplanar spin point group}\label{coplanar}
Next, we discuss the superconducting order parameters in coplanar spin space group. As discussed in Sec.\ref{method}, we classify the pairing symmetry and construct the superconducting order parameters based on the corresponding coplanar spin point group. The coplanar spin point group is the symmetry group that describes systems where the magnetic moments, $\bbb{M}(\bbb{r})$, are confined to a single plane. Unlike collinear magnetic moments, which are aligned along a specific axis, coplanar magnetic moments can orient in any direction within a fixed plane. Therefore, all magnetic moments $\bbb{M}(\bbb{r})$ are perpendicular to the same normal vector $\bbb{n}_{\text{co}}$. In general, the coplanar spin point group can be written as:
\begin{equation}
    X=G_{\text{SS}}^{p}\times G_{\text{NS}},
\end{equation}
where the $G_{\text{NS}}$ is the non-trivial spin point group without any pure spin operations, and $G_{\text{SS}}^{p}$ is a spin-only group which only contains spin operations. Without loss of generality, we assume the magnetic moments $\bbb{M}(\bbb{r})$ are perpendicular to the normal vector $\bbb{n}_{\text{co}}=\bbb{z}$. Thus, the spin-only group part $G_{\text{SS}}^{p}$ is expressed as:
\begin{equation}
    G_{\text{SS}}^{p}=\{[E||E],[\ccc{T}R_{\bbb{z}}(\pi)||E]\},
\end{equation}
where $R_{\bbb{z}}(\pi)$ is the rotation about the $\bbb{z}$ axis by $\pi$. Unlike collinear spin point group, coplanar spin point groups are finite groups. For each of them, there is a corresponding type-II magnetic point group which is isomorphic to it~\cite{schiff2023spin}. However, basis functions in coplanar spin point groups can differ from those in the corresponding type-II magnetic point group.

Here, we take the coplanar spin point group $G^{p}_{\text{SS}}\times ~^{2_{z}}4/^{1}m^{2_{x}}m^{2_{y}}m$ as an example. The coirreps and generators are shown in Table.~\ref{co-irrep-cop}. It is isomorphic to the type-II magnetic point group $4/mmm1'$. The isomorphism between these two groups can be defined by the generators of them:
\begin{equation}
    \begin{aligned}\label{isomor2}
        [R_{\bbb{z}}(\pi)||C_{4z}]~&\rightarrow~[C_{4z}||C_{4z}],\\
        [E||IC_{2z}]~&\rightarrow~[C_{2z}||IC_{2z}],\\
        [R_{\bbb{x}}(\pi)||IC_{2y}]~&\rightarrow~[C_{2y}||IC_{2y}],\\
        [\ccc{T}R_{\bbb{z}}(\pi)||E]~&\rightarrow~[\ccc{T}||E].
    \end{aligned}
\end{equation}
These two groups share the same coirreps table, and all the coirreps in both groups belong to type (a) according to the Dimmock test Eq.~(\ref{Dimmock}). The basis functions for each coirrep of $G^{p}_{\text{SS}}\times ~^{2_{z}}4/^{1}m^{2_{x}}m^{2_{y}}m$ are shown in Table.~\ref{bases-cop}. Due to the existence of the inversion symmetry $[E||I]$, the spin-singlet and triplet pairing states are decoupled. Comparing the Table.~\ref{bases-cop} and Table.~\ref{bases-4mmm1'}, we find that the basis functions for coirreps $\Gamma^{i}~(i=1,3,5,7,9)$ of $G^{p}_{\text{SS}}\times ~^{2_{z}}4/^{1}m^{2_{x}}m^{2_{y}}m$ are the same as those for the corresponding coirreps of $4/mmm1'$, and they represent spin-singlet paring states. However, the basis functions for $\Gamma^{i}~(i=2,4,6,8,10)$ differ from those for corresponding coirreps of $4/mmm1'$ and represent spin-triplet superconducting states. These results can be attributed to that the group elements of $G^{p}_{\text{SS}}\times ~^{2_{z}}4/^{1}m^{2_{x}}m^{2_{y}}m$ share the same representation matrices and spatial operations with the corresponding group elements of $4/mmm1'$, while their spin operations are different, and only the spin-triplet pairing states are affected by these spin operations.

Next, we discuss how different spin operations cause differences in the basis functions representing spin-triplet pairing states between $G^{p}_{\text{SS}}\times ~^{2_{z}}4/^{1}m^{2_{x}}m^{2_{y}}m$ and $4/mmm1'$. For example, in Eq.~(\ref{isomor2}), the generator $[R_{\bbb{z}}(\pi)||C_{4z}]$ in $G^{p}_{\text{SS}}\times ~^{2_{z}}4/^{1}m^{2_{x}}m^{2_{y}}m$ corresponds to the generator $[C_{4z}||C_{4z}]$ in $4/mmm1'$. For coirrep $\Gamma^{2}$, in $G^{p}_{\text{SS}}\times ~^{2_{z}}4/^{1}m^{2_{x}}m^{2_{y}}m$, the operator $\hat{P}_{[R_{\bbb{z}}(\pi)||C_{4z}]}$ transforms the basis function $k_{x}\tutabf{x}+k_{y}\tutabf{y}$ into $-k_{y}\tutabf{x}+k_{x}\tutabf{y}$, while in $4/mmm1'$, the operator $\hat{P}_{[C_{4z}||C_{4z}]}$ does not change the basis function $k_{x}\tutabf{x}+k_{y}\tutabf{y}$. The representation matrix of $[R_{\bbb{z}}(\pi)||C_{4z}]$ for coirrep $\Gamma^{2}$ of $G^{p}_{\text{SS}}\times ~^{2_{z}}4/^{1}m^{2_{x}}m^{2_{y}}m$ and the representation matrix of $[C_{4z}||C_{4z}]$ for coirrep $\Gamma^{2}$ of $4/mmm1'$ are both identity matrices. Besides, the matrices of other generators belonging to the unitary parts of these two groups under the basis function $k_{x}\tutabf{x}+k_{y}\tutabf{y}$ are identical to the representation matrices of these generators for $\Gamma^{2}$. Therefore, the basis function $k_{x}\tutabf{x}+k_{y}\tutabf{y}$ is forbidden for $\Gamma^{2}$ of $G^{p}_{\text{SS}}\times ~^{2_{z}}4/^{1}m^{2_{x}}m^{2_{y}}m$ but allowed for $\Gamma^{2}$ of $4/mmm1'$. On the other hand, for the coirrep $\Gamma^{2}$, the basis function $k_{z}\tutabf{z}$ is allowed in both $G^{p}_{\text{SS}}\times ~^{2_{z}}4/^{1}m^{2_{x}}m^{2_{y}}m$ and $4/mmm1'$. This is attributed to that the matrices of all generators belonging to the unitary parts of these two groups under basis function $k_{z}\tutabf{z}$ are identical to the representation matrices of these generators for $\Gamma^{2}$. A similar analysis can be applied to the coirreps $\Gamma^{4}$, $\Gamma^{6}$, and $\Gamma^{8}$. 

For the two-dimensional coirrep $\Gamma^{10}$, some basis functions allowed in $G^{p}_{\text{SS}}\times ~^{2_{z}}4/^{1}m^{2_{x}}m^{2_{y}}m$ are forbidden in $4/mmm1'$. For example, the set of basis functions $\{(k_{x}-i k_{y})\tutabf{x}, -(k_{x}+i k_{y})\tutabf{x}\}$ is allowed in $G^{p}_{\text{SS}}\times ~^{2_{z}}4/^{1}m^{2_{x}}m^{2_{y}}m$ but forbidden in $4/mmm1'$. 
The reason is as follows: the operator $\hat{P}_{[C_{4z}||C_{4z}]}$ corresponding to the generator $[C_{4z}||C_{4z}]$ of $4/mmm1'$ transforms the basis function $(k_{x}-i k_{y})\tutabf{x}$ into $(k_{y}+i k_{x})\tutabf{y}$, which is not proportional to the other basis function $-(k_{x}+i k_{y})\tutabf{x}$, thus the set of basis function $\{(k_{x}+i k_{y})\tutabf{x}, -(k_{x}+i k_{y})\tutabf{x}\}$ is forbidden for coirrep $\Gamma^{10}$ of $4/mmm1'$. However, the matrices of all generators belonging to the unitary part of $G^{p}_{\text{SS}}\times ~^{2_{z}}4/^{1}m^{2_{x}}m^{2_{y}}m$ under the set of basis functions $\{(k_{x}+i k_{y})\tutabf{x}, -(k_{x}+i k_{y})\tutabf{x}\}$ are identical to the representation matrices of these generators for $\Gamma^{10}$. Thus, this set of basis functions is allowed for $\Gamma^{10}$ of $G^{p}_{\text{SS}}\times ~^{2_{z}}4/^{1}m^{2_{x}}m^{2_{y}}m$. In addition, some basis functions are allowed for $\Gamma^{10}$ of both $G^{p}_{\text{SS}}\times ~^{2_{z}}4/^{1}m^{2_{x}}m^{2_{y}}m$ and $4/mmm1'$. For instance, under the set of basis functions \{$(k_{x}+i k_{y})\tutabf{z}$, $(k_{x}-i k_{y})\tutabf{z}$\}, the matrices of generators belonging to the unitary parts of both $G^{p}_{\text{SS}}\times ~^{2_{z}}4/^{1}m^{2_{x}}m^{2_{y}}m$ and $4/mmm1'$ are identical to these generators' representation matrices for $\Gamma^{10}$. Therefore, the set of basis function \{$(k_{x}+i k_{y})\tutabf{z}$, $(k_{x}-i k_{y})\tutabf{z}$\} is allowed for $\Gamma^{10}$ of both $G^{p}_{\text{SS}}\times ~^{2_{z}}4/^{1}m^{2_{x}}m^{2_{y}}m$ and $4/mmm1'$. The presence or absence of other basis functions for $\Gamma^{10}$ can be understood in the same manner.

\begin{table}
    \centering
\resizebox{\linewidth}{!}{
    \begin{tabular}{>{\centering\arraybackslash}m{1.5cm}|>{\centering\arraybackslash}m{6.5cm}}
        \hline \hline
    \multicolumn{2}{c}{$G^{p}_{\text{SS}}\times ~^{2_{z}}4/^{1}m^{2_{x}}m^{2_{y}}m$} \\
    \hline
     coirrep    & basis functions \\ \hline
     $\Gamma^{1}$  &   $k_{x}^2+k_{y}^{2}$; $k_{z}^2$. \\ 
     $\Gamma^{2}$  &   $k_{z}\tutabf{z}$.  \\
     $\Gamma^{3}$  &   $k_{x}k_{y}(k_{x}^2-k_{y}^{2})$  \\
     $\Gamma^{4}$  &   \makecell{($k_{x}^2-k_{y}^2)k_{z}\tutabf{x}$; $k_{x}k_{y}k_{z}\tutabf{y}$.}  \\
     $\Gamma^{5}$  &   $k_{x}^2-k_{y}^2$.  \\
     $\Gamma^{6}$  &   $k_{z}\tutabf{y}$. \\
     $\Gamma^{7}$  &   $k_{x}k_{y}$.   \\
     $\Gamma^{8}$  &   $k_{z}\tutabf{x}$.  \\
     $\Gamma^{9}$  &   \{$k_{x}k_{z}+i k_{y}k_{z}$, $-k_{x}k_{z}+i k_{y}k_{z}$\}. \\ \hline
     $\Gamma^{10}$ &  \makecell{\{$(k_{x}-i k_{y})\tutabf{x}$, $-(k_{x}+i k_{y})\tutabf{x}$\};\\ 
     \{$(k_{x}-i k_{y})\tutabf{y}$, $(k_{x}+i k_{y})\tutabf{y}$\};\\
     \{$(k_{x}+i k_{y})\tutabf{z}$, $(k_{x}-i k_{y})\tutabf{z}$\}.
     } \\ \hline
    \end{tabular}
}
    \caption{The basis functions for the coplanar group $G^{p}_{\text{SS}}\times ~^{2_{z}}4/^{1}m^{2_{x}}m^{2_{y}}m$.}
    \label{bases-cop}
\end{table}

\section{Superconductivity in noncoplanar spin point group}\label{noncoplanar}
Finally, we discuss about the superconducting order parameters in noncoplanar spin space groups. The superconducting order parameters in these materials are constructed according to the corresponding noncoplanar spin point group. The noncoplanar spin point group refers to the symmetry group of systems where the magnetic moments $\bbb{M}(\bbb{r})$ are not confined to a single plane. In general, the noncoplanar spin point group can be expressed as:
\begin{equation}
    X=G_{\text{SS}}\times G_{\text{NS}}.
\end{equation}
The spin-only group part of the noncoplanar spin point group is a trivial group $G_{\text{SS}}=\{[E||E]\}$. Therefore, noncoplanar spin point groups are all non-trivial spin point groups $G_{\text{NS}}$ without any pure spin operations. In contrast to collinear and coplanar spin point groups, which all contain anti-unitary group elements, some noncoplanar spin point groups do not contain any anti-unitary group elements. Thus, we can divide noncoplanar spin point groups into two types based on the presence or absence of anti-unitary group elements. These two types of noncoplanar spin point groups are isomorphic to type-III magnetic point group and point group, respectively. Next, we present one example for each type of noncoplanar spin point group to illustrate their basis functions.

\subsection{The noncoplanar spin point group $^{4}4/^{1}m$}

In this subsection, we choose the noncoplanar spin point group $^{4}4/^{1}m$ as an example of noncoplanar spin point groups that lack anti-unitary group elements. The noncoplanar spin point group $^{4}4/^{1}m$ is isomorphic to the point group $4/m$, and their coirreps and generators are listed in Table.~\ref{co-irrep-unitarynoncop}. The isomorphism between these two groups can be defined through their generators:
\begin{equation}
    \begin{aligned}
        [R_{\bbb{z}}(\pi/2)||C_{4z}]~&\rightarrow~[C_{4z}||C_{4z}],\\
        [E||IC_{2z}]~&\rightarrow~[C_{2z}||IC_{2z}].
    \end{aligned}
\end{equation}
The basis functions for all irreps of these two groups are presented in Table.~\ref{bases-unitarynoncop}.
\begin{table}
    \centering
\resizebox{\linewidth}{!}{
    \begin{tabular}{>{\centering\arraybackslash}m{1cm}|>{\centering\arraybackslash}m{3.5cm}|>{\centering\arraybackslash}m{3.5cm}}
        \hline \hline
    \multicolumn{2}{>{\centering\arraybackslash}m{4.5cm}|}{$^{4}4/^{1}m$} & $4/m$\\
    \hline
     irrep    & basis functions & basis functions\\ \hline
     $\Gamma^{1}$  &   \makecell{$k_{x}^2+k_{y}^{2}$;~$k_{z}^2$;\\$k_{x}\tutabf{x}+k_{y}\tutabf{y}$;~$k_{x}\tutabf{y}-k_{y}\tutabf{x}$.} & $k_{x}^2+k_{y}^{2}$;~$k_{z}^2$.
     \\ \hline
     $\Gamma^{2}$  &   $k_{z}\tutabf{z}$. & \makecell{$k_{x}\tutabf{x}+k_{y}\tutabf{y}$;~$k_{x}\tutabf{y}-k_{y}\tutabf{x}$;\\~$k_{z}\tutabf{z}$.}
     \\ \hline
     $\Gamma^{3}$  &   \makecell{$k_{x}^2-k_{y}^2$;\\$k_{x}\tutabf{x}-k_{y}\tutabf{y}$;~$k_{x}\tutabf{y}+k_{y}\tutabf{x}$.} &  $k_{x}^2-k_{y}^2$.
     \\ \hline
     $\Gamma^{4}$  &   $k_{x}k_{y}k_{z}\tutabf{z}$;~$(k_{x}^2-k_{y}^{2})k_{z}\tutabf{z}$. & $k_{x}\tutabf{x}-k_{y}\tutabf{y}$;~$k_{x}\tutabf{y}+k_{y}\tutabf{x}$.
     \\ \hline

     $\Gamma^{5}$  &   \makecell{$k_{x}k_{z}-ik_{y}k_{z}$;\\$k_{z}(\tutabf{x}-i\tutabf{y})$.} & $k_{x}k_{z}-ik_{y}k_{z}$.
     \\ \hline
     $\Gamma^{6}$  &  $(k_{x}-ik_{y})\tutabf{z}$. & $k_{z}(\tutabf{x}-i\tutabf{y})$;~$(k_{x}-ik_{y})\tutabf{z}$.
     \\ \hline
     $\Gamma^{7}$  &   \makecell{$k_{x}k_{z}+ik_{y}k_{z}$;\\ $k_{z}(\tutabf{x}+i\tutabf{y})$.}& $k_{x}k_{z}+ik_{y}k_{z}$.
     \\ \hline
     $\Gamma^{8}$  &   $(k_{x}+ik_{y})\tutabf{z}$. & $k_{z}(\tutabf{x}+i\tutabf{y})$;~$(k_{x}+ik_{y})\tutabf{z}$.
     \\ \hline
     
    \end{tabular}
}
    \caption{The basis functions for the noncoplanar spin point group $^{4}4/^{1}m$ and the point group $4/m$.}
    \label{bases-unitarynoncop}
\end{table}
The spin-singlet and triplet pairing states are decoupled in point group $4/m$, while they can coexist in the same superconducting channels of $^{4}4/^{1}m$. This is attributed to that $^{4}4/^{1}m$ does not include the inversion symmetry $[E||I]$, but the $4/m$ includes it. The group element $[E||I]$ in $4/m$ corresponds to the group element $[R_{\bbb{z}}(\pi)||I]$ in $^{4}4/^{1}m$, and they share the same representation matrices for all irreps. The differences in the spin operations between corresponding group elements of $4/m$ and $^{4}4/^{1}m$ result in discrepancies in the basis functions for each irrep of these two groups. For example, for irrep $\Gamma^{1}$, the basis functions $k_{x}\tutabf{x}+k_{y}\tutabf{y}$ and $k_{x}\tutabf{y}-k_{y}\tutabf{x}$, representing unitary $p$-wave superconducting states, are allowed in $^{4}4/^{1}m$ but forbidden in $4/m$. The reason is that the representation matrix of the group element $[R_{\bbb{z}}(\pi)||I]$ for $\Gamma^{1}$ of $^{4}4/^{1}m$ and that of the group element $[E||I]$ in $4/m$ are both identity matrices, while the basis functions $k_{x}\tutabf{x}+k_{y}\tutabf{y}$ and $k_{x}\tutabf{y}-k_{y}\tutabf{x}$ change sign under $\hat{P}_{[E||I]}$ but remain unchanged under $\hat{P}_{[R_{\bbb{z}}(\pi)||I]}$. Due to the same reason, for irreps $\Gamma^{3}$, $\Gamma^{5}$ and $\Gamma^{7}$, the basis functions representing spin-triplet pairing states are present in $^{4}4/^{1}m$ but absent in $4/m$.

For irreps $\Gamma^{2}$, $\Gamma^{4}$, $\Gamma^{6}$ and $\Gamma^{8}$, the spin-singlet pairing states are forbidden in both groups. The reason is as follows: the basis functions representing the spin-singlet superconducting states are all even scalar functions which remain unchanged under spatial inversion operation and are not affected by spin-rotation operations. However, the representation matrices of the group elements with spatial operation part being inversion $I$ for irreps $\Gamma^{i}~( i=2,4,6,8)$ in both groups are all $-1$. Therefore, spin-singlet pairing states are not allowed for these irreps in either group.

The physical meaning of all the basis functions listed in Table.~\ref{bases-unitarynoncop} has been discussed in previous sections, and their transition properties under spin and spatial operations can be analyzed in a similar manner.

\subsection{The noncoplanar spin point group $^{2_{z}}4/^{m_{x}}m^{m_{x}}m^{m_{y}}m$}

We then use the group $^{2_{z}}4/^{m_{x}}m^{m_{x}}m^{m_{y}}m$ as an example of noncoplanar spin point groups that include anti-unitary group elements to discuss the superconducting channels and basis functions within it.

The noncoplanar spin point group $^{2_{z}}4/^{m_{x}}m^{m_{x}}m^{m_{y}}m$ is isomorphic to the type-III magnetic point group $4/m'm'm'$. The coirreps and generators are listed in Table.~\ref{co-irrep-noncop}. The isomorphism between these two groups is defined as:
\begin{equation}
    \begin{aligned}
        [R_{\bbb{z}}(\pi)||C_{4z}]~&\rightarrow~[C_{4z}||C_{4z}],\\
        [E||C_{2x}]~&\rightarrow~[C_{2x}||C_{2x}],\\
        [\ccc{T}R_{\bbb{x}(\pi)}||IC_{2z}]~&\rightarrow~[\ccc{T}C_{2z}||IC_{2z}].
    \end{aligned}
\end{equation}
All the coirreps in Table.~\ref{co-irrep-noncop} belong to type (a) according to the Dimmock test given by Eq.~(\ref{Dimmock}). The basis functions for each coirrep are listed Table.~\ref{bases-non}. Due to the breaking of inversion symmetry, spin-singlet pairing states can coexist with spin-triplet pairing states in the same superconducting channel for both groups.  

From Table.~\ref{bases-non}, we find that for each coirrep, the basis functions representing spin-triplet pairing states in two groups are different, while the basis functions representing spin-singlet pairing states are the same. For example, for coirrep $\Gamma^{1}$, the basis function $k_{x}\tutabf{x}+k_{y}\tutabf{y}$ representing a unitary $p$-wave superconducting state is allowed in $4/m'm'm'$ but forbidden in $^{2_{z}}4/^{m_{x}}m^{m_{x}}m^{m_{y}}m$, while the basis functions $k_{x}^{2}+k_{y}^2$ and $k_{z}^2$ representing the $s$-wave superconducting states are allowed in both groups. The reason is as follows: under the basis function $k_{x}\tutabf{x}+k_{y}\tutabf{y}$, the matrices of generators belonging to the unitary part of the group $4/m'm'm'$ are identical to the representation matrices of these generators for $\Gamma^{1}$. In contrast, the matrix of generator $[E||C_{2x}]$ belonging to group $^{2_{z}}4/^{m_{x}}m^{m_{x}}m^{m_{y}}m$ under this basis function is $-1$, which is different from the representation matrix of this generator for $\Gamma^{1}$. Therefore, the basis function $k_{x}\tutabf{x}+k_{y}\tutabf{y}$ is allowed for $\Gamma^{1}$ of $4/m'm'm'$ but forbidden for $\Gamma^{1}$ of $^{2_{z}}4/^{m_{x}}m^{m_{x}}m^{m_{y}}m$. However, under basis functions $k_{x}^{2}+k_{y}^2$ and $k_{z}^2$, the matrices of all generators belonging to the unitary part of these two groups are identical to these generators' representation matrices for $\Gamma^{1}$. Thus, the basis functions $k_{x}^{2}+k_{y}^2$ and $k_{z}^2$ are allowed for $\Gamma^{1}$ of both groups. In addition, the basis functions $k_{x}k_{y}k_{z}\tutabf{x}$ and $k_{x}k_{y}k_{z}\tutabf{y}$, representing the unitary $f$-wave superconducting states, are allowed for $\Gamma^{1}$ of $^{2_{z}}4/^{m_{x}}m^{m_{x}}m^{m_{y}}m$ but forbidden for $\Gamma^{1}$ of $4/m'm'm'$. The reason is that the matrices of generators belonging to the unitary part of $^{2_{z}}4/^{m_{x}}m^{m_{x}}m^{m_{y}}m$ under basis functions $k_{x}k_{y}k_{z}\tutabf{x}$ and $k_{x}k_{y}k_{z}\tutabf{y}$ are identical to these generators' representation matrices for $\Gamma^{1}$. However, the operator $\hat{P}_{[C_{4z}||C_{4z}]}$ corresponding to the group element $[C_{4z}||C_{4z}]$ of $4/m'm'm'$ transforms the axial-vectors $\tutabf{x}$ and $\tutabf{y}$ into $\tutabf{y}$ and $-\tutabf{x}$, respectively. Thus, for one-dimensional coirrep $\Gamma^{1}$, basis functions $k_{x}k_{y}k_{z}\tutabf{x}$ and $k_{x}k_{y}k_{z}\tutabf{y}$ are not allowed. 

For coirrep $\Gamma^{2}$, the basis function $k_{x}^2-k_{y}^2$, representing a $d$-wave superconducting state, is allowed in both groups $^{2_{z}}4/^{m_{x}}m^{m_{x}}m^{m_{y}}m$ and $4/m'm'm'$. In contrast, the basis functions representing spin-triplet superconducting states differ between these two groups. For example, the basis function $k_{x}k_{y}k_{z}\tutabf{z}$, representing a unitary $f$-wave superconducting state, is allowed in $^{2_{z}}4/^{m_{x}}m^{m_{x}}m^{m_{y}}m$ but forbidden in $4/m'm'm'$. However, the basis function $k_{x}\tutabf{x}-k_{y}\tutabf{y}$, representing a $p$-wave superconducting state, is allowed in $4/m'm'm'$ but forbidden in $^{2_{z}}4/^{m_{x}}m^{m_{x}}m^{m_{y}}m$. These results can be understood through a similar analysis as for $\Gamma^{1}$.

For coirrep $\Gamma^{3}$, in both groups $^{2_{z}}4/^{m_{x}}m^{m_{x}}m^{m_{y}}m$ and $4/m'm'm'$, the basis function $k_{x}k_{y}(k_{x}^2-k_{y}^2)$, representing a $g$-wave superconducting state, is allowed. Conversely, the basis function $k_{z}\tutabf{z}$, representing a unitary $p$-wave superconducting state with only out-of-plane pairing components relative to the plane $z=0$, is present only in $^{2_{z}}4/^{m_{x}}m^{m_{x}}m^{m_{y}}m$. In addition, the basis function $k_{x}\tutabf{y}-k_{y}\tutabf{x}$, representing a unitary $p$-wave superconducting state with only in-plane pairing components, is present only in $4/m'm'm'$.

For coirrep $\Gamma^{4}$, the basis function $k_{x}k_{y}$ representing a $d$-wave superconducting state appears in both groups. However, similar to coirrep $\Gamma^{3}$, the basis functions $k_{z}\tutabf{x}$ and $k_{z}\tutabf{y}$, representing unitary $p$-wave superconducting states with only out-of-plane pairing components, are only allowed in $^{2_{z}}4/^{m_{x}}m^{m_{x}}m^{m_{y}}m$. Additionally, the basis function $k_{x}\tutabf{y}+k_{y}\tutabf{x}$, representing a unitary $p$-wave superconducting state with only in-plane pairing components, appears only in $4/m'm'm'$.

For two-dimensional coirrep $\Gamma^{5}$, the set of basis functions \{$k_{x}k_{z}-k_{y}k_{z}$, $k_{x}k_{z}-k_{y}k_{z}$\}, representing $d$-wave superconducting states, appears in both $^{2_{z}}4/^{m_{x}}m^{m_{x}}m^{m_{y}}m$ and $4/m'm'm'$. In contrast, the basis functions representing spin-triplet pairing states in these two groups are different. For example, the set of basis functions \{$(k_{x}-k_{y})\tutabf{x}$, $-(k_{x}+k_{y})\tutabf{x}$\}, representing unitary $p$-wave superconducting states, appears in $^{2_{z}}4/^{m_{x}}m^{m_{x}}m^{m_{y}}m$ but is absent in $4/m'm'm'$. The reason is as follows: on the one hand, the operator $\hat{P}_{[C_{4z}||C_{4z}]}$ corresponding to the generator $[C_{4z}||C_{4z}]$ of $4/m'm'm'$ transforms the basis function $(k_{x}-k_{y})\tutabf{x}$ into $(k_{y}+k_{x})\tutabf{y}$, which is not proportional to the other basis function $-(k_{x}+k_{y})\tutabf{x}$. On the other hand, the matrices of the generators belonging to the unitary part of the group $^{2_{z}}4/^{m_{x}}m^{m_{x}}m^{m_{y}}m$ under the set of basis functions \{$(k_{x}-k_{y})\tutabf{x}$, $-(k_{x}+k_{y})\tutabf{x}$\} are identical to the representation matrices of these generators for $\Gamma^{5}$. The presence or absence of other sets of basis functions for coirrep $\Gamma^{5}$ of these two groups can be understood in the same manner.
Additionally, it is worth to note that, the superconducting states represented by the set of basis functions \{$(k_{x}+k_{y})\tutabf{z}$, $(-k_{x}+k_{y})\tutabf{z}$\} in the group $^{2_{z}}4/^{m_{x}}m^{m_{x}}m^{m_{y}}m$ are equivalent to the superconducting states represented by the set of basis functions \{$(k_{x}-k_{y})\tutabf{z}$, $(k_{x}+k_{y})\tutabf{z}$\} in the group $4/m'm'm'$. This equivalence is due to the fact that these two sets of basis functions are linearly dependent on each other, and the superconducting order parameters are linear combinations of basis functions.

\begin{table}
    \centering
\resizebox{\linewidth}{!}{
    \begin{tabular}{>{\centering\arraybackslash}m{1cm}|>{\centering\arraybackslash}m{4cm}|>{\centering\arraybackslash}m{4cm}}
        \hline \hline
    \multicolumn{2}{>{\centering\arraybackslash}m{5cm}|}{$^{2_{z}}4/^{m_{x}}m^{m_{x}}m^{m_{y}}m$} & $4/m'm'm'$\\
    \hline
     coirrep    & basis functions &basis functions\\ \hline
     $\Gamma^{1}$  &   \makecell{$k_{x}^2+k_{y}^{2}$; $k_{z}^2$;\\$k_{x}k_{y}k_{z}\tutabf{x}$;~$k_{x}k_{y}k_{z}\tutabf{y}$.} & \makecell{$k_{x}^2+k_{y}^2$;$k_{z}^2$;\\ $k_{x}\tutabf{x}+k_{y}\tutabf{y}$; $k_{z}\tutabf{z}$.}\\ \hline
     $\Gamma^{2}$  &   \makecell{$k_{x}^2-k_{y}^2$;\\ $k_{x}k_{y}k_{z}\tutabf{z}$.} & \makecell{$k_{x}^2-k_{y}^2$;\\ $k_{x}\tutabf{x}-k_{y}\tutabf{y}$.} \\ \hline
     $\Gamma^{3}$  &   \makecell{$k_{x}k_{y}(k_{x}^2-k_{y}^2)$;\\$k_{z}\tutabf{z}$.} &\makecell{$k_{x}k_{y}(k_{x}^2-k_{y}^2)$;\\ $k_{x}\tutabf{y}-k_{y}\tutabf{x}$.} \\ \hline
     $\Gamma^{4}$  &   \makecell{$k_{x}k_{y}$;\\ $k_{z}\tutabf{x}$; $k_{z}\tutabf{y}$.} & \makecell{$k_{x}k_{y}$;\\$k_{x}\tutabf{y}+k_{y}\tutabf{x}$.}~\\ \hline
     $\Gamma^{5}$  &   \makecell{$\{k_{x}k_{z}-k_{y}k_{z},k_{x}k_{z}+k_{y}k_{z}\}$;\\$\{(k_{x}-k_{y})\tutabf{x},-(k_{x}+k_{y})\tutabf{x}\}$;\\ $\{(k_{x}-k_{y})\tutabf{y},-(k_{x}+k_{y})\tutabf{y}\}$;\\ $\{(k_{x}+k_{y})\tutabf{z},(-k_{x}+k_{y})\tutabf{z}\}$.} & \makecell{$\{k_{x}k_{z}-k_{y}k_{z},k_{x}k_{z}+k_{y}k_{z}\}$;\\ $\{(k_{x}-k_{y})\tutabf{z},(k_{x}+k_{y})\tutabf{z}\}$;\\ \{$k_{z}(\tutabf{x}-\tutabf{y})$,$k_{z}(\tutabf{x}+\tutabf{y})$.\}} \\ \hline
    \end{tabular}
}
    \caption{The basis functions for the noncoplanar spin point group $^{2_{z}}4/^{m_{x}}m^{m_{x}}m^{m_{y}}m$ and type-III magnetic point group $4/m'm'm'$.}
    \label{bases-non}
\end{table}

\section{Ginzburg-Landau free energy for superconducting state and its coupling with magnetic order}\label{GL}
By using basis functions, we can construct the superconducting order parameters and Ginzburg-Landau free energy to describe the properties of the superconducting states. For each coirrep, the superconducting order parameters vector $\bbb{d}(\bbb{k})$ and scalar $\psi(\bbb{k})$ are written as the linear combinations of the basis functions:
\begin{equation}
    \bbb{d}(\bbb{k})/\psi(\bbb{k})=\sum_{\nu}^{m_{i}}\eta_{\nu}\xi_{\nu}^{i},
\end{equation}
where $\xi_{\nu}^{i}$ represents the $\nu$-${th}$ basis function for the $i$-${th}$ coirrep $\Gamma^{i}$, complex numbers $\eta_{\nu}$ are the linear-combination coefficients, and $m_{i}$ is the dimension of the coirrep $\Gamma^{i}$. As discussed in Ref.~\cite{sigrist1991phenomenological}, the coefficients $\eta_{\nu}$ transform like coordinates in basis function space, and the Ginzburg-Landau free energy is constructed as an expansion in these coefficients. Moreover, since the free energy is invariant under spin point group elements, time-reversal operation, and U$(1)$ phase transformations, all terms in the free energy are real scalars and belong to the trivial representation of the spin point group. The second and fourth-order terms in the free energy can be constructed according to the trivial representation obtained from decomposing the direct products of coirreps $\Gamma\otimes\Gamma^{*}$ and $\Gamma\otimes\Gamma^{*}\otimes\Gamma\otimes\Gamma^{*}$, respectively. 

For example, we use the basis functions for coirrep $\Gamma^{12}_{\nu=1}$ of the collinear spin point group $G_{\text{SS}}^{l}\times ~^{1}4/ ^{1}m^{\overline{1}}m^{\overline{1}}m$ to construct the Ginzburg-Landau free energy.
As mentioned in Sec.\ref{bases-coll2}, the group $G_{\text{SS}}^{l}\times ~^{1}4/ ^{1}m^{\overline{1}}m^{\overline{1}}m$ belongs to type-II collinear spin point group, and there is no point group or magnetic point group which is isomorphic to it. Therefore, the Ginzburg-Landau free energy constructed using basis functions belonging to the coirrep of $G_{\text{SS}}^{l}\times ~^{1}4/ ^{1}m^{\overline{1}}m^{\overline{1}}m$ is distinct from those derived from any point group or magnetic point group. Additionally, unlike the case in the type-I collinear spin point group, the basis functions from a single two-dimensional coirrep of type-II collinear spin point group can exhibit different spin polarizations, which can lead to an intriguing phase diagram featuring distinct superconducting states with varying types of magnetic moments. The basis functions for coirrep $\Gamma^{12}_{\nu=1}$ are:
\begin{equation}\label{basg12}
\begin{aligned}
    &\{k_{z}(\tutabf{x}-i\tutabf{y}), k_{z}(\tutabf{x}+i\tutabf{y})\}; \\
    &\{k_{z}k_{x}k_{y}(k_{x}^2-k_{y}^2)(\tutabf{x}-i\tutabf{y}), -k_{z}k_{x}k_{y}(k_{x}^2-k_{y}^2)(\tutabf{x}+i\tutabf{y})\}.
\end{aligned}
\end{equation}
In Eq.(\ref{basg12}), we list two sets of the basis functions for the coirrep $\Gamma^{12}_{\nu=1}$, each with different orbital angular momenta. 
As will be discussed latter, these differences in orbital angular momentum are crucial for constructing the free energy that describes the Zeeman coupling between superconducting states and magnetic order. 
The basis functions $\{ \xi_{1}^{12,\nu=1}(\bbb{k}), \xi_{2}^{12,\nu=1}(\bbb{k}) \} \equiv \{\bbb{d}_{1}(\bbb{k}), \bbb{d}_{2}(\bbb{k})\}$ that we used to construct the free energy can be linear combinations of basis functions with lower and higher orbital angular momentum:
\begin{equation}\label{basis-high}
\begin{aligned}
        &\{\bbb{d}_{1}(\bbb{k}), \bbb{d}_{2}(\bbb{k})\}\\=&\{k_{z}(\tutabf{x}-i\tutabf{y})+ \lambda  k_{z}k_{x}k_{y}(k_{x}^2-k_{y}^2)(\tutabf{x}-i\tutabf{y}),\\
        & k_{z}(\tutabf{x}+i\tutabf{y})- \lambda  k_{z}k_{x}k_{y}(k_{x}^2-k_{y}^2)(\tutabf{x}+i\tutabf{y})
        \},
\end{aligned}
\end{equation}
where $\lambda$ is a constant. The decomposition of the direct product of coirrep $\Gamma^{12}_{\nu=1}$ and its conjugate can be expressed as:
\begin{equation}\label{decomps}
    \Gamma^{12}_{\nu=1}\otimes (\Gamma^{12}_{\nu=1})^{*}= \Gamma^{1}\oplus \Gamma^{3}\oplus \Gamma^{11}_{\nu=2}.
\end{equation}
There is one trivial representation in the decomposition given by Eq.~(\ref{decomps}). Therefore, the second-order term in the free energy is obtained accordingly:
\begin{equation}
    f_{2}=\alpha (|\eta_{1}|^2+|\eta_{2}|^2),
\end{equation}
where $\alpha$ is the coefficient of the second-order terms in the free energy. This coefficient determines the transition temperature of the superconducting states. Moreover, the fourth-order terms in the free energy can be obtained from the decomposition of the direct product:
\begin{equation}\label{dec4o}
    \Gamma^{12}_{\nu=1}\otimes (\Gamma^{12}_{\nu=1})^{*}\otimes \Gamma^{12}_{\nu=1}\otimes (\Gamma^{12}_{\nu=1})^{*}.
\end{equation}
The trivial representation $\Gamma^{1}$ appears three times during the decomposition of Eq.(\ref{dec4o}), as show in the following:
\begin{equation}
\begin{aligned}
    \Gamma^{1} \otimes \Gamma^{1}&= \Gamma^{1};\\
    \Gamma^{3} \otimes \Gamma^{3}&= \Gamma^{1};\\
    \Gamma^{11}_{\nu=2} \otimes \Gamma^{11}_{\nu=2}&= \Gamma^{1} \oplus \Gamma^{3} \oplus \Gamma^{11}_{\nu=4}.
\end{aligned}
\end{equation}
Accordingly, there are three different fourth-order terms in the free energy:
\begin{equation}
    \begin{aligned}
             &(|\eta_{1}|^2+|\eta_{2}|^2)^2;\\
             &(|\eta_{1}|^2-|\eta_{2}|^2)^2;\\
             &(|\eta_{1}\eta_{2}^{*}|^2+|\eta_{2}\eta_{1}^{*}|^2).
    \end{aligned}
\end{equation}
However, these three fourth-order terms are not linearly independent, only two of them are. Therefore, the fourth-order term in the superconducting free energy can be expressed as:
\begin{equation}
\begin{aligned}
    f_{4}&=\beta_{1}(|\eta_{1}|^2+|\eta_{2}|^2)^2+\beta_{2} (|\eta_{1}\eta_{2}^{*}|^2+|\eta_{2}\eta_{1}^{*}|^2)\\
    &=\beta_{1}(|\eta_{1}|^2+|\eta_{2}|^2)^2+2 \beta_{2} |\eta_{1}|^2|\eta_{2}|^2.
\end{aligned}
\end{equation}
The free energy for superconducting states is then derived as:
\begin{equation}\label{freesc}
    f_{\text{sc}}=\alpha (|\eta_{1}|^2+|\eta_{2}|^2)+\beta_{1}(|\eta_{1}|^2+|\eta_{2}|^2)^2+2 \beta_{2} |\eta_{1}|^2|\eta_{2}|^2,
\end{equation}
when $\alpha<0$, the superconducting states emerge. With a fixed second-order term $\alpha (|\eta_{1}|^2+|\eta_{2}|^2)$, the most stable superconducting state is determined by the coefficients of the fourth-order terms in the free energy ($\beta_{1}$ and $\beta_{2}$). Specifically, for a fixed positive $\beta_{1}$, the sign of $\beta_{2}$ determines which superconducting state is the most stable. 

When $\beta_{2}>0$, the superconducting state with the order parameter being $\eta_{1}\bbb{d}_{1}(\bbb{k})$ or $\eta_{2}\bbb{d}_{2}(\bbb{k})$ is the most stable. In this state, the Cooper pairs exhibit finite spin polarizations along the $+z$ or $-z$ axis. The magnetic moment of the superconducting state with the order parameter being $\eta_{1}\bbb{d}_{1}(\bbb{k})$ or $\eta_{2}\bbb{d}_{2}(\bbb{k})$ is given by:
\begin{equation}
    \mp 2 |\eta_{1/2}|^2(k_{z}^2+\lambda^2 k_{z}^2k_{x}^2k_{y}^2(k_{x}^2-k_{y}^2)^2 \pm 2\lambda k_{z}^2k_{x}k_{y}(k_{x}^2-k_{y})) \tutabf{z},
\end{equation}
respectively. In both case, the magnetic moments of the superconducting states do not belong to the trivial coirrep $\Gamma^{1}$ of the group $G^{l}_{SS} \times ^{1}4/^{1}m^{\overline{1}}m^{\overline{1}}m$. Instead, they are a mixture of the $s$-wave and $g$-wave magnetic moments.

When $\beta_{2}<0$, the superconducting state with the order parameter $\bbb{d}(\bbb{k})=\eta_{1} \bbb{d}_{1}(\bbb{k})+ \eta_{2} \bbb{d}_{2}(\bbb{k})$ is the most stable, where $\eta_{1}$ and $\eta_{2}$ satisfy $|\eta_{1}|=|\eta_{2}|\neq 0$. For a fixed second-order term $\alpha (|\eta_{1}|^2+|\eta_{2}|^2)$, the condition $|\eta_{1}|=|\eta_{2}|$ ensures that the fourth-order term $\beta_{2}|\eta_{1}|^2|\eta_{2}|^2$ in the free energy Eq.(\ref{freesc}) is minimized when $\beta_{2}<0$. The superconducting state with the order parameter being $\bbb{d}(\bbb{k})=\eta_{1} \bbb{d}_{1}(\bbb{k})+ \eta_{2} \bbb{d}_{2}(\bbb{k})$ is also nonunitary, but the symmetry property of the magnetic moment differs from that in $\beta_{2}>0$ case. The magnetic moment of the superconducting states for $\beta_{2}<0$ is written as:
\begin{equation}
    -4\lambda (|\eta_{1}|^2+|\eta_{2}|^2) k_{z}^2k_{x}k_{y}(k_{x}^2-k_{y}^2) \tutabf{z},
\end{equation}
which represents a $g$-wave magnetic order and belongs to the trivial coirrep $\Gamma_{1}$ of the group $G^{l}_{SS} \times ^{1}4/^{1}m^{\overline{1}}m^{\overline{1}}m$.

Next, we construct the free energy describing the coupling between superconducting states belonging to the coirrep $\Gamma^{12}_{\nu=1}$ and the magnetic order whose symmetry is described by the spin point group $G_{\text{SS}}^{l}\times ~^{1}4/ ^{1}m^{\overline{1}}m^{\overline{1}}m$. The order parameter for magnetic order is 
\begin{equation}\label{magop}
    \bbb{m}(\bbb{k})=M k_{x}k_{y}(k_{x}^2-k_{y}^2)\tutabf{z},
\end{equation}
which represents a $g$-wave magnetic order and belongs to the trivial coirrep $\Gamma_{1}$ of the group $G^{l}_{SS} \times ^{1}4/^{1}m^{\overline{1}}m^{\overline{1}}m$. 
In Eq.~(\ref{magop}), $M$ represents the coefficient of the magnetic order parameter. Accordingly, the free energy for this magnetic order can be written as:
\begin{equation}\label{freem}
    f_{\text{m}}=\alpha' M^2+\beta' M^4.
\end{equation}

Like the free energy for the superconducting state (Eq.~(\ref{freesc})) and the magnetic order (Eq.~(\ref{freem})), the free energy for the coupling between the magnetic order and the superconducting state must be a real scalar and belong to the trivial coirrep $\Gamma_{1}$ of the group $G^{l}_{SS} \times ^{1}4/^{1}m^{\overline{1}}m^{\overline{1}}m$.
In addition, the Zeeman coupling between superconducting order parameter $\bbb{d}(\bbb{k})$ and magnetic order parameter $\bbb{m}(\bbb{k})$ is expressed as:
\begin{equation} \label{cpcoff}
    f_{\text{coupling},\text{Z}}=\bbb{m}(\bbb{k})\cdot i (\bbb{d}(\bbb{k})\times \bbb{d}^{*}(\bbb{k})).
\end{equation}
Thus, to obtain the free energy for this coupling term, we use the basis functions $\{\bbb{d}_{1}(\bbb{k}), \bbb{d}_{2}(\bbb{k})\}$ listed in Eq.(\ref{basis-high}) to construct the magnetic moment for the superconducting states as:
\begin{equation}\label{scmag_1}
    \begin{aligned}
        i (\bbb{d}_{1}(\bbb{k})\times \bbb{d}_{1}(\bbb{k})^{*}+\bbb{d}_{2}(\bbb{k})\times \bbb{d}_{2}(\bbb{k})^{*})\\
         =-4\lambda k_{z}^2k_{x}k_{y}(k_{x}^2-k_{y}^2) \tutabf{z},
    \end{aligned}
\end{equation}
which belongs to the trivial coirrep $\Gamma_{1}$ of the group $G^{l}_{SS} \times ^{1}4/^{1}m^{\overline{1}}m^{\overline{1}}m$, identical to the order parameter for the $g$-wave magnetic order $\bbb{m}(\bbb{k})$. Because the coefficients $\eta_{\nu}$ transform like coordinates in basis function space $\{\bbb{d}_{1}(\bbb{k}), \bbb{d}_{2}(\bbb{k})\}$, according to the Eq.(\ref{cpcoff}) and (\ref{scmag_1}), the Zeeman coupling term between superconducting order parameter and $g$-wave magnetic order in the free energy can be expressed as:
\begin{equation}
    f_{\text{scm},\text{Z}}=\gamma_{1} M (|\eta_{1}|^2+|\eta_{2}|^2),
\end{equation}
where $\gamma_{1}$ is the Zeeman coupling strength between magnetic order and the magnetic moment of superconducting states. $f_{\text{scm},\text{Z}}$ is a second-order term for the free energy of superconductivity and will either increase or decrease the transition temperature of the superconducting states, depending on the sign of the coefficient $(\gamma_{1} M)$ for this coupling term.

Besides the Zeeman coupling term in the free energy, there are higher-order terms which couple the superconducting states and magnetic order parameters. Such as the fourth-order term in the free energy:
\begin{equation}\label{fsc-m4}
    f_{\text{scm},4}=\gamma_{2} M^2 (|\eta_{1}|^2+|\eta_{2}|^2),
\end{equation}
where $\gamma_2$ is the coupling strength. Similar to the Zeeman term, the coupling term in Eq.~(\ref{fsc-m4}) is also a second-order term for the free energy of superconductivity, it will also affect the transition temperature of the superconducting states. 

As for the sixth-order term in the free energy, one possible form combines the second-order terms of superconductivity with the fourth-order terms of magnetism and is expressed as:
\begin{equation}\label{fsc-m6}
   \begin{aligned}
        f_{\text{scm},6,1}=&\gamma'_{3} M^4 (|\eta_{1}|^2+|\eta_{2}|^2),
   \end{aligned}
\end{equation}
where $\gamma^{'}_{3}$ is the coupling strength. This term will affect the transition temperature of the superconducting states.
Other sixth-order terms are combinations of the second-order terms of magnetic orders and the fourth-order terms of superconductivity. In the free energy, these can be written as:
\begin{equation}\label{fsc-m6}
   \begin{aligned}
        f_{\text{scm},6,2}=&\gamma_{3} M^2 (|\eta_{1}|^2+|\eta_{2}|^2)^2+ \gamma_{4} M^2 (|\eta_{1}\eta_{2}^{*}|^2+|\eta_{2}\eta_{1}^{*}|^2),\\
                  =&\gamma_{3} M^2 (|\eta_{1}|^2+|\eta_{2}|^2)^2+ 2\gamma_{4} M^2 (|\eta_{1}|^2|\eta_{2}|^2),
   \end{aligned}
\end{equation}
where $\gamma_3$ and $\gamma_4$ are the coupling strengths. The terms shown in Eq.~(\ref{fsc-m6}) are the fourth-order terms in the free energy of superconductivity, and the term with coefficient $\gamma_4$ may drive phase transitions between different superconducting states.

\section{Application to realistic system}\label{application}
In this section, we discuss the application of our theory to two realistic systems, one is the electric-field-induced altermagnetism in monolayer FeSe~\cite{mazin2023induced}, the other is the all-in-all-out (AIAO) order on kagome lattice~\cite{sachdev1992,Yang2024}. The symmetries of both systems can be described by spin space groups. The basis functions are constructed by using the coirreps of the corresponding spin point groups.

\subsection{The monolayer FeSe with an applied out-of-plane electric field}
\begin{figure}[b]
    \centering
    \includegraphics[width=0.4\textwidth]{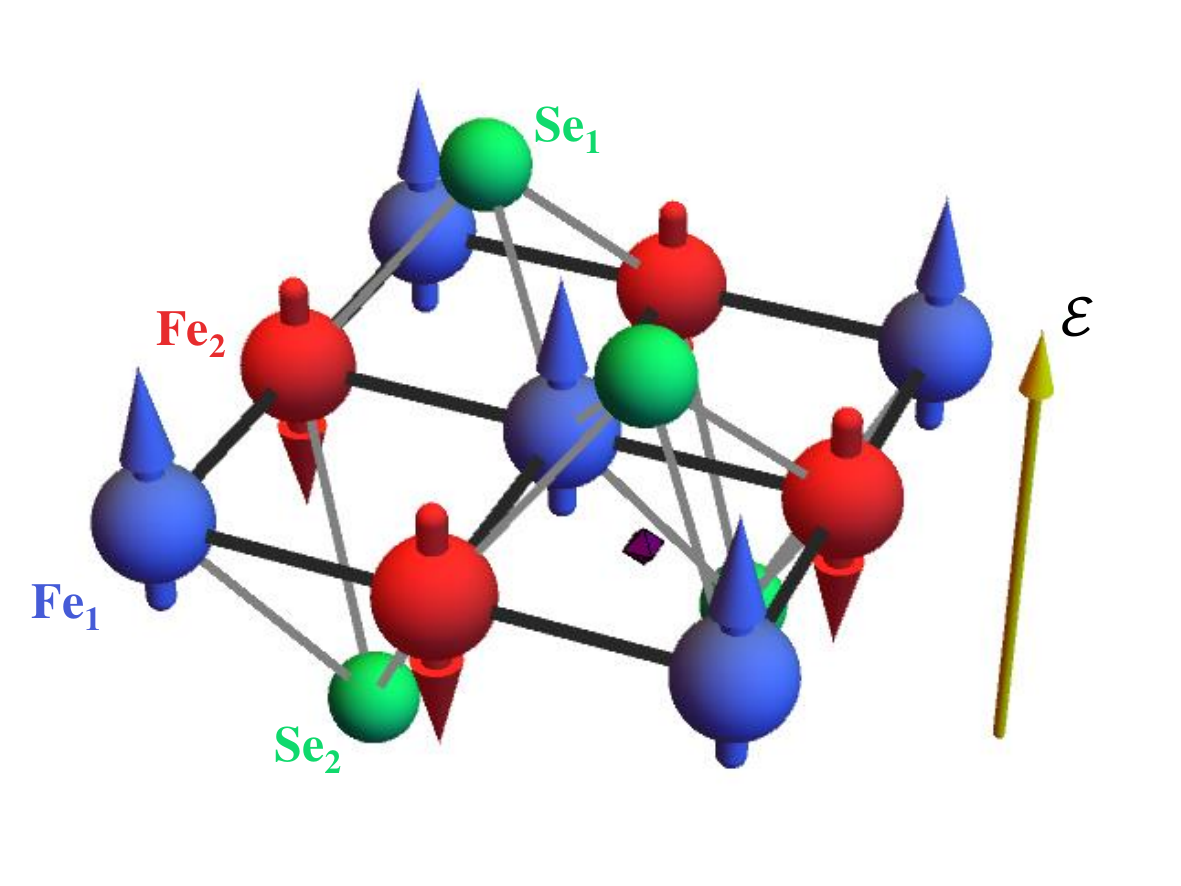}
    \caption{The real space structure and spin magnetic moments $\bbb{M}(\bbb{r})$ of the electric-field-gated monolayer FeSe. The yellow arrow indicates the applied electric field $\ccc{E}$. The green spheres represent the Se atoms. The blue and red spheres represent two nonequivalent Fe atoms (labeled as Fe$_{1}$ and Fe$_{2}$, respectively). The spin magnetic moments of this altermagnetic system are depicted by the blue and red arrows. It is important to note that, due to the presence of the electron field, the Se atoms above the Fe-plane (labeled as Se$_{1}$) are closer to the Fe-plane than the Se atoms below the Fe-plane (labeled as Se$_{2}$). This fact leads to the broken of the $\ccc{P}\ccc{T}$ symmetry $[\ccc{T}||I]$ and the spatial symmetry $[E||IC_{4z}]$. However, the symmetries $[\ccc{T}||C_{4z}]$ and $[E||IC_{2y}]$ are preserved. Their rotation center is marked by the purple octahedron.}
    \label{FeSe_ele}
\end{figure}

As discussed in Ref.~\cite{mazin2023induced}, altermagnetism can be induced in monolayer FeSe by applying an out-of-plane electric field. The real space pattern and the spin magnetic moments $\bbb{M}(\bbb{r})$ of the electric-field-gated monolayer FeSe are illustrated in Fig.~\ref{FeSe_ele}. The $[\ccc{T}||I]$ symmetry is broken by the electric field $\ccc{E}$. However, the symmetries $[\ccc{T}||C_{4z}]$ and $[E||IC_{2y}]$ are preserved, and their rotation center is marked by the purple octahedron.
Accordingly, we find that the spin point group corresponding to the spin space group that describes the symmetry of the electric-field-gated monolayer FeSe is $G^{l}_{SS}\times~^{\overline{1}}4^{1}m^{\overline{1}}m$. The group $G^{l}_{SS}\times~^{\overline{1}}4^{1}m^{\overline{1}}m$ belongs to the type-II collinear spin point groups. The generators and coirreps of it are listed in the Table.~\ref{co-irrep-appl2}. All the coirreps belong to type (a) according to the Dimmock test Eq.~(\ref{Dimmock}). By applying the approach shown in the flowchart Fig.~\ref{flowchart}, the basis functions for each superconducting channel can be constructed. The results are shown in Table.~\ref{bases-appl2}.
\begin{table}[t]
    \centering
        \begin{tabular}{>{\centering\arraybackslash}m{1.5cm}|>{\centering\arraybackslash}m{6.5cm}}
    \hline\hline
    \multicolumn{2}{>{\centering\arraybackslash}m{8cm}}{$G^{cl}_{SS}\times~^{\overline{1}}4^{1}m^{\overline{1}}m$}\\
    \hline
     coirrep    & basis functions \\ \hline
     $\Gamma^{1}$  &   $k_{x}^2+k_{y}^{2}$; $k_{z}^2$; $(k_{x}^2-k_{y}^2)k_{z}\tutabf{z}$.  \\ \hline
     $\Gamma^{2}$  &   $k_{z}\tutabf{z}$;~$k_{x}^2-k_{y}^2$. \\ \hline
     $\Gamma^{3}$  &    $k_{x}k_{y}(k_{x}^2-k_{y}^2)$;~$k_{x}k_{y}k_{z}\tutabf{z}$  \\ \hline
     $\Gamma^{4}$  &    $k_{x}k_{y}$;~$k_{x}k_{y}(k_{x}^2-k_{y}^2)k_{z}\tutabf{z}$.  \\ \hline
     $\Gamma^{5}$  &   \makecell{\{$k_{y}k_{z}$, $-k_{x}k_{z}$\}; \\ \{$k_{y}\tutabf{z}$, $k_{x}\tutabf{z}$\}. }   \\ \hline
     \multicolumn{2}{c}{$\nu=1$}\\ \hline
     $\Gamma^{6}_{\nu=1}$  &   \makecell{ \{$k_{z}(\tutabf{x}-i\tutabf{y})$, $k_{z}(\tutabf{x}+i\tutabf{y})$\}.}   \\ \hline
     $\Gamma^{7}_{\nu=1}$  &   \{$k_{x}k_{y}k_{z}(\tutabf{x}-i\tutabf{y})$, $-k_{x}k_{y}k_{z}(\tutabf{x}+i\tutabf{y})$\}.   \\ \hline
     \multicolumn{2}{c}{$\mu=1$}\\ \hline
     $\Gamma^{8}_{\mu=1}$  &   \makecell{ \{$k_{y}(\tutabf{x}-i\tutabf{y})$, $-k_{x}(\tutabf{x}+i\tutabf{y})$\}. }
       \\ \hline
     \multicolumn{2}{c}{$\mu=-1$}\\ \hline
     $\Gamma^{8}_{\mu=-1}$  &   \makecell{ \{$k_{y}(\tutabf{x}+i\tutabf{y})$, $-k_{x}(\tutabf{x}-i\tutabf{y})$\}. }
       \\ \hline
       
    \end{tabular}
    \caption{The basis functions for the collinear spin point group $G^{l}_{SS}\times~^{\overline{1}}4^{1}m^{\overline{1}}m$. Here, $\nu \in \mathbb{N}/\{0\}$ and $\mu \in \mathbb{Z}/\{0\}$.}
    \label{bases-appl2}
\end{table}
Due to the absence of inversion symmetry, spin-singlet and triplet pairing states can coexist in the same superconducting channels. Next, we briefly introduce the superconducting states represented by basis functions for each channel.

For coirreps $\Gamma^{i}$ with $i=1,2,...,5$, there is no $\nu$ or $\mu$ index, which means the basis functions do not represent nonunitary triplet paring states. For coirrep $\Gamma^{1}$, the scalar basis functions $k_{x}^2+k_{y}^{2}$ and $k_{z}^{2}$ both represent s-wave superconducting states. In the same superconducting channel, the vector basis function $(k_{x}^2-k_{y}^2)k_{z}\tutabf{z}$ represents a unitary $f$-wave superconducting state.

For coirrep $\Gamma^{2}$, the scalar basis function $k_{x}^{2}-k_{y}^{2}$ represents a $d$-wave superconducting state, and the vector basis function $k_{z}\tutabf{z}$ represents a unitary $p$-wave superconducting state with only out-of-plane pairing components. Moreover, for coirrep $\Gamma^{3}$, the scalar basis function $k_{x}k_{y}(k_{x}^{2}-k_{y}^{2})$ represents a $g$-wave superconducting state with only in-plane pairing components, and the vector basis function $k_{x}k_{y}k_{z}\tutabf{z}$ represents a unitary $f$-wave superconducting state. For coirrep $\Gamma^{4}$, the scalar basis function $k_{x}k_{y}$ represents a $d$-wave superconducting state, and the vector function $k_{x}k_{y}(k_{x}^{2}-k_{y}^{2})k_{z}\tutabf{z}$ represents a unitary $h$-wave superconducting state. 

For two-dimensional coirreps $\Gamma^{5}$, the set of scalar basis functions \{$k_{y}k_{z}$, $-k_{x}k_{z}$\} represents $d$-wave superconducting states with out-of-plane pairing components. In addition, the set of vector basis functions \{$k_{y}\tutabf{z}$, $k_{x}\tutabf{z}$\} represents unitary $p$-wave superconducting states with only in-plane pairing components.

However, for coirreps $\Gamma^{6}$, $\Gamma^{7}$ and $\Gamma^{8}$, there are nonzero $\nu$ or $\mu$ indices. As discussed in Sec.~\ref{collinear}, the $z$-components of Cooper pairs' spin angular momentum of the superconducting states represented by the basis functions in these superconducting channels are nonzero. Therefore, only the basis functions representing nonunitary spin-triplet pairing states are allowed in these channels. For coirrep $\Gamma^{6}_{\nu=1}$, the two basis functions \{$k_{z}(\tutabf{x}-i\tutabf{y})$, $k_{z}(\tutabf{x}+i\tutabf{y})$\} represent nonunitary $p$-wave superconducting states with opposite spin polarizations. Similarly, for coirrep $\Gamma^{7}_{\nu=1}$, the two basis functions \{$k_{x}k_{y}k_{z}(\tutabf{x}-i\tutabf{y})$, $-k_{x}k_{y}k_{z}(\tutabf{x}+i\tutabf{y})$\} represent nonunitary $f$-wave superconducting states with opposite spin polarizations. 

Additionally, the basis functions for coirrep $\Gamma^{8}_{\mu=1}$ are \{$k_{y}(\tutabf{x}-i\tutabf{y})$, $-k_{x}(\tutabf{x}+i\tutabf{y})$\} which represent two nonunitary $p$-wave superconducting states with $z$-components of Cooper pairs' spin angular momentum being $\pm \hbar$. When the $\mu$ index changes sign, the $z$-components of Cooper pairs' spin angular momentum of superconducting states represented by the two basis functions are exchanged. Thus, the basis functions for coirrep $\Gamma^{8}_{\mu=-1}$ are \{$k_{y}(\tutabf{x}+i\tutabf{y})$, $-k_{x}(\tutabf{x}-i\tutabf{y})$\}. Although both sets of basis functions for coirreps $\Gamma^{8}_{\mu=1}$ and $\Gamma^{8}_{\mu=-1}$ represent nonunitary $p$-wave superconducting states, the specific forms of the superconducting order parameters for these two coirreps are different in general because these two sets of basis functions are linearly independent.

\subsection{All-in-all-out order on kagome lattice}
\begin{figure}[b]
    \centering
    \includegraphics[width=0.4\textwidth]{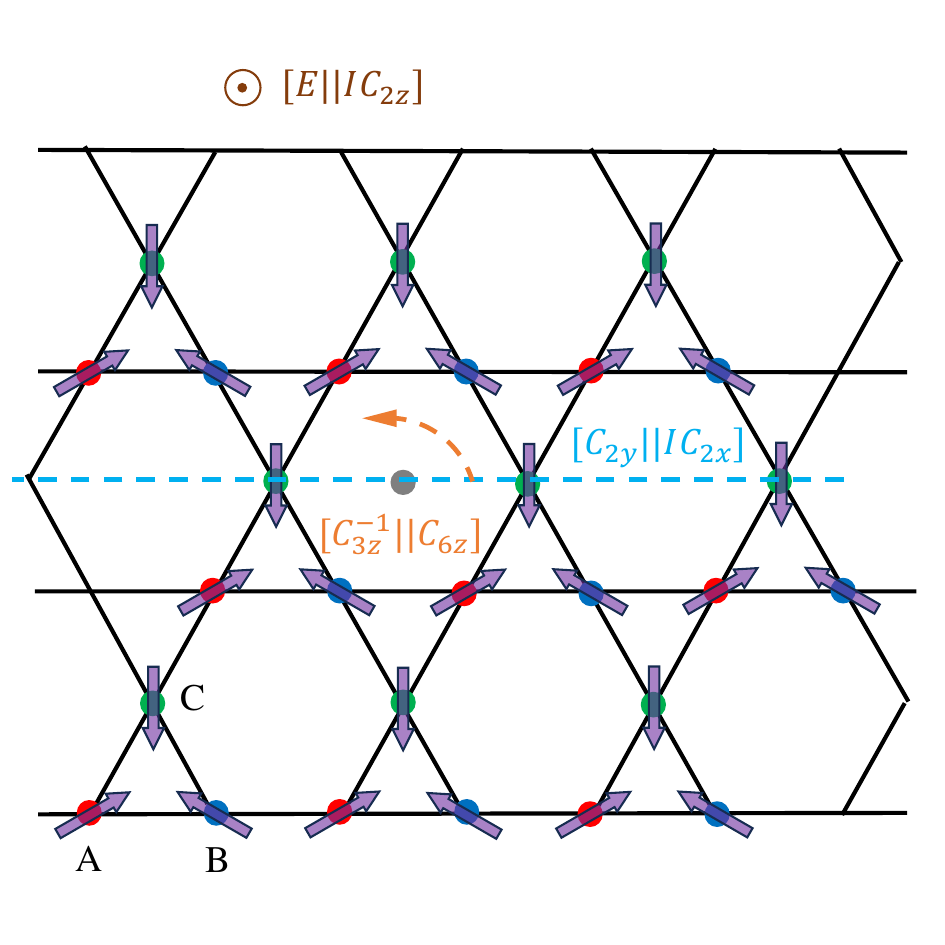}
    \caption{The real space structure and spin magnetic moments of all-in-all-out (AIAO) order on kagome lattice. The red, blue, and green lattices label the A,B and C sublattices, respectively. The arrows represent the local spin magnetic moments on each sublattice. The group elements labeled in the diagram are the generators of the nontrivial spin point group part $^{3_{z}}6/^{1}m^{2_{x}}m^{2_{xy}}m$ of this coplanar spin point group. The cyan dashed line indicates the normal direction of the spatial mirror ($IC_{2x}$) plane. The brown dot marks the normal direction of another spatial mirror ($IC_{2z}$) plane of the AIAO order. The dashed orange arrow represents the direction of both the spatial and spin rotations about the $\bbb{z}$-axis.}
    \label{aiao}
\end{figure}

The real space structure and spin magnetic moments of the AIAO order on kagome lattice are illustrated in Fig.~\ref{aiao}. According to Fig.~\ref{aiao}, we conclude that the spin point group corresponding to the spin space group which describes the symmetry of this AIAO order is $G_{\text{SS}}^{p}\times~^{3_{z}}6/^{1}m^{2_{x}}m^{2_{xy}}m$. This spin point group belongs to the coplanar spin point groups and is isomorphic to the type-II magnetic point group $6/mmm1'$. 
The generators of the spin point group $G_{\text{SS}}^{p}\times~^{3_{z}}6/^{1}m^{2_{x}}m^{2_{xy}}m$ and type-II magnetic point group $6/mmm1'$ are listed in Table.~\ref{co-irrep-appl1}. By using the generators of these two groups, the isomorphism between them can be expressed as:
\begin{equation}
    \begin{aligned}
        [R_{\bbb{z}}(4\pi/3)||C_{6z}]~&\rightarrow~[C_{6z}||C_{6z}],\\
        [E||IC_{2z}]~&\rightarrow~[C_{2z}||IC_{2z}],\\
        [R_{\bbb{y}}(\pi)||IC_{2x}]~&\rightarrow~[C_{2x}||IC_{2x}],\\
        [\ccc{T}R_{\bbb{z}}(\pi)||E]~&\rightarrow~[\ccc{T}||E].
    \end{aligned}
\end{equation}
Therefore, they share the same coirrep table (see Table.~\ref{co-irrep-appl1}). Additionally, all coirreps belong to type (a) according to the Dimmock test Eq.~(\ref{Dimmock}). The basis functions for each coirrep of these two groups are listed in Table.~\ref{bases-appl1}. Due to the existence of the inversion symmetry $[E||I]$, the spin-singlet and triplet pairing states are decoupled in both groups. 

We start by discussing the spin-singlet pairing states. For the coirreps $\Gamma^{i}$ with $i=1,3,5,7,9,11$, the basis functions are all scalar functions representing the spin-singlet superconducting states. In Table.~\ref{bases-appl1}, the spin point group $G_{\text{SS}}^{p}\times~^{3_{z}}6/^{1}m^{2_{x}}m^{2_{xy}}m$ and the type-II magnetic point group $6/mmm1'$ share the same basis functions for coirreps $\Gamma^{i}$ with $i=1,3,5,7,9,11$. The reason is that the generators belonging to the unitary part of $G_{\text{SS}}^{p}\times~^{3_{z}}6/^{1}m^{2_{x}}m^{2_{xy}}m$ share the same spatial operations and representation matrices with the corresponding generators in $6/mmm1'$, and the forms of scalar basis functions are only determined by the spatial operations and representation matrices of the unitary group elements in these two groups.

The physical meanings of the basis functions for each coirrep $\Gamma^{i}~(i=1,3,5,7,9,11)$ are as follows: The basis functions $k_{x}^2+k_{y}^{2}$ and $k_{z}^2$ for coirrep $\Gamma^{1}$ are invariant under all spatial and spin operations and represent $s$-wave superconducting states. For coirrep $\Gamma^{3}$, the basis function $\xi^{3}=k_{x}k_{y}[3(k_{x}^2-k_{y}^2)^2-4k_{x}^2k_{y}^2]$ represents an $i$-wave superconducting state. It is invariant under the transformation operator $\hat{P}_{[C_{6z}||C_{6z}]}$ but changes sign under the transformation operator $\hat{P}_{[C_{2x}||IC_{2x}]}$:
\begin{equation}
    \begin{aligned}
        \hat{P}_{[C_{6z}||C_{6z}]}\xi^{3}=\xi^{3},\\
        \hat{P}_{[C_{2x}||IC_{2x}]}\xi^{3}=-\xi^{3}.
    \end{aligned}
\end{equation}
Moreover, the basis function $\xi^{5}=(3k_{x}^2k_{y}-k_{y}^3)k_{z}$ for coirrep $\Gamma^{5}$ and the basis function $\xi^{7}=(k_{x}^3-3k_{y}^2k_{x})k_{z}$ for coirrep $\Gamma^{7}$ both represent $g$-wave superconducting states. The transformation operator $\hat{P}_{[C_{6z}||C_{6z}]}$ changes their sign:
\begin{equation}
    \hat{P}_{[C_{6z}||C_{6z}]}\xi^{5/7}=-\xi^{5/7}.
\end{equation}

As for the two-dimensional coirreps $\Gamma^{9}$ and $\Gamma^{11}$, the sets of basis functions \{$\xi^{9}_{1}$, $\xi^{9}_{2}$\}$=$\{$(k_{x}+ik_{y})^2$, $(k_{x}-ik_{y})^2$\} and \{$\xi^{11}_{1}$, $\xi^{11}_{2}$\}$=$\{$k_{x}k_{z}-ik_{y}k_{z}$, $-k_{x}k_{z}-ik_{y}k_{z}$\} both represent $d\pm i d$ superconducting states. However, the superconducting states represented by the basis functions \{$\xi^{9}_{1}$, $\xi^{9}_{2}$\} have only in-plane pairing components relative to the plane $z=0$, while the superconducting states represented by the basis functions \{$\xi^{11}_{1}$, $\xi^{11}_{2}$\} contain out-of-plane paring components.

\begin{table}
    \centering
\resizebox{\linewidth}{!}{
    \begin{tabular}{>{\centering\arraybackslash}m{1cm}|>{\centering\arraybackslash}m{4.2cm}|>{\centering\arraybackslash}m{4.2cm}}
        \hline \hline
    \multicolumn{2}{>{\centering\arraybackslash}m{5cm}|}{$G_{\text{SS}}^{p}\times~^{3_{z}}6/^{1}m^{2_{x}}m^{2_{xy}}m$} & $6/mmm1'$\\
    \hline
     coirrep    & basis functions &basis functions\\ \hline
     $\Gamma^{1}$  &   $k_{x}^2+k_{y}^{2}$; $k_{z}^2$. & \makecell{$k_{x}^2+k_{y}^2$;$k_{z}^2$.}\\ \hline
     $\Gamma^{2}$  &   $k_{z}\tutabf{z}$. & \makecell{$k_{x}\tutabf{x}+k_{y}\tutabf{y}$;$k_{z}\tutabf{z}$.} \\ \hline
     $\Gamma^{3}$  &   $k_{x}k_{y}[3(k_{x}^2-k_{y}^2)^2-4k_{x}^2k_{y}^2]$. & $k_{x}k_{y}[3(k_{x}^2-k_{y}^2)^2-4k_{x}^2k_{y}^2]$. \\ \hline
     $\Gamma^{4}$  &   $2k_{x}k_{y}k_{z}\tutabf{x}+(k_{x}^2-k_{y}^2)k_{z}\tutabf{y}$. & $k_{x}\tutabf{y}-k_{y}\tutabf{x}$.~\\ \hline
     $\Gamma^{5}$  &   $(3k_{x}^2k_{y}-k_{y}^3)k_{z}$. & $(3k_{x}^2k_{y}-k_{y}^3)k_{z}$. \\ \hline
     $\Gamma^{6}$  &  $k_{x}\tutabf{y}-k_{y}\tutabf{x}$.  & \makecell{$(k_{y}^3-3k_{x}^2k_{y})\tutabf{z}$;\\$2k_{x}k_{y}k_{z}\tutabf{x}+(k_{x}^2-k_{y}^2)k_{z}\tutabf{y}$.} \\ \hline
     $\Gamma^{7}$  &   $(k_{x}^3-3k_{y}^2k_{x})k_{z}$. & $(k_{x}^3-3k_{y}^2k_{x})k_{z}$.  \\ \hline
     $\Gamma^{8}$  &   $k_{x}\tutabf{x}+k_{y}\tutabf{y}$. &\makecell{$(k_{x}^3-3k_{x}k_{y}^2)\tutabf{z}$;\\$-2k_{x}k_{y}k_{z}\tutabf{y}+(k_{x}^2-k_{y}^2)k_{z}\tutabf{x}$.} \\ \hline
     $\Gamma^{9}$  &   \{$(k_{x}+ik_{y})^2$, $(k_{x}-ik_{y})^2$\}. &\{$(k_{x}+ik_{y})^2$, $(k_{x}-ik_{y})^2$\}. \\ \hline
     $\Gamma^{10}$ &  \makecell{\{$k_{z}(\tutabf{x}-i\tutabf{y})$, $k_{z}(\tutabf{x}+i\tutabf{y})$\}.
     } & \{$(k_{x}+ik_{y})(\tutabf{x}+i\tutabf{y})$, $(k_{x}-ik_{y})(\tutabf{x}-i\tutabf{y})$\}.  \\ \hline
     $\Gamma^{11}$  &   \{$k_{x}k_{z}-ik_{y}k_{z}$, $-k_{x}k_{z}-ik_{y}k_{z}$\}. &\{$k_{x}k_{z}-ik_{y}k_{z}$, $-k_{x}k_{z}-ik_{y}k_{z}$\}. \\ \hline
     $\Gamma^{12}$ &  \makecell{\{$(k_{x}+ik_{y})(\tutabf{x}+i\tutabf{y})$, \\$(k_{x}-ik_{y})(\tutabf{x}-i\tutabf{y})$\};\\ \{$(k_{x}-ik_{y})\tutabf{z}$, $(k_{x}+ik_{y})\tutabf{z}$\}.
     } & \makecell{\{$(k_{x}-i k_{y})\tutabf{z}$, $(k_{x}+i k_{y})\tutabf{z}$\};\\ \{$k_{z}(\tutabf{x}-i\tutabf{y})$, $k_{z}(\tutabf{x}+i\tutabf{y})$\}.}  \\ \hline
    \end{tabular}
}
    \caption{The basis functions for the coplanar spin point group $G_{\text{SS}}^{p}\times~^{3_{z}}6/^{1}m^{2_{x}}m^{2_{xy}}m$ and the type-II magnetic point group $6/mmm1'$.}
    \label{bases-appl1}
\end{table}

Next, we discuss the spin-triplet pairing states. The basis functions for coirreps $\Gamma^{i}$ with $i=2,4,6,8,10,12$ are all vector functions representing the spin-triplet superconducting states. For coirrep $\Gamma^{2}$, the basis function $k_{x}\tutabf{x}+k_{y}\tutabf{y}$, representing the unitary $p$-wave superconducting state with only in-plane pairing components, is forbidden in the spin point group $G_{\text{SS}}^{p}\times~^{3_{z}}6/^{1}m^{2_{x}}m^{2_{xy}}m$ but allowed in the type-II magnetic point group $6/mmm1'$. The reason is as follows: the operator $\hat{P}_{[R_{\tutabf{z}}(4\pi/3)||C_{6z}]}$ transforms the basis function $k_{x}\tutabf{x}+k_{y}\tutabf{y}$ into $-k_{x}\tutabf{x}-k_{y}\tutabf{y}$, while the corresponding operator $\hat{P}_{[C_{6z}||C_{6z}]}$ in the type-II magnetic point group does not change this basis function. The representation matrix of generator $[R_{\tutabf{z}}(4\pi/3)||C_{6z}]$ for $\Gamma^{2}$ of $G_{\text{SS}}^{p}\times~^{3_{z}}6/^{1}m^{2_{x}}m^{2_{xy}}m$ and that of $[C_{6z}||C_{6z}]$ for $\Gamma^{2}$ of $6/mmm1'$ are both identity matrices. In addition, the matrices of other generators belonging to the unitary part of $6/mmm1'$ under the basis function $k_{x}\tutabf{x}+k_{y}\tutabf{y}$ are identical to the representation matrices of these generators for $\Gamma^{2}$. Thus, the basis function $k_{x}\tutabf{x}+k_{y}\tutabf{y}$ is allowed for $\Gamma^{2}$ only in $6/mmm1'$. However, under the basis function $k_{z}\tutabf{z}$, the matrices of generators belonging to the unitary part of both groups are identical to the representation matrices of these generators for $\Gamma^{2}$. This indicates that the basis function $k_{z}\tutabf{z}$, representing the unitary $p$-wave superconducting state with only out-of-plane pairing components, is allowed for $\Gamma^{2}$ of both groups.

In the same manner, for coirrep $\Gamma^{4}$, the presence of the basis function $k_{x}\tutabf{y}-k_{y}\tutabf{x}$ in $6/mmm1'$ and absence of this basis function in $G_{\text{SS}}^{p}\times~^{3_{z}}6/^{1}m^{2_{x}}m^{2_{xy}}m$ can be understood. The basis function for coirrep $\Gamma^{4}$ of $G_{\text{SS}}^{p}\times~^{3_{z}}6/^{1}m^{2_{x}}m^{2_{xy}}m$ is $2k_{x}k_{y}k_{z}\tutabf{x}+(k_{x}^2-k_{y}^2)k_{z}\tutabf{y}$ which represents a unitary $f$-wave superconducting state and remains unchanged under the operator $\hat{P}_{[R_{\bbb{z}}(4\pi/3)||C_{6z}]}$. In addition, since the basis function $2k_{x}k_{y}k_{z}\tutabf{x}+(k_{x}^2-k_{y}^2)k_{z}\tutabf{y}$ changes sign under the operator $\hat{P}_{[C_{6z}||C_{6z}]}$, and the representation matrix of group element $[C_{6z}||C_{6z}]$ for $\Gamma^{4}$ of $6/mmm1'$ is identity matrix, this basis function is not allowed for coirrep $\Gamma^{4}$ of group $6/mmm1'$.

In contrast to the coirreps $\Gamma^{2}$ and $\Gamma^{4}$, for coirreps $\Gamma^{6}$ and $\Gamma^{8}$, the basis functions $k_{x}\tutabf{y}-k_{y}\tutabf{x}$ and $k_{x}\tutabf{x}+k_{y}\tutabf{y}$, representing unitary $p$-wave superconducting states, are allowed only in $G_{\text{SS}}^{p}\times~^{3_{z}}6/^{1}m^{2_{x}}m^{2_{xy}}m$ but forbidden in $6/mmm1'$. However, the basis functions representing unitary $f$-wave superconducting states allowed in $6/mmm1'$ are absent in $G_{\text{SS}}^{p}\times~^{3_{z}}6/^{1}m^{2_{x}}m^{2_{xy}}m$ for both coirreps $\Gamma^{6}$ and $\Gamma^{8}$. These results can be understood in the same manner as for coirrep $\Gamma^{2}$ and $\Gamma^{4}$.

The presence or absence of the basis functions in two-dimensional coirreps can also be understood by similar analysis. For the two-dimensional coirrep $\Gamma^{10}$, in group $G_{\text{SS}}^{p}\times~^{3_{z}}6/^{1}m^{2_{x}}m^{2_{xy}}m$, the basis functions \{$k_{z}(\tutabf{x}-i\tutabf{y})$, $k_{z}(\tutabf{x}+i\tutabf{y})$\} which represent nonunitary $p$-wave superconducting states with only out-of-plane pairing components are allowed. This is attributed to that the matrices of all generators belonging to the unitary part of $G_{\text{SS}}^{p}\times~^{3_{z}}6/^{1}m^{2_{x}}m^{2_{xy}}m$ under this set of basis functions are identical to the representation matrices of these generators for $\Gamma^{10}$. However, this set of basis functions \{$k_{z}(\tutabf{x}-i\tutabf{y})$, $k_{z}(\tutabf{x}+i\tutabf{y})$\} is forbidden in $6/mmm1'$. The reason is that the matrix of group element $[C_{6z}||C_{6z}]$ under this set of basis functions is different from the representation matrix of group element $[C_{6z}||C_{6z}]$ for $\Gamma^{10}$ as shown in Table.~\ref{co-irrep-appl1}:
\begin{equation}
    D([C_{6z}||C_{6z}])=\begin{bmatrix} e^{i\frac{\pi}{3}}&0\\ 0&e^{-i\frac{\pi}{3}}\end{bmatrix}\neq D^{10}([C_{6z}||C_{6z}]),
\end{equation}
where the $D^{10}([C_{6z}||C_{6z}])$ is the representation matrix of group element $[C_{6z}||C_{6z}]$ for $\Gamma^{10}$. 

In contrast, the basis functions \{$(k_{x}+i k_{y})(\tutabf{x}+i\tutabf{y})$, $(k_{x}-i k_{y})(\tutabf{x}-i\tutabf{y})$\} which represent nonunitary $p$-wave superconducting states with only in-plane paring components are allowed for coirrep $\Gamma^{10}$ only in $6/mmm1'$. This is attributed to that the matrices of all generators belonging to the unitary part of group $6/mmm1'$ under the set of basis functions \{$(k_{x}+i k_{y})(\tutabf{x}+i\tutabf{y})$, $(k_{x}-i k_{y})(\tutabf{x}-i\tutabf{y})$\} are identical to the representation matrices of these generators for $\Gamma^{10}$. However, the matrix of generator $[R_{\bbb{z}}(4\pi/3)||C_{6z}]$ belonging to $G_{\text{SS}}^{p}\times~^{3_{z}}6/^{1}m^{2_{x}}m^{2_{xy}}m$ under this set of basis function is:
\begin{equation}
    D([R_{\bbb{z}}(4\pi/3)||C_{6z}])=\begin{bmatrix} e^{i\frac{\pi}{3}}&0\\ 0&e^{-i\frac{\pi}{3}}\end{bmatrix},
\end{equation}
which is different from the representation matrix of group element $[R_{\bbb{z}}(4\pi/3)||C_{6z}]$ for $\Gamma^{10}$ of $G_{\text{SS}}^{p}\times~^{3_{z}}6/^{1}m^{2_{x}}m^{2_{xy}}m$. Therefore, the set of basis functions \{$(k_{x}+i k_{y})(\tutabf{x}+i\tutabf{y})$, $(k_{x}-i k_{y})(\tutabf{x}-i\tutabf{y})$\} is forbidden for coirrep $\Gamma^{10}$ of $G_{\text{SS}}^{p}\times~^{3_{z}}6/^{1}m^{2_{x}}m^{2_{xy}}m$.

Interestingly, for two-dimensional coirrep $\Gamma^{12}$, there are both unitary and nonunitary spin-triplet superconducting states in these two groups. The basis functions \{$(k_{x}-i k_{y})\tutabf{z}$, $(k_{x}+i k_{y})\tutabf{z}$\}, representing unitary $p$-wave superconducting states, are present in both groups. However, the basis functions \{$(k_{x}+i k_{y})(\tutabf{x}+i\tutabf{y})$, $(k_{x}-i k_{y})(\tutabf{x}-i\tutabf{y})$\}, representing nonunitary $p$-wave superconducting states with only in-plane paring components, are allowed only in group $G_{\text{SS}}^{p}\times~^{3_{z}}6/^{1}m^{2_{x}}m^{2_{xy}}m$. In contrast, the basis functions \{$k_{z}(\tutabf{x}-i\tutabf{y})$, $k_{z}(\tutabf{x}+i\tutabf{y})$\}, representing nonunitary $p$-wave superconducting states with only out-of-plane pairing components, are allowed only in group $6/mmm1'$. These results can be understood by a similar analysis as for coirrep $\Gamma^{10}$.

\section{Discussion and Conclusion}\label{con}
In this article, we apply a general approach to construct the superconducting order parameters in all types of spin point groups. For collinear spin point groups, there is no magnetic point group that is isomorphic to them. Thus the classification of superconducting channels and basis functions differ from that of any magnetic point group. One important difference is that for some coirreps of collinear spin point groups, there are nonzero indices $\sigma$, $\nu$ or $\mu$ which label the nonzero $z$-components of Cooper pairs' spin angular momentum. In superconducting channels with nonzero indices $\sigma$, spin-singlet pairing states are forbidden, resulting in the absence of charge-2e superconducting states for some channels of centrosymmetric collinear spin point groups. Moreover, the number of electron operators in the superconducting gap function should be at least $2\sigma$ for channels with nonzero index $\sigma$. For example, when $\sigma = 2$, the constraint on the Cooper pairs' spin angular momentum requires that the superconducting gap function $\Delta(\bbb{k})$ be the mean value of at least four electron operators, i.e., charge-4e superconducting states. A similar conclusion applies to superconducting channels with indices $\nu$ or $\mu$ as well.

In contrast to the collinear spin point groups, for each coplanar spin point group, there is a corresponding type-II magnetic point group which is isomorphic to it. However, although the coplanar spin point group and the corresponding type-II magnetic point group share the same coirrep table and, consequently, the same superconducting channels, the basis functions for each channel of these two groups may be different. Similarly, for each noncoplanar spin point group, there is a type-III magnetic point group or point group which is isomorphic to it, but the basis functions for each channel of these two groups may differ.

Moreover, by using basis functions obtained for each superconducting channel, we can construct the possible superconducting order parameters and the Ginzburg-Landau free energy of the superconducting states. The relationship between the coefficients of the fourth-order terms in the free energy determines which superconducting state is the most stable.

To be more practical, we also apply our theory to two realistic systems whose symmetries are described by the spin space group: one is the electric-field-gated monolayer FeSe, and the other is the AIAO order on the kagome lattice. Our theory provides a rigorous description of the possible superconducting states in materials described by spin space groups, enabling further investigation of superconductivity in these materials.
It is worth to note that although magnetic orders in magnetic materials can induce orbital effects that suppress superconducting states~\cite{clogston1962upper,chandrasekhar1962note,huebener2013vortices,huebener2019path}, under certain conditions, these orbital effects can be mitigated, allowing the superconducting states to persist. For example, in a two-dimensional system, the motion of electrons is confined to in-plane directions. As a result, the orbital effects induced by in-plane magnetic orders are negligible, allowing for the coexistence of the superconducting states with the in-plane magnetic orders. 

Interestingly, when magnetic orders coexist with superconducting states, the coupling between the magnetic and superconducting order parameters can be derived. These coupling terms in the free energy enable the study of feedback effects induced by magnetic orders or spin fluctuations of the magnetic orders on the superconducting states~\cite{anderson1973anisotropic,brinkman1974spin,kuroda1975paramagnon,Amin2020, konno1989superconductivity, chamon2001p, joynt2002superconducting, dahl2007derivation, mukherjee2014macroscopic, feng2020phenomenological, nayak2021pairing, machida2021nonunitary, machida2023violation, rosuel2023field}. This framework facilitates further studies of the interplay between superconductivity and magnetic orders across a wide range of materials protected by spin space groups.

\bibliography{spg.bib}

\begin{widetext}

\appendix
\section{The transformation properties of the gap function under spin rotation operations, time-reversal symmetry and point group operations}\label{app-a}
\setcounter{equation}{0}
\renewcommand\theequation{A\arabic{equation}}
\setcounter{table}{0}
\renewcommand\thetable{A\arabic{table}}
In this appendix, we briefly review the transformation properties of the gap function under spin-rotation operations, time-reversal symmetry, and point group operations.

In Nambu basis $\hat{\Psi}_{\bbb{k}}=(\hat{c}_{\bbb{k},\uparrow},\hat{c}_{\bbb{k},\downarrow},\hat{c}^{\dagger}_{-\bbb{k}\uparrow},\hat{c}^{\dagger}_{-\bbb{k}\downarrow})^{T}$ (where the superscript $T$ means transpose), the mean-field Hamiltonian for single-orbital superconductivity is $\hat{H}^{\text{MF}}_{\text{SC}}=\sum_{\bbb{k}}\hat{\Psi}^{\dagger}_{\bbb{k}}H^{\text{MF}}(\bbb{k})\hat{\Psi}_{\bbb{k}}$, where the matrix $H^{\text{MF}}(\bbb{k})$ can be expressed as:
\begin{equation}
    H^{\text{MF}}(\bbb{k})=\left(\begin{array}{cc}H_{0}(\bbb{k})&\Delta(\bbb{k})\\\Delta^{\dagger}(\bbb{k})&-H_{0}^{*}(-\bbb{k})\end{array} \right).
\end{equation}
The single-orbital gap function can be written as:
\begin{equation}
\Delta(\bbb{k})=(\bbb{d}(\bbb{k})\cdot \bm{\sigma}+\psi(\bbb{k}))i\sigma_{2},
\end{equation}
where $\bm{\sigma}$ indicates $(\sigma_{1},\sigma_{2},\sigma_{3})$ and the $\sigma_{i} (i=1,2,3)$ are Pauli matrices. The spin-rotation operation $\hat{g}_{s}$ transforms the electron creation operator as~\cite{ueda1985p}:
\begin{equation}
\hat{g}_{s} \hat{c}_{\bbb{k},\sigma}^{\dagger}\equiv [D_{1/2}(\hat{g}_{s})]_{\sigma',\sigma}\hat{c}_{\bbb{k},\sigma'}^{\dagger}=(e^{-i\frac{\phi}{2} \bbb{n}\cdot{\bm{\sigma}}})_{\sigma',\sigma}\hat{c}_{\bbb{k},\sigma'}^{\dagger},
\end{equation}
where $\bbb{n}$ is rotation axis and $\phi$ is rotation angle. The pairing terms in the Hamiltonian can be expressed as: $\hat{\Delta}=\sum_{\bbb{k},\sigma\sigma'}\hat{c}_{\bbb{k},\sigma}^{\dagger}\Delta(\bbb{k})_{\sigma\sigma'}\hat{c}_{-\bbb{k},\sigma'}^{\dagger}+ h.c.$, and the spin-rotation operation $\hat{g}_{s}$ transform the gap function $\Delta(\bbb{k})$ as:
\begin{equation}
\begin{aligned}
\hat{g}_{s}\Delta(\bbb{k})=D_{1/2}(\hat{g}_{s})\Delta(\bbb{k})D_{1/2}^{T}(\hat{g}_{s}).
\end{aligned}
\end{equation}
By using the following two identities:
\begin{equation}
    \begin{aligned}
        D_{1/2}(\hat{g}_{s})(\bbb{m}\cdot \bm{\sigma})D_{1/2}^{-1}(\hat{g}_{s})&=(R_{\bbb{n}}(\phi) \bbb{m})\cdot \bm{\sigma},\\
        D_{1/2}(\hat{g}_{s})(i\sigma_{2})D_{1/2}^{T}(\hat{g}_{s})&=i\sigma_{2},
    \end{aligned}
\end{equation}
where $R_{\bbb{n}}(\phi)$ represents the matrix of the three-dimensional rotation with respect to the axis $\bbb{n}$ by an angle $\phi$, $\bbb{m}$ is an arbitrary three-dimensional vector, it can be shown that the spin-rotation operation $\hat{g}_{s}$ acting on the gap function $\Delta(\bbb{k})$ is equivalent to a three-dimensional rotation about the same rotation axis $\bbb{n}$ by the same angle $\phi$ acting on the vector $\bbb{d}(\bbb{k})$, or the identity operation acting on the scalar $\psi(\bbb{k})$~\cite{balian1963superconductivity,leggett1975theoretical,sigrist1991phenomenological}. For convenience, we define the basis function $\xi(\bbb{k},\bbb{r})$ as given in the Eq.~(\ref{basisf}) of the main text. Then the acting of spin-rotation operation $\hat{g}_{s}$ on the vector $\bbb{d}(\bbb{k})$ can be represented by the three-dimensional rotation $R^{-1}_{\bbb{n}}(\phi)$ acting on the vector formed by axial-vectors $\bbb{r}=(\tutabf{x},\tutabf{y},\tutabf{z})$.

The time-reversal operation $\ccc{T}$ is represented as $-i\sigma_{2}K$ in spin subspace, where $K$ denotes the complex conjugation operator. The change in the gap function under time-reversal can also be expressed in terms of the changes in the vector $\bbb{d}(\bbb{k})$ and $\psi(\bbb{k})$:
\begin{equation}\label{Tt}
\begin{aligned}
     \ccc{T}\bbb{d}(\bbb{k})=-\bbb{d}^{*}(-\bbb{k}),\\
     \ccc{T}\psi(\bbb{k})=\psi^{*}({-\bbb{k}}).
\end{aligned}
\end{equation}
In addition, the anticommutative property of the Fermion operator in the pairing terms of the Hamiltonian requires that the gap function $\Delta(\bbb{k})$ satisfies:
\begin{equation}
    \Delta(\bbb{k})=-\Delta^{T}(-\bbb{k}).
\end{equation}
Then we can conclude that:
\begin{equation}\label{Aext}
    \begin{aligned}
        \bbb{d}(\bbb{k})=-\bbb{d}(-\bbb{k}),\\
        \psi(\bbb{k})=\psi(-\bbb{k}).
    \end{aligned}
\end{equation}
Combining Eq.~(\ref{Tt}) with the Eq.~(\ref{Aext}), the changes of the vector $\bbb{d}(\bbb{k})$ and $\psi(\bbb{k})$ under time-reversal operator are:
\begin{equation}
    \begin{aligned}
        \ccc{T}\bbb{d}(\bbb{k})=\bbb{d}^{*}(\bbb{k}),\\
     \ccc{T}\psi(\bbb{k})=\psi^{*}({\bbb{k}}).
    \end{aligned}
\end{equation}

The point group operation $\hat{g}_{r}\in \ccc{G}$ only acts on the momentum $\bbb{k}$ of the gap functions and transforms the vector $\bbb{d}(\bbb{k})$ and $\psi(\bbb{k})$ as:
\begin{equation}
\begin{aligned}
    \hat{g}_{r}\bbb{d}(\bbb{k})=\bbb{d}(D^{-1}(\hat{g}_{r})\bbb{k}),\\
    \hat{g}_{r}\psi(\bbb{k})=\psi(D^{-1}(\hat{g}_{r})\bbb{k}),
\end{aligned}
\end{equation}
where the $D(\hat{g}_{r})$ is the matrix of group element $\hat{g}_{r}$ for the faithful representation of the point group $\ccc{G}$.

\section{coirreps of spin point groups}
\setcounter{equation}{0}
\renewcommand\theequation{B\arabic{equation}}
\setcounter{table}{0}
\renewcommand\thetable{B\arabic{table}}

\subsection{The coirreps of the collinear spin point group $G_{\text{SS}}^{l}\times ~^{1}4/^{1}m^{1}m^{1}m$}

\begin{table}[h]\label{co-irrep-coll}
\caption{The coirreps of the collinear spin point group $G_{\text{SS}}^{l}\times ~^{1}4/^{1}m^{1}m^{1}m$. $R_{\bbb{z}}(\theta)$ means a rotation about $\bbb{z}$ axis by an angle $\theta$ (counterclockwise). $\bbb{n}_{\bot}$ is an arbitrary axis which is perpendicular to the direction of the spin magnetic moments $\bbb{M}(\bbb{r})$ of the system whose symmetry is described by $G_{\text{SS}}^{l}\times ~^{1}4/^{1}m^{1}m^{1}m$. Here we choose $\bbb{M}(\bbb{r}) \parallel \bbb{z}$. $R_{\bbb{n}_{\bot}}(\pi)$ is a rotation about the axis $\bbb{n}_{\bot}$ by $\pi$. In this paper, we choose $\bbb{n}_{\bot}=\bbb{x}$. $\ccc{T}$ is time-reversal operator. $\sigma \in \mathbb{Z}$.}
\begin{tabular}{ >{\centering\arraybackslash}m{3cm}|>{\centering\arraybackslash}m{2cm}|>{\centering\arraybackslash}m{3cm}|>{\centering\arraybackslash}m{3cm}|>{\centering\arraybackslash}m{3cm}|>{\centering\arraybackslash}m{3cm}
}
\hline \hline
$G_{\text{SS}}^{l}\times ~^{1}4/^{1}m^{1}m^{1}m$ & $[R_{\bbb{z}}(\theta)||E]$ & $[R_{\bbb{z}}(\theta)||C_{4z}]$ & $[R_{\bbb{z}}(\theta)||IC_{2z}]$ & $[R_{\bbb{z}}(\theta)||IC_{2y}]$ & $[\ccc{T} R_{\bbb{n_{\bot}}}(\pi) R_{\bbb{z}}(\theta)||E]$ \\ \hline 
$\Gamma^{1}_{\sigma}$                                  & $e^{i\sigma \theta}$          & $e^{i\sigma \theta}$               & $e^{i\sigma \theta}$                & $e^{i\sigma \theta}$                & $e^{-i\sigma \theta}$                                                 \\ \hline
$\Gamma^{2}_{\sigma}$                                  & $e^{i\sigma \theta}$          & $e^{i\sigma \theta}$               & $-e^{i\sigma \theta}$               & $-e^{i\sigma \theta}$               &   $e^{-i\sigma \theta}$                                                  \\ \hline
$\Gamma^{3}_{\sigma}$                                  & $e^{i\sigma \theta}$          & $e^{i\sigma \theta}$               & $e^{i\sigma \theta}$                & $-e^{i\sigma \theta}$                & $e^{-i\sigma \theta}$                                                 \\ \hline
$\Gamma^{4}_{\sigma}$                                  & $e^{i\sigma \theta}$          & $e^{i\sigma \theta}$               & $-e^{i\sigma \theta}$                & $e^{i\sigma \theta}$                & $e^{-i\sigma \theta}$                                                 \\ \hline
$\Gamma^{5}_{\sigma}$                                  & $e^{i\sigma \theta}$          & $-e^{i\sigma \theta}$               & $e^{i\sigma \theta}$                & $e^{i\sigma \theta}$                & $e^{-i\sigma \theta}$                                                 \\ \hline
$\Gamma^{6}_{\sigma}$                                  & $e^{i\sigma \theta}$          & $-e^{i\sigma \theta}$               & $-e^{i\sigma \theta}$                & $-e^{i\sigma \theta}$                & $e^{-i\sigma \theta}$                                                 \\ \hline
$\Gamma^{7}_{\sigma}$                                  & $e^{i\sigma \theta}$          & $-e^{i\sigma \theta}$               & $e^{i\sigma \theta}$                & $-e^{i\sigma \theta}$                & $e^{-i\sigma \theta}$                                                 \\ \hline
$\Gamma^{8}_{\sigma}$                                  & $e^{i\sigma \theta}$          & $-e^{i\sigma \theta}$               & $-e^{i\sigma \theta}$                & $e^{i\sigma \theta}$                & $e^{-i\sigma \theta}$                                                 \\ \hline
$\Gamma^{9}_{\sigma}\vspace{2pt}$ &  $\begin{bmatrix} e^{i\sigma \theta} &0\\0& e^{i\sigma \theta} \end{bmatrix} $ 

& $\begin{bmatrix} -i e^{i\sigma \theta} &0\\0& i e^{i\sigma \theta} \end{bmatrix} $ 

&$\begin{bmatrix} -e^{i\sigma \theta} &0\\0& -e^{i\sigma \theta} \end{bmatrix} $ 

&$\begin{bmatrix} 0 &-e^{i\sigma \theta}\\-e^{i\sigma \theta}& 0 \end{bmatrix} $ 

&$\begin{bmatrix} 0 &e^{-i\sigma \theta}\\e^{-i\sigma \theta}& 0 \end{bmatrix} \vspace{2pt}$\\ \hline

$\Gamma^{10}_{\sigma}\vspace{2pt}$ & $\begin{bmatrix} e^{i\sigma \theta} &0\\0& e^{i\sigma \theta} \end{bmatrix}$ 

&$\begin{bmatrix} -i e^{i\sigma \theta} &0\\0& i e^{i\sigma \theta} \end{bmatrix} $ 

& $\begin{bmatrix} e^{i\sigma \theta} &0\\0& e^{i\sigma \theta} \end{bmatrix} $ 

& $\begin{bmatrix} 0 &e^{i\sigma \theta}\\e^{i\sigma \theta}& 0 \end{bmatrix} $ 

& $\begin{bmatrix} 0 &e^{-i\sigma \theta}\\e^{-i\sigma \theta}& 0 \end{bmatrix} \vspace{2pt}$ \\ \hline

\end{tabular}
\end{table}

\newpage
\subsection{The coirreps of the type-II magnetic point group $4/mmm1'$}

\begin{table}[h]\label{co-irrep-cop}\label{co-irrep-4mmm1}
    \centering
        \caption{The coirreps of the type-II magnetic point group $4/mmm1'$.}
    \begin{tabular}{>{\centering\arraybackslash}m{2cm}|>{\centering\arraybackslash}m{3cm}|>{\centering\arraybackslash}m{3cm}|>{\centering\arraybackslash}m{3cm}|>{\centering\arraybackslash}m{3cm}|>{\centering\arraybackslash}m{3cm}}
     \hline \hline
$4/mmm1'$ &  $[E||E]$ & $[C_{4z}||C_{4z}]$ &$[C_{2z}||IC_{2z}]$ & $[C_{2y}||IC_{2y}]$ &$[\ccc{T}||E]$ \\ \hline
$\Gamma^{1}$                                  & 1  & 1               & 1            & 1                & 1                                    \\ \hline
$\Gamma^{2}$                                  & 1  & 1               & -1            & -1                & 1                                    \\ \hline
$\Gamma^{3}$                                    & 1  & 1               & 1            & -1                & 1                                    \\ \hline
$\Gamma^{4}$                                 & 1  & 1               & -1            & 1                & 1                                    \\ \hline
$\Gamma^{5}$                                       & 1  & -1               & 1            & 1                & 1                                    \\ \hline
$\Gamma^{6}$                                    & 1  & -1               & -1            & -1                & 1                                    \\ \hline
$\Gamma^{7}$                                   & 1  & -1               & 1            & -1                & 1                                    \\ \hline
$\Gamma^{8}$                                     & 1  & -1               & -1            & 1                & 1                                    \\ \hline
$\Gamma^{9}$ & $\begin{bmatrix} 1 &0\\0& 1 \end{bmatrix}$ 

&$\begin{bmatrix} -i &0\\ 0& i\end{bmatrix} $ 

& $\begin{bmatrix} -1 &0 \\0& -1 \end{bmatrix} $ 

& $\begin{bmatrix} 0 &-1 \\-1& 0 \end{bmatrix} $ 

& $\begin{bmatrix} 0 &1\\1& 0 \end{bmatrix} \vspace{2pt}$ \\ \hline

$\Gamma^{10}$ &  $\begin{bmatrix} 1 &0\\0& 1 \end{bmatrix}$ 

&$\begin{bmatrix} -i &0\\ 0& i\end{bmatrix} $ 

& $\begin{bmatrix} 1 &0 \\0& 1 \end{bmatrix} $ 

& $\begin{bmatrix} 0 &1 \\1 & 0 \end{bmatrix} $ 

& $\begin{bmatrix} 0 &1\\1& 0 \end{bmatrix} \vspace{2pt}$ \\ \hline

    \end{tabular}
\end{table}

\subsection{The coirreps of the collinear spin point group $G_{\text{SS}}^{l}\times ~^{1}4/^{1}m^{\overline{1}}m^{\overline{1}}m$}

\begin{table}[h]\label{co-irrep-coll2}

\caption{The coirreps of the collinear spin point group $G_{\text{SS}}^{l}\times ~^{1}4/^{1}m^{\overline{1}}m^{\overline{1}}m$, $\nu \in \mathbb{N}/\{0\}$. The meanings of the other notations are as defined in the caption of Table.~\ref{co-irrep-coll}.}
\resizebox{\linewidth}{!}{
\begin{tabular}{>{\centering\arraybackslash}m{3cm}|>{\centering\arraybackslash}m{4.8cm}|>{\centering\arraybackslash}m{4.8cm}|>{\centering\arraybackslash}m{4.8cm}|>{\centering\arraybackslash}m{4.8cm}|>{\centering\arraybackslash}m{4.8cm}}
\hline \hline 
$G_{\text{SS}}^{l}\times ~^{1}4/^{1}m^{\overline{1}}m^{\overline{1}}m$ & $[R_{\bbb{z}}(\theta)||E]$ & $[R_{\bbb{z}}(\theta)||C_{4z}]$ & $[R_{\bbb{z}}(\theta)||IC_{2z}]$ & $[R_{\bbb{n_{\bot}}}(\pi)R_{\bbb{z}}(\theta)||IC_{2y}]$ & $[\ccc{T} R_{\bbb{n_{\bot}}}(\pi) R_{\bbb{z}}(\theta)||E]$ \\ \hline 
$\Gamma^{1}$                                  & 1          & 1              & 1                & 1                & 1           \\ \hline
$\Gamma^{2}$                                  & 1          & 1              & -1                & -1                & 1           \\ \hline
$\Gamma^{3}$                                  & 1          & 1              & 1                & -1                & 1           \\ \hline
$\Gamma^{4}$                                  & 1          & 1              & -1                & 1                & 1           \\ \hline
$\Gamma^{5}$                                  & 1          & -1              & 1                & 1                & 1           \\ \hline
$\Gamma^{6}$                                  & 1          & -1              & -1                & -1                & 1           \\ \hline
$\Gamma^{7}$                                  & 1          & -1              & 1                & -1                & 1           \\ \hline
$\Gamma^{8}$                                  & 1          & -1              & -1                & 1                & 1           \\ \hline
$\Gamma^{9}$ & $\begin{bmatrix} 1 &0\\0& 1 \end{bmatrix}$ 

&$\begin{bmatrix} -i &0\\0& i \end{bmatrix} $ 

& $\begin{bmatrix} -1 &0\\0& -1 \end{bmatrix} $ 

& $\begin{bmatrix} 0 &-1\\-1& 0 \end{bmatrix} $ 

& $\begin{bmatrix} 0 &1\\1& 0 \end{bmatrix} \vspace{2pt}$ \\ \hline

$\Gamma^{10}$ & $\begin{bmatrix} 1 &0\\0& 1 \end{bmatrix}$ 

&$\begin{bmatrix} -i &0\\0& i \end{bmatrix} $ 

& $\begin{bmatrix} 1 &0\\0& 1 \end{bmatrix} $ 

& $\begin{bmatrix} 0 &1\\1& 0 \end{bmatrix} $ 

& $\begin{bmatrix} 0 &1\\1& 0 \end{bmatrix} \vspace{2pt}$ \\ \hline

$\Gamma^{11}_{\nu}$ &  $\begin{bmatrix} e^{i\nu \theta} &0\\0& e^{-i\nu \theta} \end{bmatrix} $ 

& $\begin{bmatrix} e^{i\nu \theta} &0\\0& e^{-i\nu \theta} \end{bmatrix} $

&$\begin{bmatrix} e^{i\nu \theta} &0\\0& e^{-i\nu \theta} \end{bmatrix} $ 

&$\begin{bmatrix} 0 &e^{-i\nu \theta}\\e^{i\nu \theta}& 0 \end{bmatrix} $ 

&$\begin{bmatrix} e^{-i\nu \theta}&0\\0&e^{i\nu \theta} \end{bmatrix} \vspace{2pt}$ \\ \hline

$\Gamma^{12}_{\nu}$ &  $\begin{bmatrix} 

e^{i\nu \theta} &0\\0& e^{-i\nu \theta} \end{bmatrix} $ 

& $\begin{bmatrix} e^{i\nu \theta} &0\\0& e^{-i\nu \theta} \end{bmatrix} $

&$\begin{bmatrix} -e^{i\nu \theta} &0\\0& -e^{-i\nu \theta} \end{bmatrix} $ 

&$\begin{bmatrix} 0 &e^{-i\nu \theta}\\e^{i\nu \theta}& 0 \end{bmatrix} $ 

&$\begin{bmatrix} e^{-i\nu \theta}&0\\0&e^{i\nu \theta} \end{bmatrix} \vspace{2pt}$ \\ \hline

$\Gamma^{13}_{\nu}$ &  
$\begin{bmatrix} e^{i\nu \theta} &0\\0& e^{-i\nu \theta} \end{bmatrix} $ 

& $\begin{bmatrix} -e^{i\nu \theta} &0\\0& -e^{-i\nu \theta} \end{bmatrix} $

&$\begin{bmatrix} e^{i\nu \theta} &0\\0& e^{-i\nu \theta} \end{bmatrix} $ 

&$\begin{bmatrix} 0 &e^{-i\nu \theta}\\e^{i\nu \theta}& 0 \end{bmatrix} $ 

&$\begin{bmatrix} e^{-i\nu \theta}&0\\0&e^{i\nu \theta} \end{bmatrix} \vspace{2pt}$ \\ \hline

$\Gamma^{14}_{\nu}$ &  
$\begin{bmatrix} e^{i\nu \theta} &0\\0& e^{-i\nu \theta} \end{bmatrix} $ 

& $\begin{bmatrix} -e^{i\nu \theta} &0\\0& -e^{-i\nu \theta} \end{bmatrix} $

&$\begin{bmatrix} -e^{i\nu \theta} &0\\0& -e^{-i\nu \theta} \end{bmatrix} $ 

&$\begin{bmatrix} 0 &e^{-i\nu \theta}\\e^{i\nu \theta}& 0 \end{bmatrix} $ 

&$\begin{bmatrix} e^{-i\nu \theta}&0\\0&e^{i\nu \theta} \end{bmatrix} \vspace{2pt}$ \\ \hline

$\Gamma^{15}_{\nu}$ &  
$\begin{bmatrix} e^{i\nu \theta} &0&0&0\\0& e^{-i\nu \theta}&0&0\\0&0&e^{i\nu \theta}&0\\0&0&0&e^{-i\nu \theta} \end{bmatrix} $ 

& $\begin{bmatrix} -i e^{i\nu \theta} &0&0&0\\0& ie^{-i\nu \theta}&0&0\\0&0&ie^{i\nu \theta}&0\\0&0&0&-ie^{-i\nu \theta} \end{bmatrix} $

&$\begin{bmatrix} -e^{i\nu \theta} &0&0&0\\0& -e^{-i\nu \theta}&0&0\\0&0&-e^{i\nu \theta}&0\\0&0&0&-e^{-i\nu \theta} \end{bmatrix} $ 

&$\begin{bmatrix} 0 &-e^{-i\nu \theta}&0&0\\-e^{i\nu \theta}& 0&0&0\\0&0&0&-e^{-i\nu \theta}\\0&0&-e^{i\nu \theta}&0 \end{bmatrix} $ 

&$\begin{bmatrix} 0&0&e^{-i\nu \theta}&0\\0&0&0&e^{i\nu \theta}\\e^{-i\nu \theta}&0&0&0\\0&e^{i\nu \theta}&0&0 \end{bmatrix} \vspace{2pt}$ \\ \hline

$\Gamma^{16}_{\nu}$ &  
$\begin{bmatrix} e^{i\nu \theta} &0&0&0\\0& e^{-i\nu \theta}&0&0\\0&0&e^{i\nu \theta}&0\\0&0&0&e^{-i\nu \theta} \end{bmatrix} $ 

& $\begin{bmatrix} -i e^{i\nu \theta} &0&0&0\\0& ie^{-i\nu \theta}&0&0\\0&0&ie^{i\nu \theta}&0\\0&0&0&-ie^{-i\nu \theta} \end{bmatrix} $

&$\begin{bmatrix} e^{i\nu \theta} &0&0&0\\0& e^{-i\nu \theta}&0&0\\0&0&e^{i\nu \theta}&0\\0&0&0&e^{-i\nu \theta} \end{bmatrix} $ 

&$\begin{bmatrix} 0 &-e^{-i\nu \theta}&0&0\\-e^{i\nu \theta}& 0&0&0\\0&0&0&-e^{-i\nu \theta}\\0&0&-e^{i\nu \theta}&0 \end{bmatrix} $ 

&$\begin{bmatrix} 0&0&e^{-i\nu \theta}&0\\0&0&0&e^{i\nu \theta}\\e^{-i\nu \theta}&0&0&0\\0&e^{i\nu \theta}&0&0 \end{bmatrix} \vspace{2pt}$ \\ \hline

\end{tabular}
}
\end{table}

\newpage

\subsection{The coirreps of the coplanar spin point group $G^{p}_{\text{SS}}\times ~^{2_{z}}4/^{1}m^{2_{x}}m^{2_{y}}m $}

\begin{table}[h]\label{co-irrep-cop}
    \centering
        \caption{The coirreps of the coplanar spin point group $G^{p}_{\text{SS}}\times ~^{2_{z}}4/^{1}m^{2_{x}}m^{2_{y}}m $.}
    \begin{tabular}{>{\centering\arraybackslash}m{4cm}|>{\centering\arraybackslash}m{2.5cm}|>{\centering\arraybackslash}m{2.5cm}|>{\centering\arraybackslash}m{2.5cm}|>{\centering\arraybackslash}m{2.5cm}|>{\centering\arraybackslash}m{2.5cm}}
     \hline \hline
\makecell{$G^{p}_{\text{SS}}\times ~^{2_{z}}4/^{1}m^{2_{x}}m^{2_{y}}m$ }& \makecell{$[E||E]$ } & \makecell{$[R_{\bbb{z}}(\pi)||C_{4z}]$ } &\makecell{ $[E||IC_{2z}]$ } & \makecell{$[R_{\bbb{x}}(\pi)||IC_{2y}]$ } &\makecell{ $[\ccc{T} R_{\bbb{z}}(\pi) ||E]$} \\ \hline \hline

$\Gamma^{1}$                                  & 1  & 1               & 1            & 1                & 1                                    \\ \hline
$\Gamma^{2}$                                  & 1  & 1               & -1            & -1                & 1                                    \\ \hline
$\Gamma^{3}$                                    & 1  & 1               & 1            & -1                & 1                                    \\ \hline
$\Gamma^{4}$                                 & 1  & 1               & -1            & 1                & 1                                    \\ \hline
$\Gamma^{5}$                                       & 1  & -1               & 1            & 1                & 1                                    \\ \hline
$\Gamma^{6}$                                    & 1  & -1               & -1            & -1                & 1                                    \\ \hline
$\Gamma^{7}$                                   & 1  & -1               & 1            & -1                & 1                                    \\ \hline
$\Gamma^{8}$                                     & 1  & -1               & -1            & 1                & 1                                    \\ \hline
$\Gamma^{9}$ & $\begin{bmatrix} 1 &0\\0& 1 \end{bmatrix}$ 

&$\begin{bmatrix} -i &0\\ 0& i\end{bmatrix} $ 

& $\begin{bmatrix} -1 &0 \\0& -1 \end{bmatrix} $ 

& $\begin{bmatrix} 0 &-1 \\-1& 0 \end{bmatrix} $ 

& $\begin{bmatrix} 0 &1\\1& 0 \end{bmatrix} \vspace{2pt}$ \\ \hline

$\Gamma^{10}$ &  $\begin{bmatrix} 1 &0\\0& 1 \end{bmatrix}$ 

&$\begin{bmatrix} -i &0\\ 0& i\end{bmatrix} $ 

& $\begin{bmatrix} 1 &0 \\0& 1 \end{bmatrix} $ 

& $\begin{bmatrix} 0 &1 \\1 & 0 \end{bmatrix} $ 

& $\begin{bmatrix} 0 &1\\1& 0 \end{bmatrix} \vspace{2pt}$ \\ \hline

    \end{tabular}
\end{table}

\subsection{The irreps of the noncoplanar spin point group $^{4}4/^{1}m$ and the point group $4/m$}

\begin{table}[h]\label{co-irrep-unitarynoncop}
    \centering
        \caption{The irreps of the noncoplanar spin point group $^{4}4/^{1}m$ and the point group $4/m$.}
    \begin{tabular}{>{\centering\arraybackslash}m{4cm}|>{\centering\arraybackslash}m{3cm}|>{\centering\arraybackslash}m{3cm}|>{\centering\arraybackslash}m{3cm}}
     \hline \hline
\makecell{$^{4}4/^{1}m$  }& \makecell{$[E||E]$ } & \makecell{$[R_{\bbb{z}}(\pi/2)||C_{4z}]$ } &\makecell{ $[E||IC_{2z}]$ } \\ \hline \hline
$4/m$ &  $[E||E]$ & $[C_{4z}||C_{4z}]$ &$[C_{2z}||IC_{2z}]$ \\ \hline 
$\Gamma^{1}$                                  & 1  & 1               & 1                                                                \\ \hline
$\Gamma^{2}$                                  & 1  & 1               & -1                                                                \\ \hline
$\Gamma^{3}$                                       & 1  & -1               & 1                                                            \\ \hline
$\Gamma^{4}$                                    & 1  & -1               & -1                                                             \\ \hline
$\Gamma^{5}$                                  & 1  & $i$               & -1                                                               \\ \hline
$\Gamma^{6}$                                  & 1  & $i$               & 1                                                              \\ \hline
$\Gamma^{7}$                                       & 1  & -$i$               & -1                                                           \\ \hline
$\Gamma^{8}$                                    & 1  & -$i$              & 1                                                             \\ \hline

    \end{tabular}
\end{table}

\subsection{The coirreps of the noncoplanar spin point group $^{2_{z}}4/^{m_{x}}m^{m_{x}}m^{m_{y}}m$ and the type-III magnetic point group $4/m'm'm'$}

\begin{table}[h]\label{co-irrep-noncop}
    \centering
        \caption{The coirreps of the noncoplanar spin point group $^{2_{z}}4/^{m_{x}}m^{m_{x}}m^{m_{y}}m$ and the type-III magnetic point group $4/m'm'm'$.}
    \begin{tabular}{>{\centering\arraybackslash}m{3cm}|>{\centering\arraybackslash}m{2cm}|>{\centering\arraybackslash}m{2cm}|>{\centering\arraybackslash}m{2cm}|>{\centering\arraybackslash}m{3cm}}
     \hline
\makecell{$^{2_{z}}4/^{m_{x}}m^{m_{x}}m^{m_{y}}m$  }& \makecell{$[E||E]$ } & \makecell{$[R_{\bbb{z}}(\pi)||C_{4z}]$ } &\makecell{ $[E||C_{2x}]$ } &\makecell{ $[\ccc{T} R_{\bbb{x}}(\pi) ||IC_{2z}]$} \\ \hline \hline
$4/m'm'm'$ &  $[E||E]$ & $[C_{4z}||C_{4z}]$ &$[C_{2x}||C_{2x}]$ &$[\ccc{T}C_{2z}||IC_{2z}]$ \\ \hline 
$\Gamma^{1}$                                  & 1  & 1               & 1            & 1                                                    \\ \hline
$\Gamma^{2}$                                  & 1  & -1               & 1            & 1                                                    \\ \hline
$\Gamma^{3}$                                       & 1  & 1               & -1            & 1                                                  \\ \hline
$\Gamma^{4}$                                    & 1  & -1               & -1            & 1                                                   \\ \hline

$\Gamma^{5}$ &  $\begin{bmatrix} 1 &0\\0& 1 \end{bmatrix}$ 

&$\begin{bmatrix} 0 &-1\\ 1& 0\end{bmatrix} $ 

& $\begin{bmatrix} 0 &-1 \\-1 & 0 \end{bmatrix} $ 

& $\begin{bmatrix} -1 &0\\0& -1 \end{bmatrix} \vspace{2pt}$ \\ \hline

    \end{tabular}
\end{table}

\newpage

\subsection{The coirreps of the collinear spin point group $G_{\text{SS}}^{l}\times ~^{\overline{1}}4^{1}m^{\overline{1}}m$}

\begin{table}[h]\label{co-irrep-appl2}

\caption{The coirreps of the collinear spin point group $G_{\text{SS}}^{l}\times ~^{\overline{1}}4^{1}m^{\overline{1}}m$, here $\nu \in \mathbb{N}/\{0\}$ and $\mu \in \mathbb{Z}/\{0\}$. The meanings of the other notations are as defined in the caption of Table.~\ref{co-irrep-coll}.}

\begin{tabular}{>{\centering\arraybackslash}m{3cm}|>{\centering\arraybackslash}m{3cm}|>{\centering\arraybackslash}m{3cm}|>{\centering\arraybackslash}m{3cm}|>{\centering\arraybackslash}m{3cm}}
\hline \hline 
$G_{\text{SS}}^{l}\times ~^{\overline{1}}4^{1}m^{\overline{1}}m$ & $[R_{\bbb{z}}(\theta)||E]$ & $[R_{\bbb{z}}(\theta)||IC_{2y}]$ & $[R_{\bbb{n_{\bot}}}(\pi)R_{\bbb{z}}(\theta)||C_{4z}]$ & $[\ccc{T} R_{\bbb{n_{\bot}}}(\pi) R_{\bbb{z}}(\theta)||E]$ \\ \hline \hline
$\Gamma^{1}$                                  & 1          & 1              & 1                & 1                          \\ \hline
$\Gamma^{2}$                                  & 1          & 1              & -1                           & 1           \\ \hline
$\Gamma^{3}$                                  & 1          & -1              & 1                           & 1           \\ \hline
$\Gamma^{4}$                                  & 1          & -1              & -1                          & 1           \\ \hline
$\Gamma^{5}$ & $\begin{bmatrix} 1 &0\\0& 1 \end{bmatrix}$ 

&$\begin{bmatrix} -1 &0\\0& 1 \end{bmatrix} $ 

& $\begin{bmatrix} 0 &-1\\1& 0 \end{bmatrix} $ 

& $\begin{bmatrix} 1 &0\\0& 1 \end{bmatrix} \vspace{2pt}$ \\ \hline
$\Gamma^{6}$ &  $\begin{bmatrix} e^{i\nu \theta} &0\\0& e^{-i\nu \theta} \end{bmatrix} $ 

& $\begin{bmatrix} e^{i\nu \theta} &0\\0& e^{-i\nu \theta} \end{bmatrix} $

&$\begin{bmatrix} 0 &e^{-i\nu \theta}\\e^{i\nu \theta}& 0 \end{bmatrix} $ 

&$\begin{bmatrix} e^{i\nu \theta}&0\\0&e^{-i\nu \theta} \end{bmatrix} \vspace{2pt}$ \\ \hline

$\Gamma^{7}$ &  $\begin{bmatrix} e^{i\nu \theta} &0\\0& e^{-i\nu \theta} \end{bmatrix} $ 

& $\begin{bmatrix} -e^{i\nu \theta} &0\\0& -e^{-i\nu \theta} \end{bmatrix} $

&$\begin{bmatrix} 0 &e^{-i\nu \theta}\\e^{i\nu \theta}& 0 \end{bmatrix} $ 

&$\begin{bmatrix} e^{i\nu \theta}&0\\0&e^{-i\nu \theta} \end{bmatrix} \vspace{2pt}$ \\ \hline

$\Gamma^{8}$ &  $\begin{bmatrix} e^{i\mu \theta} &0\\0& e^{-i\mu \theta} \end{bmatrix} $ 

& $\begin{bmatrix} -e^{i\mu \theta} &0\\0& e^{-i\mu \theta} \end{bmatrix} $

&$\begin{bmatrix} 0 &-e^{-i\mu \theta}\\e^{i\mu \theta}& 0 \end{bmatrix} $ 

&$\begin{bmatrix} e^{i\mu \theta}&0\\0&e^{-i\mu \theta} \end{bmatrix} \vspace{2pt}$ \\ \hline

\end{tabular}

\end{table}

\subsection{The coirreps of the spin point group $G^{p}_{\text{SS}}\times~^{3_{z}}6/^{1}m^{2_{x}}m^{2_{xy}}m$ and the type-II magnetic point group $6/mmm1'$}
\begin{table}[h]\label{co-irrep-appl1}
    \centering
        \caption{The coirreps of the coplanar spin point group $G^{p}_{\text{SS}}\times~^{3_{z}}6/^{1}m^{2_{x}}m^{2_{xy}}m$ and the type-II magnetic point group $6/mmm1'$.}
    \begin{tabular}{>{\centering\arraybackslash}m{4cm}|>{\centering\arraybackslash}m{2cm}|>{\centering\arraybackslash}m{3cm}|>{\centering\arraybackslash}m{2cm}|>{\centering\arraybackslash}m{2cm}|>{\centering\arraybackslash}m{2cm}}
     \hline \hline
\makecell{$G_{\text{SS}}^{p}\times~^{3_{z}}6/^{1}m^{2_{x}}m^{2_{xy}}m$ }& \makecell{$[E||E]$ } & \makecell{$[R_{\bbb{z}}(4\pi/3)||C_{6z}]$ } &\makecell{ $[E||IC_{2z}]$ } & \makecell{$[R_{\bbb{y}}(\pi)||IC_{2x}]$ } &\makecell{ $[\ccc{T} R_{\bbb{z}}(\pi) ||E]$} \\ \hline \hline
$6/mmm1'$ &  $[E||E]$ & $[C_{6z}||C_{6z}]$ &$[C_{2z}||IC_{2z}]$ & $[C_{2x}||IC_{2x}]$ &$[\ccc{T}||E]$ \\ \hline
$\Gamma^{1}$                                  & 1  & 1               & 1            & 1                & 1                                    \\ \hline
$\Gamma^{2}$                                  & 1  & 1               & -1            & -1                & -1                                    \\ \hline
$\Gamma^{3}$                                       & 1  & 1               & 1            & -1                & 1                                    \\ \hline
$\Gamma^{4}$                                    & 1  & 1               & -1            & 1                & -1                                    \\ \hline
$\Gamma^{5}$                                    & 1  & -1               & -1            & 1                & 1                                    \\ \hline
$\Gamma^{6}$                                 & 1  & -1               & 1            & -1                & -1                                    \\ \hline
$\Gamma^{7}$                                   & 1  & -1               & -1            & -1                & 1                                    \\ \hline
$\Gamma^{8}$                                     & 1  & -1               & 1            & 1                & -1                                    \\ \hline
$\Gamma^{9}$ & $\begin{bmatrix} 1 &0\\0& 1 \end{bmatrix}$ 

&$\begin{bmatrix} e^{-i\frac{2\pi}{3}}&0\\ 0&e^{i\frac{2\pi}{3}}\end{bmatrix} $  

& $\begin{bmatrix} 1 &0 \\0& 1 \end{bmatrix} $ 

& $\begin{bmatrix} 0 &1 \\1& 0 \end{bmatrix} $ 

& $\begin{bmatrix} 0 &1\\1& 0 \end{bmatrix} \vspace{2pt}$ \\ \hline

$\Gamma^{10}$ & $\begin{bmatrix} 1 &0\\0& 1 \end{bmatrix}$ 

&$\begin{bmatrix} e^{-i\frac{2\pi}{3}}&0\\ 0&e^{i\frac{2\pi}{3}}\end{bmatrix} $ 

& $\begin{bmatrix} -1 &0 \\0& -1 \end{bmatrix} $ 

& $\begin{bmatrix} 0 &-1 \\-1& 0 \end{bmatrix} $ 

& $\begin{bmatrix} 0 &-1\\-1& 0 \end{bmatrix} \vspace{2pt}$ \\ \hline

$\Gamma^{11}$ & $\begin{bmatrix} 1 &0\\0& 1 \end{bmatrix}$ 

&$\begin{bmatrix} e^{i\frac{\pi}{3}}&0\\ 0&e^{-i\frac{\pi}{3}}\end{bmatrix} $  

& $\begin{bmatrix} -1 &0 \\0& -1 \end{bmatrix} $ 

& $\begin{bmatrix} 0 &1 \\1& 0 \end{bmatrix} $ 

& $\begin{bmatrix} 0 &1\\1& 0 \end{bmatrix} \vspace{2pt}$ \\ \hline

$\Gamma^{12}$ &  $\begin{bmatrix} 1 &0\\0& 1 \end{bmatrix}$ 

&$\begin{bmatrix} e^{i\frac{\pi}{3}}&0\\ 0&e^{-i\frac{\pi}{3}}\end{bmatrix} $  

& $\begin{bmatrix} 1 &0 \\0& 1 \end{bmatrix} $ 

& $\begin{bmatrix} 0 &-1 \\-1& 0 \end{bmatrix} $ 

& $\begin{bmatrix} 0 &-1\\-1& 0 \end{bmatrix} \vspace{2pt}$ \\ \hline

    \end{tabular}
\end{table}

\newpage

\section{The spin angular momentum of Cooper pairs}\label{magnetic_moment}
\setcounter{equation}{0}
\renewcommand\theequation{C\arabic{equation}}
\setcounter{table}{0}
\renewcommand\thetable{C\arabic{table}}

In this appendix, we discuss the relationship between the phase factors $e^{i\sigma \theta}$ of the representation matrices for the continuous spin-rotation operations $[R_{\bbb{z}}(\theta)||E]$ and the spin angular momentum of Cooper pairs, by using the examples from type-I collinear spin point groups. 

First of all, we consider the coirrep $\Gamma^{4}_{\sigma=0}$ of the collinear spin point group $G_{\text{SS}}^{l}\times ~^{1}4/ ^{1}m^{1}m^{1}m$ as an example for the case $\sigma=0$. The basis function for $\Gamma^{4}_{\sigma=0}$ is $k_{z}\tutabf{z}$, and the corresponding order parameter of superconductivity is $\bbb{d}(\bbb{k})=d_{0}k_{z}\tutabf{z}$, where the $d_{0}$ is a real constant. According to the definition of the spin polarization of the superconducting states~\cite{balian1963superconductivity,leggett1975theoretical,sigrist1991phenomenological}:
\begin{equation}\label{supermagnet}
    \bbb{q}(\bbb{k})=i\bbb{d}(\bbb{k})\times\bbb{d}^{*}(\bbb{k}),
\end{equation}
we can conclude that the spin polarization of the superconducting state which is represented by the order parameter $\bbb{d}(\bbb{k})$ is zero. In addition, by using the Eq.~(\ref{basistrans}) given in the main text, we can derive out the transformation of the basis function $k_{z}\tutabf{z}$ under the spin-rotation operator $\hat{P}_{[R_{\bbb{z}}(\theta)||E]}$:
\begin{equation}
    \hat{P}_{[R_{\bbb{z}}(\theta)||E]} k_{z}\tutabf{z}=k_{z}\tutabf{z}.
\end{equation}
The basis functions for other coirreps in the $\sigma=0$ case of type-I collinear spin point groups can also be analyzed in a similar manner. Therefore, we conclude that the trivial phase factor of the representation matrices of group elements $[R_{\bbb{z}}(\theta)||E]$ indicates that the superconducting states do not exhibit any spin polarization. In other words, the $z$-component of the Cooper pairs' spin angular momentum is zero.

Next, we consider the coirrep $\Gamma^{4}_{\sigma=1}$ of collinear spin point group $G_{\text{SS}}^{l}\times ~^{1}4/ ^{1}m^{1}m^{1}m$ as an example for the $\sigma=1$ case. The basis function for coirrep $\Gamma^{4}_{\sigma=1}$ is $k_{z}(\tutabf{x}-i\tutabf{y})$. Thus, the vector order parameter $\bbb{d}'(\bbb{k})$ of the corresponding superconducting state in this channel is proportional to it:
\begin{equation}
    \bbb{d}'(\bbb{k})=d'_{0}k_{z}(\tutabf{x}-i\tutabf{y}),
\end{equation}
where the $d'_{0}$ is a real constant. According to the Eq.~(\ref{supermagnet}), the spin polarization of this superconducting state is $\bbb{q}'(\bbb{k})=-2k_{z}^{2}d_{0}'^2\tutabf{z}$, which is directed along the negative direction of the $\bbb{z}$ axis. This fact indicates that the $z$-component of the Cooper pairs' spin angular momentum in this superconducting state is $-\hbar$. On the other hand, by using the Eq.~(\ref{basistrans}) from the main text, we can obtain the transformation of the basis function $k_{z}(\tutabf{x}-i\tutabf{y})$ under the spin-rotation operator $\hat{P}_{[R_{\bbb{z}}(\theta)||E]}$:
\begin{equation}\label{phasefactor}
\begin{aligned}
    \hat{P}_{[R_{\bbb{z}}(\theta)||E]} k_{z}(\tutabf{x}-i\tutabf{y})&=k_{z}[(\cos(\theta)\tutabf{x}+\sin(\theta)\tutabf{y}) -i(-\sin(\theta)\tutabf{x}+\cos(\theta)\tutabf{y})]\\&=e^{i\theta}k_{z}(\tutabf{x}-i\tutabf{y}).
\end{aligned}
\end{equation}
This analysis can also be applied to other coirreps with $\sigma=1$ of type-I collinear spin point groups.
Thus, the nontrivial phase factor $e^{i\theta}$ of the representation matrices of the group elements $[R_{\bbb{z}}(\theta)||E]$ indicates that the $z$-component of the Cooper pairs' spin angular momentum is $-\hbar$.

Finally, by combining the results for the $\sigma=0$ and $\sigma=1$ cases, we can derive that, for type-I collinear spin point groups, different $\sigma$ in the phase factors of the representation matrices of $[R_{\bbb{z}}(\theta)||E]$ correspond to different $z$-components of spin angular momentum of Cooper pairs in superconductivity. Simultaneously, $\sigma$ labels different superconducting channels. Thus, the basis functions representing superconducting states with Cooper pairs having different $z$-components of spin angular momentum belong to different superconducting channels.

\end{widetext}

\end{document}